\documentclass{lmcs} %%% last changed 2014-08-20
\pdfoutput=1
\usepackage[utf8]{inputenc}

\usepackage{lastpage}
\lmcsdoi{20}{2}{10}
\lmcsheading{}{\pageref{LastPage}}{}{}%
{Mar.~09,~2023}{May~20,~2024}{}


\usepackage{hyperref}
\usepackage{cleveref}

\usepackage{amsmath}
\usepackage{mathtools}
\usepackage{xspace}

\usepackage{amsfonts}
\usepackage{amssymb}
\usepackage{bm}
\usepackage[braket, qm]{qcircuit}
\usepackage{caption}
\usepackage{subcaption}
\usepackage{stmaryrd}
\usepackage{multicol}
\usepackage{pgfplots}
\usepackage{tikz}
\usepackage{circuitikz}
\usepackage{calc}
\usepackage{graphicx}
\usepackage{adjustbox}
\usepackage[T1]{fontenc}

\usetikzlibrary{decorations.markings}
\usetikzlibrary{matrix,backgrounds,folding}
\usetikzlibrary{shapes.geometric}
\usetikzlibrary{math} 
\usetikzlibrary{positioning}
\usetikzlibrary{arrows}
\usetikzlibrary{arrows.meta}
\usetikzlibrary{decorations.pathreplacing,calligraphy,backgrounds}
\usetikzlibrary{trees}
\usetikzlibrary{shapes}

\pgfdeclarelayer{edgelayer}
\pgfdeclarelayer{nodelayer}
\pgfsetlayers{background,edgelayer,nodelayer,main}

\newcommand{\tikzformula}[1]{
\scalebox{0.75}{
\input{./ZX_diagrams/#1.tikz}
}
}

\tikzstyle{every picture}=[baseline=-0.25em]
\tikzstyle{identity}=[rectangle,fill=white,dashed,draw=black,xscale=1,yscale=1,minimum width=0.3cm,minimum height=0.3cm]
\tikzstyle{zxnode}=[shape=circle, minimum width=.3cm, inner sep=0.5pt, font=\footnotesize, draw=black]
\tikzstyle{z_dot}=[zxnode ,fill=green!30]
\tikzstyle{x_dot}=[zxnode ,fill=red!30]
\tikzstyle{h_box}=[rectangle,fill=yellow!30,draw=black,xscale=1,yscale=1,font=\small,inner sep=0.75pt,minimum width=0.15cm,minimum height=0.15cm]

\tikzstyle{plus-edge}=[decoration={markings,mark=at position 0.7 with
    {\arrow[scale=2.2,>=stealth]{Triangle}}},postaction={decorate}]
\tikzstyle{brace} = [thick, decorate, 
decoration = {calligraphic brace, raise = 2pt, amplitude = 4pt} ]
\tikzstyle{underbrace} = [decoration={brace,mirror,raise=0.0cm}, decorate]
\tikzstyle{red_edge}=[draw=red] 
\tikzstyle{blue_edge}=[draw=blue] 
\tikzstyle{red_arrow}=[plus-edge, draw=red] 

\tikzstyle{classical}=[rectangle, draw, very thick, fill=green!30, text width=4.2cm, align=center]
\tikzstyle{quantum}=[rectangle, draw, very thick, fill=red!30, text width=4cm, align=center]
\tikzstyle{none}=[minimum width=.1cm, inner sep=0.5pt]
\tikzstyle{white_node}=[fill=white, minimum width=.1cm, inner sep=0.5pt]
\tikzstyle{diag}=[shape=circle, minimum width=.5cm, inner sep=0.5pt, font=\footnotesize, draw=black]
\tikzstyle{big_diag}=[rectangle, fill=white, draw=black, xscale=1, yscale=1, font=\small,inner sep=0.75pt, minimum width=2cm, minimum height=0.5cm]

\newcommand{\interp}[1]{\left\llbracket #1 \right\rrbracket}
\newcommand{\equal}[2][~~]{#1 \underset{\substack{#2}}{=} #1}
\newcommand{\controlizer}[1]{C\left(#1\right)}
\newcommand{\circled}[1]{\tikz[baseline=(char.base)]{\node[shape=circle,draw,inner sep=2pt] (char) {#1};}}

\newcommand{\callrule}[2]{\hyperlink{r:#1}{\textnormal{(#2)}}\xspace}

\newcommand{\soo}{\callrule{rules}{S1}}
\newcommand{\stt}{\callrule{rules}{S2}}

\newcommand{\bo}{\callrule{rules}{B1}}
\newcommand{\bt}{\callrule{rules}{B2}}
\newcommand{\kt}{\callrule{rules}{K}}

\newcommand{\h}{\callrule{rules}{H}}
\newcommand{\supp}{\callrule{rules}{SUP}}
\newcommand{\gket}{
\begin{tikzpicture}
	\begin{pgfonlayer}{nodelayer}
		\node [style={z_dot}] (0) at (0, 0.25) {};
		\node [style=none] (1) at (0, 0) {};
	\end{pgfonlayer}
	\begin{pgfonlayer}{edgelayer}
		\draw (0) to (1.center);
	\end{pgfonlayer}
\end{tikzpicture}

}
\newcommand{\rbeta}{
\begin{tikzpicture}
	\begin{pgfonlayer}{nodelayer}
		\node [style={x_dot}] (0) at (0, 0.1) {$\beta$};
		\node [style=none] (1) at (0, -0.3) {};
	\end{pgfonlayer}
	\begin{pgfonlayer}{edgelayer}
		\draw (0) to (1.center);
	\end{pgfonlayer}
\end{tikzpicture}

}
\newcommand{\gbeta}{
\begin{tikzpicture}
	\begin{pgfonlayer}{nodelayer}
		\node [style={z_dot}] (0) at (0, 0.25) {$\beta$};
		\node [style=none] (1) at (0, -0.25) {};
	\end{pgfonlayer}
	\begin{pgfonlayer}{edgelayer}
		\draw (0) to (1.center);
	\end{pgfonlayer}
\end{tikzpicture}

}
\newcommand{\rminusbeta}{
\begin{tikzpicture}
	\begin{pgfonlayer}{nodelayer}
		\node [style={x_dot}] (0) at (0, 0.25) {$-\beta$};
		\node [style=none] (1) at (0, -0.25) {};
	\end{pgfonlayer}
	\begin{pgfonlayer}{edgelayer}
		\draw (0) to (1.center);
	\end{pgfonlayer}
\end{tikzpicture}

}
\newcommand{\gminusbeta}{
\begin{tikzpicture}
	\begin{pgfonlayer}{nodelayer}
		\node [style={z_dot}] (0) at (0, 0.25) {$-\beta$};
		\node [style=none] (1) at (0, -0.25) {};
	\end{pgfonlayer}
	\begin{pgfonlayer}{edgelayer}
		\draw (0) to (1.center);
	\end{pgfonlayer}
\end{tikzpicture}

}
\newcommand{\dtriangle}{
\begin{tikzpicture}
	\begin{pgfonlayer}{nodelayer}
		\node [style=none] (0) at (0, 0.5) {};
		\node [style=none] (2) at (-0.25, 0) {};
		\node [style=none] (3) at (0.25, 0) {};
		\node [style=none] (4) at (0, -0.5) {};
	\end{pgfonlayer}
	\begin{pgfonlayer}{edgelayer}
		\draw [style=plus-edge] (4.center) to (0.center);
	\end{pgfonlayer}
\end{tikzpicture}

}
\newcommand{\normalizer}{
\begin{tikzpicture}
	\begin{pgfonlayer}{nodelayer}
		\node [style={x_dot}] (6) at (0, 0.25) {};
		\node [style={z_dot}] (7) at (0, -0.25) {};
	\end{pgfonlayer}
	\begin{pgfonlayer}{edgelayer}
		\draw (6) to (7);
		\draw [bend right=45] (6) to (7);
		\draw [bend left=45] (6) to (7);
	\end{pgfonlayer}
\end{tikzpicture}

}

\newcommand{\diagi}{
\begin{tikzpicture}
	\begin{pgfonlayer}{nodelayer}
		\node [style={z_dot}] (0) at (0, 0.25) {\small$\pi$};
		\node [style={x_dot}] (1) at (0, -0.25) {\tiny$\frac{\pi}{2}$};
	\end{pgfonlayer}
	\begin{pgfonlayer}{edgelayer}
		\draw (0) to (1);
	\end{pgfonlayer}
\end{tikzpicture}

}
\newcommand{\minusi}{
\begin{tikzpicture}
	\begin{pgfonlayer}{nodelayer}
		\node [style={z_dot}] (0) at (0, 0.25) {\small$\pi$};
		\node [style={x_dot}] (1) at (0, -0.25) {\tiny$-\frac{\pi}{2}$};
	\end{pgfonlayer}
	\begin{pgfonlayer}{edgelayer}
		\draw (0) to (1);
	\end{pgfonlayer}
\end{tikzpicture}

}
\newcommand{\gdot}{
\begin{tikzpicture}
	\begin{pgfonlayer}{nodelayer}
		\node [style={z_dot}] (0) at (0, 0) {};
	\end{pgfonlayer}
\end{tikzpicture}

}
\newcommand{\ketzero}{
\begin{tikzpicture}
	\begin{pgfonlayer}{nodelayer}
		\node [style={x_dot}] (0) at (-0.5, 0.25) {};
		\node [style={z_dot}] (1) at (-0.5, -0.25) {};
		\node [style={x_dot}] (2) at (0, 0.25) {};
		\node [style=none] (3) at (0, -0.25) {};
	\end{pgfonlayer}
	\begin{pgfonlayer}{edgelayer}
		\draw (0) to (1);
		\draw [bend left=45] (0) to (1);
		\draw [bend right=45] (0) to (1);
		\draw (2) to (3.center);
	\end{pgfonlayer}
\end{tikzpicture}

}
\newcommand{\ketone}{
\begin{tikzpicture}
	\begin{pgfonlayer}{nodelayer}
		\node [style={x_dot}] (0) at (-0.5, 0.25) {};
		\node [style={z_dot}] (1) at (-0.5, -0.25) {};
		\node [style={x_dot}] (2) at (0, 0.25) {$\pi$};
		\node [style=none] (3) at (0, -0.25) {};
	\end{pgfonlayer}
	\begin{pgfonlayer}{edgelayer}
		\draw (0) to (1);
		\draw [bend left=45] (0) to (1);
		\draw [bend right=45] (0) to (1);
		\draw (2) to (3.center);
	\end{pgfonlayer}
\end{tikzpicture}

}

\DeclareMathOperator*{\Mat}{\mathcal{M}}
\DeclareMathOperator*{\ZX}{ZX}

\begin{document}

\title[Addition and Differentiation of ZX-diagrams]{Addition and Differentiation of ZX-diagrams}
\titlecomment{The work is an extended version of \cite{my_paper_2}}

\author[E.~Jeandel]{Emmanuel Jeandel\lmcsorcid{https://orcid.org/0000-0001-7236-2906}}[a, b, c, d]

\author[S.~Perdrix]{Simon Perdrix\lmcsorcid{https://orcid.org/0000-0002-1808-2409}}[a, b, c, d]

\author[M.~Veshchezerova]{Margarita Veshchezerova\lmcsorcid{https://orcid.org/0000-0001-6881-368X}}[e]

% affiliations
\address{LORIA, Campus scientifique, BP 239, 54506, Vandoeuvre-lès-Nancy Cedex, France}
\email{emmanuel.jeandel@loria.fr, simon.perdrix@loria.fr}
\address{CNRS, Délégation Centre-Est, 17 rue Notre-Dame des Pauvres, BP 10075, 54519, Vandoeuvre-lès-Nancy Cedex, France}
\address{Université de Lorraine, 34 cours Léopold, 54052, Nancy Cedex, France}
\address{Inria MOCQUA, 615 Rue du Jardin-Botanique, 54600 Villers-lès-Nancy, France}

\address{Terra Quantum AG, St. Gallerstrasse 16A, 9400 Rorschach, Switzerland}
\email{mv@terraquantum.swiss}

\begin{abstract}
The ZX-calculus is a powerful  framework  for reasoning in quantum computing. It provides in particular a compact representation of matrices of interest. A peculiar property of the ZX-calculus is the absence of a formal sum allowing the linear combinations of arbitrary ZX-diagrams. The universality of the formalism guarantees however that for any two ZX-diagrams, the sum of their interpretations can be represented by a ZX-diagram. We introduce a general, inductive definition of the addition of ZX-diagrams, relying on the construction of controlled diagrams. Based on this addition technique, we provide an inductive differentiation of ZX-diagrams.

Indeed, given a ZX-diagram with variables in the description of its angles, one can differentiate the diagram according to one of these variables. Differentiation is ubiquitous in quantum mechanics and quantum  computing (e.g. for solving optimization problems). Technically, the differentiation of ZX-diagrams is strongly related to summation as witnessed by the product rules. 

We also introduce an alternative, non-inductive, differentiation technique rather based on the isolation of the variables. Finally, we apply our results to deduce a diagram for an Ising Hamiltonian.
\end{abstract}

\maketitle

\section{Introduction}

ZX-calculus, originally introduced in \cite{zx_original}, is a \textit{graphical language} that allows reasoning about quantum computing. In this language complex computations on qubits are represented with \textit{diagrams} made out of \textit{elementary generators}\footnote{We use the TikZit software https://tikzit.github.io to draw ZX-diagrams}. Each diagram corresponds to a linear transformation between Hilbert spaces of qubit states. A compact set of \textit{rewrite rules} allows to transform diagrams into equivalent ones. The notable advantage of ZX-calculus compared to other representations (e.g. linear maps, circuits, and tensor networks) is that in this language the computations may be done \textit{entirely graphically}. A general introduction to the language alongside the overview of the main applications is available in \cite{zx_introduction}. 

Due to its flexibility, $\ZX$-calculus is widely used to address different problems of quantum computing. For instance, ZX-calculus allowed the derivation of important results in the field of \textit{measurement-based quantum computing (MBQC)} \cite{MBQC_circuit, MBQC_universality}. It was also successfully applied to circuit optimization \cite{phase_gadget, t_count, t_count_2, pivoting_simplification}, design and verification of  error-correction color codes \cite{color_code}, and to the analysis and compilation of \textit{surface codes} \cite{surface_code, lattice_surgery_1, fault_tolerant_compilation}.  

However, the applications of the ZX-calculus to the rapidly growing field of variational algorithms \cite{VQA} such as \textit{Quantum Approximate Optimization Algorithm (QAOA)} \cite{QAOA_original}, \textit{Variational Quantum Eigensolver (VQE)} \cite{VQE_original} and variational quantum machine learning are so far limited. Nevertheless, as variational algorithms do not require error correction, the incoming emergence of \textit{NISQ} devices makes them an object of particular attention \cite{Preskill_NISQ}. We believe that the reason why they are still unexplored with the means of ZX-calculus is the absence of a convenient way to \textbf{differentiate parametrized diagrams}. Indeed, the basic building blocks of variational algorithms are parametrized circuits, and the search for optimal parameter values is a crucial part of these algorithms. In theory, finding optimal parameters may be NP-hard \cite{Training_is_hard}. In practice, the search is usually done by classical numerical optimization methods and most of them use derivatives \cite{parameter_opt_compare}.

The main difficulty for differentiation of $\ZX$-diagrams comes from the \textit{product rule}: $\partial fg = \partial f g + f \partial g$ that involves adding two terms. This rule is crucial to evaluate the derivative of the sequential composition and the tensor product. As ZX-diagrams are defined inductively with these two compositions, the derivative of a complex diagram could be computed out of the derivatives of its parts if we were allowed to use the product rule, and therefore sum diagrams.

The works \cite{barren_plateau_zx, zx_differentation_toumi} use explicit sums of diagrams (usually referred to as \textit{bags of diagrams}) to represent the derivative of diagrams with multiple occurrences of the parameter. The major disadvantage of this approach is that there are no rules to manipulate sums of $\ZX$-diagrams. Therefore, we can't fully exploit the power of graphical computation (while it's still possible to use rewrite rules on summands \cite{tobias}). In this work, we suggest an approach where \textit{the derivative of a parametrized $\ZX$-diagram is another $\ZX$-diagram}. Hence we avoid the extension of the signature with formal sums. In order to tackle sums that appear in the product rule, we introduce an original technique to perform the addition of diagrams entirely in the $\ZX$-calculus. For this purpose, we use special diagrams called controlled states \cite{normal_form_nancy}. We suggest an inductive way to represent every $\ZX$-diagram by such a state. As we know how to sum controlled states \cite{normal_form_nancy} the addition for arbitrary diagrams follows. An inductive definition of the derivative is obtained by an explicit diagrammatic representation of the product rules. 

In an attempt to give a ready-to-use toolbox for differentiation, we provide an easy and convenient way to compute the derivative for the family of linear diagrams $\ZX(\beta)$ \cite{normal_form_nancy}, that is diagrams where angles might depend linearly (with integer coefficients) on one parameter $\beta$. Most of the circuits that are used for variational algorithms belong to $\ZX(\beta)$ and we believe that our formulas will make their analysis much simpler.

A definition for a derivative similar to our formulas was obtained in the independent work \cite{zx_differention_harny}. This work uses \textit{W-spiders} \cite{w_spider} to handle product rules. In contrast to our result, the diagrammatic differentiation presented in \cite{zx_differention_harny} maps to a ZX-diagram a diagram from another language called \textit{algebraic ZX-calculus} \cite{normal_form_algebraic}. The algebraic ZX-calculus is convenient to represent arbitrary complex numbers, therefore their differentiation procedure can handle more general families of diagrams than $\ZX(\bm{\beta})$. This advantage comes with the cost of abandoning the legacy of the vanilla ZX-calculus which is by far the most popular graphical calculus for quantum computing.

In the end, we show how our result together with the Stone's theorem \cite{Stone_theorem} allows finding a $\ZX$-diagram representing an \textit{Ising Hamiltonian}. In variational algorithms Ising Hamiltonians are typically used in the definition of the loss function for parameter optimization. In practical applications Hamiltonians are usually given as a sum of local terms $H = \sum_{i=1}^k H_i$
where each $H_i$ acts only on a limited number of qubits. Small individual terms $H_i$ can be relatively easily represented as diagrams. For instance, for the Hamiltonian $H = \sum_{u, v}Z_u Z_v$ diagrams for individual terms are trivial: $H_{u, v} = 
\begin{tikzpicture}
	\begin{pgfonlayer}{nodelayer}
		\node [style={x_dot}] (0) at (0, 0.25) {$\pi$};
		\node [style={x_dot}] (1) at (0, -0.25) {$\pi$};
		\node [style=none] (2) at (-0.5, 0.25) {};
		\node [style=none] (3) at (0.5, 0.25) {};
		\node [style=none] (4) at (-0.5, -0.25) {};
		\node [style=none] (5) at (0.5, -0.25) {};
	\end{pgfonlayer}
	\begin{pgfonlayer}{edgelayer}
		\draw (2.center) to (0);
		\draw (0) to (3.center);
		\draw (1) to (5.center);
		\draw (1) to (4.center);
	\end{pgfonlayer}
\end{tikzpicture}

$. However, in the traditional framework there is no graphical procedure that constructs the overall sum $H$ out of such elementary blocks. 

The ability to perform addition directly leads to a procedure for differentiation. On the other direction, with Stone's theorem some specific kinds of sums might be straightforwardly expressed as diagrams if we can compute derivatives. We demonstrate how to combine our formula for the derivatives and the theorem to find a diagrammatic representation for example Hamiltonian.

\paragraph*{Structure of the paper.} In \cref{section:ZX-calculus}, we give a brief introduction to the $\ZX$-calculus followed by some useful lemmas. In \cref{section:related_work}, we provide a general overview of addition and differentiation in quantum computing. In particular, we show how addition may be represented in circuit notation. We continue by giving an overview of the potential benefits coming from using $\ZX$-diagrams to reason about derivatives. This section, alongside additional examples and explanations, constitutes the major extension of the current work with respect to the previous version \cite{my_paper_2}.

In the next \cref{section:addition} we introduce an inductive procedure for the transformation of a diagram into a controlled form. We show how this procedure leads to an algorithm for the addition of $\ZX$-diagrams. In \cref{section:differentiation}, we provide two approaches for the differentiation, including an inductive definition (directly inspired by the product rules) and compact formulas for derivatives of linear diagrams $\ZX(\beta)$ where $\beta$ is a vector of parameters. In \cref{section:hamiltonian_diagram}, we show how to apply our formulas to obtain a diagram for an Ising Hamiltonian. 

\section{ZX-calculus}\label{section:ZX-calculus}

\subsection{Syntax and Semantics}

The ZX-diagrams %form a compact closed PROP \cite{} 
are generated by green spiders $
\input{./ZX_diagrams/generators/gdot-s.tikz}
$, red spiders $
\input{./ZX_diagrams/generators/rdot-s.tikz}
$ and Hadamard $
\begin{tikzpicture}
	\begin{pgfonlayer}{nodelayer}
		\node [style=none] (0) at (0, 0.25) {};
		\node [style=none] (1) at (0, -0.25) {};
		\node [style={h_box}] (2) at (0, 0) {};
	\end{pgfonlayer}
	\begin{pgfonlayer}{edgelayer}
		\draw (0.center) to (2);
		\draw (2) to (1.center);
	\end{pgfonlayer}
\end{tikzpicture}

$, where both kinds of spiders have an arbitrary number of inputs/outputs and are decorated with \textit{angles}. ZX-diagrams are also made of wires: the identity $
\begin{tikzpicture}
	\begin{pgfonlayer}{nodelayer}
		\node [style=none] (0) at (0, 0.25) {};
		\node [style=none] (1) at (0, -0.25) {};
	\end{pgfonlayer}
	\begin{pgfonlayer}{edgelayer}
		\draw (0.center) to (1.center);
	\end{pgfonlayer}
\end{tikzpicture}

$, the swap $\!\!\tikzformula{generators/swap-s}\!\!$ and also the possibility to bend wires with a cup $
\begin{tikzpicture}
	\begin{pgfonlayer}{nodelayer}
		\node [style=none] (1) at (-0.25, 0.175) {};
		\node [style=none] (3) at (0.25, 0.175) {};
	\end{pgfonlayer}
	\begin{pgfonlayer}{edgelayer}
		\draw [in=270, out=-90, looseness=2.00] (1.center) to (3.center);
	\end{pgfonlayer}
\end{tikzpicture}

$ and a cap $
\begin{tikzpicture}
	\begin{pgfonlayer}{nodelayer}
		\node [style=none] (1) at (-0.25, -0.1) {};
		\node [style=none] (3) at (0.25, -0.1) {};
	\end{pgfonlayer}
	\begin{pgfonlayer}{edgelayer}
		\draw [in=-270, out=90, looseness=2.00] (1.center) to (3.center);
	\end{pgfonlayer}
\end{tikzpicture}

$. A wire can connect two nodes or, alternatively, have free ends that point either up or down. Wires with free ends pointing towards the top (towards the bottom) of the diagram are called \textit{inputs} (\textit{outputs}). We denote by $D: n \rightarrow m$ a diagram with $n$ inputs and $m$ outputs.  Finally, the empty diagram is denoted with $
\begin{tikzpicture}[scale=0.2]
	\begin{pgfonlayer}{nodelayer}
		\node [style=identity] (0) at (0, 0) {}; 
	\end{pgfonlayer}
\end{tikzpicture}

$. 

\begin{defi} %Given a group $\mathcal G$, 
ZX-diagrams are inductively defined as follows: for $n,m\in \mathbb N$ and $\alpha \in  \mathbb R/2\pi\mathbb Z$,%\mathcal G$,
\[\begin{array}{cclccclccclcccl}

\input{./ZX_diagrams/generators/gdot.tikz}
 &:& n\to m &\quad &
\input{./ZX_diagrams/generators/rdot.tikz}
 &:& n\to m &\quad&
\begin{tikzpicture}
	\begin{pgfonlayer}{nodelayer}
		\node [style=none] (0) at (0, 0.25) {};
		\node [style=none] (1) at (0, -0.25) {};
		\node [style={h_box}] (2) at (0, 0) {};
	\end{pgfonlayer}
	\begin{pgfonlayer}{edgelayer}
		\draw (0.center) to (2);
		\draw (2) to (1.center);
	\end{pgfonlayer}
\end{tikzpicture}

\!\!\!\!\!&:& 1\to 1 &\quad& 
\begin{tikzpicture}
	\begin{pgfonlayer}{nodelayer}
		\node [style=none] (0) at (0, 0.25) {};
		\node [style=none] (1) at (0, -0.25) {};
	\end{pgfonlayer}
	\begin{pgfonlayer}{edgelayer}
		\draw (0.center) to (1.center);
	\end{pgfonlayer}
\end{tikzpicture}

 &:&1 \to 1\\[0.6cm]

\begin{tikzpicture}
	\begin{pgfonlayer}{nodelayer}
		\node [style=none] (1) at (-0.25, 0.175) {};
		\node [style=none] (3) at (0.25, 0.175) {};
	\end{pgfonlayer}
	\begin{pgfonlayer}{edgelayer}
		\draw [in=270, out=-90, looseness=2.00] (1.center) to (3.center);
	\end{pgfonlayer}
\end{tikzpicture}

~&:&2\to 0 &\quad &
\begin{tikzpicture}
	\begin{pgfonlayer}{nodelayer}
		\node [style=none] (1) at (-0.25, -0.1) {};
		\node [style=none] (3) at (0.25, -0.1) {};
	\end{pgfonlayer}
	\begin{pgfonlayer}{edgelayer}
		\draw [in=-270, out=90, looseness=2.00] (1.center) to (3.center);
	\end{pgfonlayer}
\end{tikzpicture}

~&:&0\to 2 &\quad &\tikzformula{generators/swap-s}\!\!\!\!\!&:&2\to 2& \quad&   
\begin{tikzpicture}[scale=0.2]
	\begin{pgfonlayer}{nodelayer}
		\node [style=identity] (0) at (0, 0) {}; 
	\end{pgfonlayer}
\end{tikzpicture}

&:&0\to 0
\end{array}\]
are ZX-diagrams, and for any ZX-diagrams $D_0 :a \to b$, $D_1 :b\to c$, and $D_2:c\to d$, $D_1\circ D_0 : a \to c$  and $D_0\otimes D_2 : a+c\to b+d$ are ZX-diagrams. Pictorially: \vspace{0.1cm}

\centerline{
${
\input{./ZX_diagrams/calculus/D1.tikz}
}\circ {
\input{./ZX_diagrams/calculus/D0.tikz}
}={
\input{./ZX_diagrams/calculus/compD.tikz}
}\qquad$ and $\qquad  {
\input{./ZX_diagrams/calculus/D0.tikz}
}\otimes {
\input{./ZX_diagrams/calculus/D2.tikz}
}={
\input{./ZX_diagrams/calculus/tensorD.tikz}
}$\vspace{0.1cm}}

\end{defi}

As in most works on ZX-calculus, an empty spider denotes the spider with $\alpha = 0$.
A diagram with no input/output  is called a \emph{scalar}. To compactly write scalar factors, we introduce syntactic sugar $[-]^{\otimes n}$. For any scalar $d:0\rightarrow 0$ the notation $d^{\otimes n}$ corresponds to $ \underbrace{d \otimes \dots \otimes d}_n$.

Semantically, ZX-diagrams % $D:n \rightarrow m$ 
are standardly interpreted as linear maps, and thus they can be used to represent quantum evolutions. 

\begin{defi}For any ZX-diagram $D:n\to m$, let $\interp D \in \mathcal M_{2^m,2^n}(\mathbb C)$ be inductively defined as: 
$\interp {D_1\circ D_0} = \interp {D_1}\circ \interp{D_0}$, $\interp {D_0\otimes D_2} = \interp {D_0}\otimes  \interp{D_2}$,  and 
\begin{align*}
&\interp{
\input{./ZX_diagrams/generators/gdot.tikz}
} = |0\rangle^{\otimes m}\langle0|^{\otimes n} + e^{i\alpha} |1\rangle^{\otimes m}\langle1|^{\otimes n},  \interp{
\input{./ZX_diagrams/generators/rdot.tikz}
} =|+\rangle^{\otimes m}\langle+|^{\otimes n} + e^{i\alpha} |-\rangle^{\otimes m}\langle-|^{\otimes n}\\
&\interp{~
\begin{tikzpicture}
	\begin{pgfonlayer}{nodelayer}
		\node [style=none] (0) at (0, 0.25) {};
		\node [style=none] (1) at (0, -0.25) {};
	\end{pgfonlayer}
	\begin{pgfonlayer}{edgelayer}
		\draw (0.center) to (1.center);
	\end{pgfonlayer}
\end{tikzpicture}

~} = |0\rangle\langle 0| + |1\rangle \langle 1|, \quad \interp{
\begin{tikzpicture}
	\begin{pgfonlayer}{nodelayer}
		\node [style=none] (0) at (0, 0.25) {};
		\node [style=none] (1) at (0, -0.25) {};
		\node [style={h_box}] (2) at (0, 0) {};
	\end{pgfonlayer}
	\begin{pgfonlayer}{edgelayer}
		\draw (0.center) to (2);
		\draw (2) to (1.center);
	\end{pgfonlayer}
\end{tikzpicture}

} = |+\rangle\langle 0| + |-\rangle \langle 1| \nonumber, \quad \interp{
\begin{tikzpicture}[scale=0.2]
	\begin{pgfonlayer}{nodelayer}
		\node [style=identity] (0) at (0, 0) {}; 
	\end{pgfonlayer}
\end{tikzpicture}

} = 1 \nonumber  \\
&\interp{
\begin{tikzpicture}
	\begin{pgfonlayer}{nodelayer}
		\node [style=none] (1) at (-0.25, 0.175) {};
		\node [style=none] (3) at (0.25, 0.175) {};
	\end{pgfonlayer}
	\begin{pgfonlayer}{edgelayer}
		\draw [in=270, out=-90, looseness=2.00] (1.center) to (3.center);
	\end{pgfonlayer}
\end{tikzpicture}

} = \langle 00| + \langle 11|, \quad
\interp{
\begin{tikzpicture}
	\begin{pgfonlayer}{nodelayer}
		\node [style=none] (1) at (-0.25, -0.1) {};
		\node [style=none] (3) at (0.25, -0.1) {};
	\end{pgfonlayer}
	\begin{pgfonlayer}{edgelayer}
		\draw [in=-270, out=90, looseness=2.00] (1.center) to (3.center);
	\end{pgfonlayer}
\end{tikzpicture}

} = |00\rangle + |11\rangle, \quad
\interp{\tikzformula{generators/swap-s}} = \sum_{i, j \in \{0, 1\}} |ij\rangle\langle ji| \nonumber
\end{align*}
where \emph{bra-ket} notations are used: $|0\rangle = {1\choose 0}$, $|1\rangle = {0\choose 1}$, $|+\rangle = \frac{|0\rangle + |1\rangle}{\sqrt{2}}$, $|-\rangle = \frac{|0\rangle - |1\rangle}{\sqrt{2}}$, $|xy\rangle = |x\rangle\otimes|y\rangle$ and $\langle x|= |x\rangle^\dagger$ is the adjoint (complex conjugate) of $|x\rangle$. 
\end{defi}

\begin{exa}\label{lemma:zx_scalars} Complex numbers can be represented with diagrams:
\begin{align}
\interp{
\begin{tikzpicture}
	\begin{pgfonlayer}{nodelayer}
		\node [style={z_dot}] (0) at (0, 0) {};
	\end{pgfonlayer}
\end{tikzpicture}

} = 2, 
\qquad \interp{
\begin{tikzpicture}
	\begin{pgfonlayer}{nodelayer}
		\node [style={z_dot}] (0) at (0, 0) {$\alpha$};
	\end{pgfonlayer}
\end{tikzpicture}

} = 1 + e^{i\alpha}, 
\qquad \interp{
\begin{tikzpicture}
	\begin{pgfonlayer}{nodelayer}
		\node [style={x_dot}] (0) at (0, 0.25) {};
		\node [style={z_dot}] (1) at (0, -0.25) {};
	\end{pgfonlayer}
	\begin{pgfonlayer}{edgelayer}
		\draw (0) to (1);
	\end{pgfonlayer}
\end{tikzpicture}

} = \sqrt{2},
\qquad \interp{
\begin{tikzpicture}
	\begin{pgfonlayer}{nodelayer}
		\node [style={x_dot}] (0) at (0, 0.25) {$\pi$};
		\node [style={z_dot}] (1) at (0, -0.25) {$\alpha$};
	\end{pgfonlayer}
	\begin{pgfonlayer}{edgelayer}
		\draw (0) to (1);
	\end{pgfonlayer}
\end{tikzpicture}

} = \sqrt{2} e^{i\alpha},
\qquad \interp{
\begin{tikzpicture}
	\begin{pgfonlayer}{nodelayer}
		\node [style={x_dot}] (6) at (0, 0.5) {};
		\node [style={z_dot}] (7) at (0, -0.25) {};
	\end{pgfonlayer}
	\begin{pgfonlayer}{edgelayer}
		\draw (6) to (7);
		\draw [bend right=45] (6) to (7);
		\draw [bend left=45] (6) to (7);
	\end{pgfonlayer}
\end{tikzpicture}

}  = \frac{1}{\sqrt{2}} \label{eq:scalar_diagrams}
\end{align}
\end{exa}

%\subsection{Variables and constants}
Sometimes it is meaningful to consider diagrams with angles from a restricted sub-group $\mathcal{G}$ of $\mathbb{R}/2\pi\mathbb{Z}$. Such restrictions lead to \textit{fragments} of the language, denoted $\ZX_{\mathcal G}$-calculus \cite{normal_form_nancy}. 
The standard interpretation associates to each $\ZX_{\mathcal G}$-diagram $D:n \rightarrow m$ a matrix $\interp{D}\in\mathcal M_{2^m,2^n}(\mathcal R_{\mathcal G})$ with elements in the ring $\mathcal{R}_{\mathcal{G}} = \mathbb{Z}\left[\frac{1}{\sqrt{2}}, e^{i\mathcal{G}}\right]$ - the smallest ring that contains $\mathbb{Z}$, $\frac{1}{\sqrt{2}}$ and $\{e^{ia}| a\in \mathcal{G}\}$ \cite{normal_form_nancy}. 
 
In particular, the $\frac \pi 2$- (resp. $\pi$-) fragment\footnote{I.e. the fragment of diagrams which angles are in the group generated by $\frac \pi 2$ (resp. $\pi$)}, also called Clifford (resp. real Clifford) fragment, enjoys nice properties \cite{clifford_completeness_Backens, real_clifford_completeness} but is not universal for quantum computing, even approximately. Furthermore, any quantum computation that can be expressed in this fragment can be efficiently simulated on a classical  computer. As soon as the group contains the angle $\frac \pi 4$, the corresponding fragment is approximatively universal  for quantum computing: any $2^n\times 2^n$ unitary transformation can be approximated by a ZX-diagram from this fragment with arbitrary precision. In particular the $\frac \pi 4$-fragment, also called `Clifford+T' fragment has be extensively studied \cite{clifford_t_completeness_nancy, jeandel2017cyclotomic, clifford_t_completeness_oxford}. Other finitely generated fragments have been considered in \cite{normal_form_nancy}.
 
Notice that for any sub-group $\mathcal G$  of $\mathbb{R}/2\pi\mathbb{Z}$ that contains $\frac \pi 4$, $\ZX_{\mathcal G}$-diagrams are \emph{universal} \cite{normal_form_nancy} in the sense that for any matrix $M\in \mathcal M_{2^m,2^n}(\mathcal R_{\mathcal G})$ there exists a $\ZX_{\mathcal G}$-diagram $D:n\to m$ such that $\interp D = M$. 

\subsection{The calculus}

\begin{figure*}[!htb]
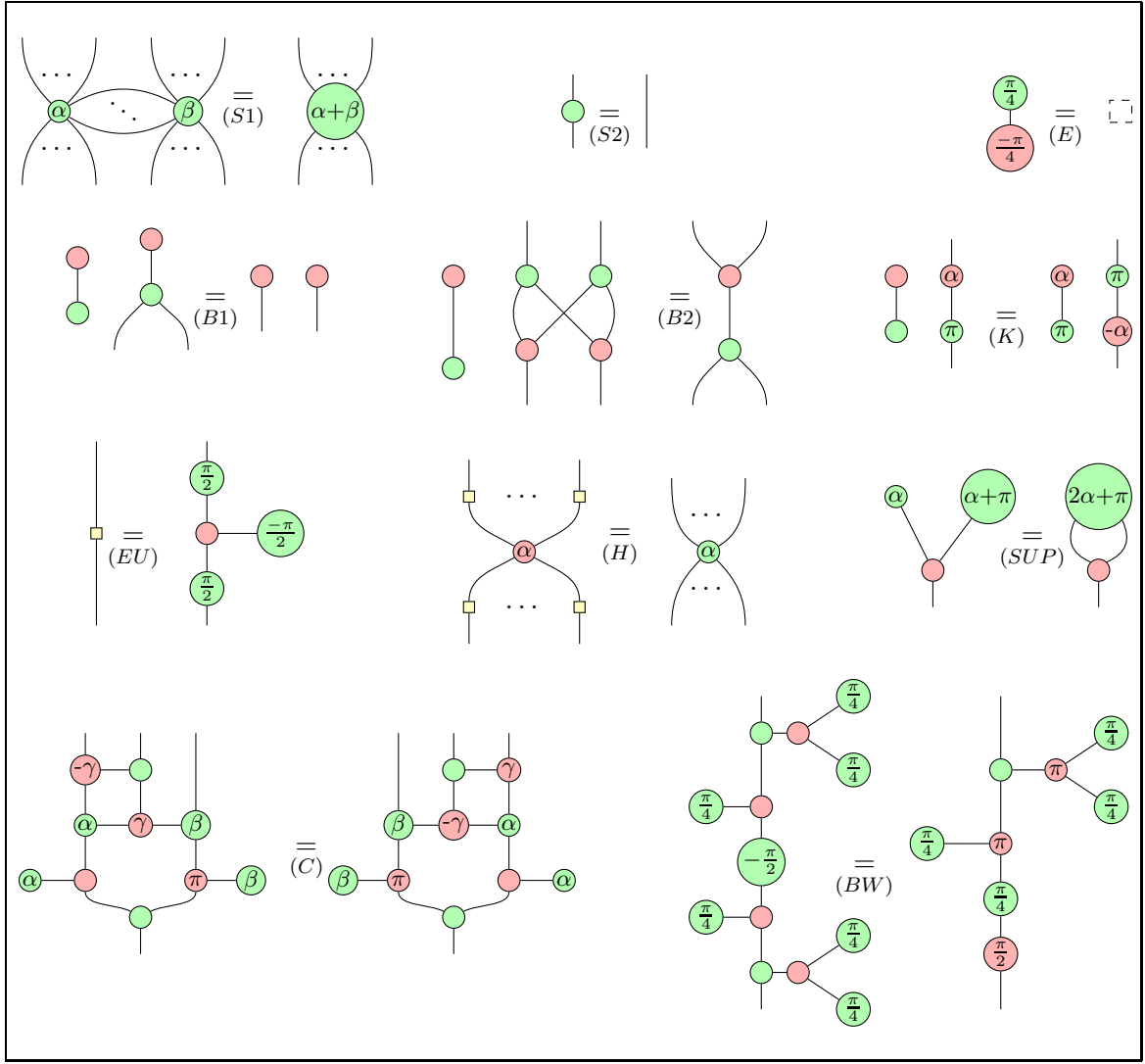

 \centering
 \hypertarget{r:rules}{}
 \begin{tabular}{|c@{$\qquad\quad~$}c@{$\quad\quad~$}c@{~}|}
   \hline
   && \\
   $
\input{./ZX_diagrams/axioms/s1.tikz}
$ & $
\input{./ZX_diagrams/axioms/s2.tikz}
$ & $
\input{./ZX_diagrams/axioms/e.tikz}
$ \\
   && \\
   $
\input{./ZX_diagrams/axioms/b1.tikz}
$ & $
\input{./ZX_diagrams/axioms/b2.tikz}
$ & $
\input{./ZX_diagrams/axioms/k.tikz}
$ \\
   &&\\
   $
\input{./ZX_diagrams/axioms/eu.tikz}
$ & $
\input{./ZX_diagrams/axioms/h.tikz}
$ & $
\input{./ZX_diagrams/axioms/sup.tikz}
$ \\
   && \\
   \multicolumn{3}{|c|}{ $
\input{./ZX_diagrams/axioms/c.tikz}
 \quad\qquad 
\input{./ZX_diagrams/axioms/bw.tikz}
$}\\
   && \\
   \hline
  \end{tabular}
 \caption[]{ Axioms for ZX as presented in \cite{normal_form_nancy}. All rules stay true flipped upside down and with inverted colors. Families of equations are given using `dots': $\dots$ means any number of wires, $\ddots$ means at least one wire.}
 \label{fig:ZX_rules}
\end{figure*}

Two ZX-diagrams may have the same interpretation, as a consequence the language is equipped with a set of rewrite rules (\cref{fig:ZX_rules}) that allows to transform diagrams.

In addition,  ZX-diagrams can be deformed at will: all wires may be bent in any manner that keeps intact the order of inputs and outputs. It is also allowed to arbitrarily change the order of wires for greed and red spiders and the Hadamard. Corresponding transformation rules are aggregated under the paradigm \textit{Only topology matters}:

\vspace{0.3cm}
\centerline{$
\input{./ZX_diagrams/bent-wire-1.tikz}
$}

\vspace{0.3cm}
\centerline{$
\input{./ZX_diagrams/bent-wire-2.tikz}
$}

As a direct consequence of the axiom \soo , these rules also hold for non-empty spiders. They also imply that whenever a wire has no free ends, we can draw it in any way, f.e. as a horizontal line.

We denote $\ZX \vdash D_1 = D_2$ if $D_1$ may be transformed to $D_2$ by local application of rewriting rules.

The ZX-calculus is sound, i.e. the rules preserve the semantics: if $\ZX \vdash D_1 = D_2$ then $\interp {D_1} = \interp{D_2}$. The converse property is called completeness. The set of rules (\cref{fig:ZX_rules}) was proven complete for the $\frac \pi 4$-fragment \cite{normal_form_nancy}, and a single extra-rule makes the language complete for arbitrary diagrams \cite{completeness_nancy}. Notice that alternative sets of rules have been shown to be complete for general ZX-diagrams \cite{completeness_oxford, vilmart_optimal_axioms}. We choose to consider the rules of \cref{fig:ZX_rules} as they have been used to study diagrams with parameters in \cite{completeness_nancy}.

\subsection{Useful lemmas}

Theorems and demonstrations in this work extensively use triangle: $ 
\begin{tikzpicture}
	\begin{pgfonlayer}{nodelayer}
		\node [style=none] (0) at (0, 0.5) {};
		\node [style=none] (2) at (-0.25, 0) {};
		\node [style=none] (3) at (0.25, 0) {};
		\node [style=none] (4) at (0, -0.5) {};
	\end{pgfonlayer}
	\begin{pgfonlayer}{edgelayer}
		\draw [style=plus-edge] (4.center) to (0.center);
	\end{pgfonlayer}
\end{tikzpicture}

$ - a syntactic sugar introduced in \cite{clifford_t_completeness_nancy}. It corresponds to a non-unitary transformation: $\interp{
\begin{tikzpicture}
	\begin{pgfonlayer}{nodelayer}
		\node [style=none] (0) at (0, 0.5) {};
		\node [style=none] (2) at (-0.25, 0) {};
		\node [style=none] (3) at (0.25, 0) {};
		\node [style=none] (4) at (0, -0.5) {};
	\end{pgfonlayer}
	\begin{pgfonlayer}{edgelayer}
		\draw [style=plus-edge] (4.center) to (0.center);
	\end{pgfonlayer}
\end{tikzpicture}

} = |0\rangle\langle 0| + |0\rangle\langle 1| + |1\rangle \langle 1|$. 
The triangle may be written in terms of red and green spiders as: 

\begin{align}
&
\begin{tikzpicture}
	\begin{pgfonlayer}{nodelayer}
		\node [style=none] (0) at (0, 0.5) {};
		\node [style=none] (2) at (-0.25, 0) {};
		\node [style=none] (3) at (0.25, 0) {};
		\node [style=none] (4) at (0, -0.5) {};
	\end{pgfonlayer}
	\begin{pgfonlayer}{edgelayer}
		\draw [style=plus-edge] (4.center) to (0.center);
	\end{pgfonlayer}
\end{tikzpicture}

 = 
\input{./ZX_diagrams/generators/triangle_in_spiders.tikz}
 \label{d:triangle}
\end{align}

We also use multiple lemmas introduced in the work \cite{normal_form_nancy}:

\begin{multicols}{3}

\begin{lem}\label{lemma:sqrt-2-sqrt-1-over-2}
$$
\input{./ZX_diagrams/old_lemmas/sqrt-2-sqrt-1-over-2.tikz}
$$
\end{lem}

\begin{lem}\label{lemma:phase-scalars}
$$
\input{./ZX_diagrams/old_lemmas/phase-scalars.tikz}
$$
\end{lem}

\begin{lem}\label{lemma:pi-push}
$$
\input{./ZX_diagrams/old_lemmas/pi-push.tikz}
$$
\end{lem}

\begin{lem}\label{lemma:zero-triangle-down}
$$
\input{./ZX_diagrams/old_lemmas/zero-triangle-down.tikz}
$$
\end{lem}

\begin{lem}\label{lemma:zero-triangle-up}
$$
\input{./ZX_diagrams/old_lemmas/zero-triangle-up.tikz}
$$
\end{lem}

\begin{lem}\label{lemma:pi-triangle-down}
$$
\input{./ZX_diagrams/old_lemmas/pi-triangle-down.tikz}
$$
\end{lem}

\begin{lem}\label{lemma:pi-triangle-up}
$$
\input{./ZX_diagrams/old_lemmas/pi-triangle-up.tikz}
$$
\end{lem}

\begin{lem}\label{lemma:pi-triangle-push}
$$
\input{./ZX_diagrams/old_lemmas/pi-triangle-push.tikz}
$$
\end{lem}

\begin{lem}\label{lemma:cs-plus-commutation}
$$
\input{./ZX_diagrams/old_lemmas/cs-plus-commutation.tikz}
$$
\end{lem}

\begin{lem}\label{lemma:cs-two-push}
$$
\input{./ZX_diagrams/old_lemmas/cs-two-push.tikz}
$$
\end{lem}

\begin{lem}\label{lemma:green-dots-with-opposed-triangles}
$$
\input{./ZX_diagrams/old_lemmas/green-dots-with-opposed-triangles.tikz}
$$
\end{lem}

\begin{lem}\label{lemma:h-link-on-triangle}
$$
\input{./ZX_diagrams/old_lemmas/h-link-on-triangle.tikz}
$$
\end{lem}

\begin{lem}\label{lemma:opposite-triangles}
$$
\input{./ZX_diagrams/old_lemmas/opposite-triangles.tikz}
$$
\end{lem}

\begin{lem}\label{lemma:opposite-triangles-pi}
$$
\input{./ZX_diagrams/old_lemmas/opposite-triangles-pi.tikz}
$$
\end{lem}

\begin{lem}\label{lemma:same-direction-triangles-link-removal}
$$
\input{./ZX_diagrams/old_lemmas/same-direction-triangles-link-removal.tikz}
$$
and
$$
\input{./ZX_diagrams/old_lemmas/same-direction-triangles-link-removal-2.tikz}
$$
\end{lem}

\begin{lem}\label{lemma:tranzistor-swap-legs}
$$
\input{./ZX_diagrams/old_lemmas/tranzistor-swap-legs.tikz}
$$
\end{lem}

\begin{lem}\label{lemma:triangle-and-extra-link}
$$
\input{./ZX_diagrams/old_lemmas/triangle-and-extra-link.tikz}
$$
\end{lem}

\begin{lem}\label{lemma:two-pi-triangle-is-identity}
$$
\input{./ZX_diagrams/old_lemmas/two-pi-triangle-is-identity.tikz}
$$
\end{lem}

\end{multicols}

\begin{lem}\label{lemma:triangle-cycle}
$$
\input{./ZX_diagrams/old_lemmas/triangle-cycle.tikz}
$$
\end{lem}

We also introduce some new lemmas that will be useful in the following sections. The proofs serve as an illustration of the graphical computation process.

\begin{multicols}{2}

\begin{lem}\label{lemma:derivative-base-induction}
$$
\begin{tikzpicture}
	\begin{pgfonlayer}{nodelayer}
		\node [style={x_dot}] (24) at (-2.75, 0.5) {\small$\pi$};
		\node [style={z_dot}] (25) at (-1.75, 0.5) {\small$\pi$};
		\node [style={z_dot}] (26) at (-0.75, 0.5) {\small$\pi$};
		\node [style={x_dot}] (33) at (-1.25, 0) {\small $\beta$};
		\node [style=none] (35) at (-1.25, -0.5) {};
		\node [style={x_dot}] (55) at (-0.25, 0.5) {};
		\node [style=none] (56) at (0, 0) {$=$};
		\node [style={z_dot}] (58) at (1, 0.5) {\small$\pi$};
		\node [style={x_dot}] (59) at (1, 0) {\small $\beta$};
		\node [style=none] (60) at (1, -0.5) {};
		\node [style={z_dot}] (61) at (0.5, 0.5) {};
		\node [style={x_dot}] (62) at (0.5, -0.25) {};
	\end{pgfonlayer}
	\begin{pgfonlayer}{edgelayer}
		\draw [style=plus-edge] (25) to (24);
		\draw [style=plus-edge] (26) to (25);
		\draw (25) to (33);
		\draw (33) to (26);
		\draw (33) to (35.center);
		\draw (26) to (55);
		\draw (58) to (59);
		\draw (59) to (60.center);
		\draw (61) to (62);
	\end{pgfonlayer}
\end{tikzpicture}

$$
\end{lem}

\begin{lem}\label{lemma:factor-beta}
$$
\begin{tikzpicture}
	\begin{pgfonlayer}{nodelayer}
		\node [style=none] (126) at (0, 0) {$=$};
		\node [style={x_dot}] (129) at (-1.5, 0) {$\beta$};
		\node [style={z_dot}] (133) at (-1.5, -0.5) {};
		\node [style={x_dot}] (134) at (-0.5, -0.5) {$\pi$};
		\node [style=none] (135) at (-1.5, -1) {};
		\node [style={z_dot}] (141) at (0.5, -0.25) {};
		\node [style={x_dot}] (142) at (1.5, -0.25) {$\pi$};
		\node [style={x_dot}] (143) at (0.5, -0.75) {$\beta$};
		\node [style=none] (144) at (0.5, -1) {};
		\node [style=none] (145) at (-1, 1.25) {};
		\node [style=none] (172) at (-0.5, -1) {};
		\node [style=none] (173) at (1.5, -1) {};
		\node [style={z_dot}] (176) at (-1, 0.75) {};
		\node [style={x_dot}] (177) at (-0.5, 0) {$\beta$};
		\node [style=none] (179) at (1, 1.25) {};
		\node [style={z_dot}] (180) at (1, 0.75) {};
	\end{pgfonlayer}
	\begin{pgfonlayer}{edgelayer}
		\draw (129) to (133);
		\draw (133) to (135.center);
		\draw (133) to (134);
		\draw (141) to (143);
		\draw (141) to (142);
		\draw (143) to (144.center);
		\draw (134) to (172.center);
		\draw (142) to (173.center);
		\draw [style=plus-edge] (129) to (176);
		\draw (179.center) to (180);
		\draw (145.center) to (176);
		\draw [style=plus-edge] (176) to (177);
		\draw (177) to (134);
		\draw [style=plus-edge] (141) to (180);
		\draw [style=plus-edge] (180) to (142);
	\end{pgfonlayer}
\end{tikzpicture}

$$
\end{lem}

\end{multicols}

\begin{lem}\label{lemma:change_direction_y1}
$$
\begin{tikzpicture}
	\begin{pgfonlayer}{nodelayer}
		\node [style={z_dot}] (0) at (-2.75, 0.5) {$\pi$};
		\node [style={x_dot}] (1) at (-2.25, 0) {};
		\node [style=none] (3) at (-1.25, -0.5) {};
		\node [style={x_dot}] (7) at (-3.25, 0.5) {};
		\node [style={z_dot}] (10) at (-1.75, 0.5) {$\pi$};
		\node [style={x_dot}] (11) at (-1.25, 0) {};
		\node [style=none] (12) at (-2.25, -0.5) {};
		\node [style={z_dot}] (13) at (-0.75, 0.5) {};
		\node [style={x_dot}] (14) at (-0.25, 0.5) {$\pi$};
		\node [style={z_dot}] (15) at (0.75, 0.5) {};
		\node [style={x_dot}] (16) at (1.25, 0) {};
		\node [style=none] (17) at (1.25, -0.5) {};
		\node [style={x_dot}] (18) at (0.25, 0.5) {$\pi$};
		\node [style={z_dot}] (19) at (1.75, 0.5) {$\pi$};
		\node [style={x_dot}] (20) at (2.25, 0) {};
		\node [style=none] (21) at (2.25, -0.5) {};
		\node [style={z_dot}] (22) at (2.75, 0.5) {$\pi$};
		\node [style=none] (23) at (2.75, 1) {};
		\node [style=none] (24) at (-0.25, 1) {};
		\node [style=none] (25) at (0, 0) {$=$};
	\end{pgfonlayer}
	\begin{pgfonlayer}{edgelayer}
		\draw (0) to (1);
		\draw (0) to (7);
		\draw [in=-180, out=0, looseness=1.25] (1) to (3.center);
		\draw (10) to (11);
		\draw [style=plus-edge, in=-180, out=0] (10) to (13);
		\draw (13) to (14);
		\draw (13) to (11);
		\draw [in=0, out=-180, looseness=1.25] (11) to (12.center);
		\draw [style=plus-edge, in=-180, out=0] (0) to (10);
		\draw (10) to (1);
		\draw (16) to (15);
		\draw (18) to (15);
		\draw (17.center) to (16);
		\draw (20) to (19);
		\draw [style=plus-edge, in=0, out=180] (22) to (19);
		\draw (23.center) to (22);
		\draw (20) to (22);
		\draw (21.center) to (20);
		\draw [style=plus-edge, in=0, out=180] (19) to (15);
		\draw (16) to (19);
		\draw (24.center) to (14);
	\end{pgfonlayer}
\end{tikzpicture}

$$
\end{lem}

\begin{proof}[Proof (lemma \ref{lemma:derivative-base-induction})]
\begin{align}

\begin{tikzpicture}
	\begin{pgfonlayer}{nodelayer}
		\node [style={x_dot}] (24) at (-1.5, 0.5) {\small$\pi$};
		\node [style={z_dot}] (25) at (-0.5, 0.5) {\small$\pi$};
		\node [style={z_dot}] (26) at (0.5, 0.5) {\small$\pi$};
		\node [style={x_dot}] (33) at (0, 0) {\small $\beta$};
		\node [style=none] (35) at (0, -0.5) {};
		\node [style={x_dot}] (55) at (1, 0.5) {};
	\end{pgfonlayer}
	\begin{pgfonlayer}{edgelayer}
		\draw [style=plus-edge] (25) to (24);
		\draw [style=plus-edge] (26) to (25);
		\draw (25) to (33);
		\draw (33) to (26);
		\draw (33) to (35.center);
		\draw (26) to (55);
	\end{pgfonlayer}
\end{tikzpicture}

\equal{\bo \\ \ref{lemma:pi-triangle-up} \\ \ref{lemma:sqrt-2-sqrt-1-over-2} }

\begin{tikzpicture}
	\begin{pgfonlayer}{nodelayer}
		\node [style={z_dot}] (25) at (0, 0.5) {\small$\pi$};
		\node [style={x_dot}] (33) at (0, 0) {\small $\beta$};
		\node [style=none] (35) at (0, -0.5) {};
		\node [style={x_dot}] (55) at (0.75, 0.5) {};
		\node [style={x_dot}] (56) at (0.75, 0) {};
	\end{pgfonlayer}
	\begin{pgfonlayer}{edgelayer}
		\draw (25) to (33);
		\draw (33) to (35.center);
		\draw [style=plus-edge] (55) to (25);
		\draw (33) to (56);
	\end{pgfonlayer}
\end{tikzpicture}

\equal{\ref{lemma:zero-triangle-down}}

\begin{tikzpicture}
	\begin{pgfonlayer}{nodelayer}
		\node [style={z_dot}] (25) at (0, 0.5) {\small$\pi$};
		\node [style={x_dot}] (33) at (0, 0) {\small $\beta$};
		\node [style=none] (35) at (0, -0.5) {};
		\node [style={z_dot}] (36) at (0.5, 0.5) {};
		\node [style={x_dot}] (37) at (0.5, 0) {};
	\end{pgfonlayer}
	\begin{pgfonlayer}{edgelayer}
		\draw (25) to (33);
		\draw (33) to (35.center);
		\draw (36) to (37);
	\end{pgfonlayer}
\end{tikzpicture}

\end{align}
\qedhere
\end{proof}

\begin{proof}[Proof (lemma \ref{lemma:change_direction_y1})]
The left hand side is:
\begin{align}

\input{./ZX_diagrams/new_lemmas/change_direction_y1_proof/lhs1.tikz}
 
\equal{\bo \\ \ref{lemma:zero-triangle-down} }

\input{./ZX_diagrams/new_lemmas/change_direction_y1_proof/lhs2.tikz}

\end{align}
For the right hand side we get:
\begin{align}

\input{./ZX_diagrams/new_lemmas/change_direction_y1_proof/rhs1.tikz}
 
\equal{\bo \\ \ref{lemma:pi-triangle-up} } 

\input{./ZX_diagrams/new_lemmas/change_direction_y1_proof/rhs2.tikz}

\equal{\ref{lemma:pi-push} } 
\input{./ZX_diagrams/new_lemmas/change_direction_y1_proof/rhs3.tikz}
 
\equal{\ref{lemma:pi-push} \\ \ref{lemma:pi-triangle-push}}

\input{./ZX_diagrams/new_lemmas/change_direction_y1_proof/rhs4.tikz}

\equal{\ref{lemma:cs-plus-commutation}} 
\input{./ZX_diagrams/new_lemmas/change_direction_y1_proof/rhs5.tikz}

\end{align}
\qedhere
\end{proof}

\begin{proof}[Proof (lemma \ref{lemma:factor-beta})]

To prove the lemma we adopt a slightly different technique than a usual sequence of rewrites. The technique consists in verification that the diagrams from the left hand side and the right hand side lead to the same result when applied on a set of \textit{basis states} $\{|0\rangle, |1\rangle\}$ corresponding to diagrams $\ketzero$ and $\ketone$. As was pointed out in \cite{dodo_book} for two matrices $M_1, M_2: \mathbb{C}^{2^n} \rightarrow \mathbb{C}^{2^m}$ and a basis $\{|b_1\rangle, \dots, |b_{2^n}\rangle\}$ in $\mathbb{C}^{2^n}$ the equality on basis input is equivalent to the equality of matrices, i.e.
\begin{align}
\forall i \in [1, \dots, 2^n]: \; M_1|b_i\rangle = M_2|b_i\rangle \iff M_1 \equiv M_2
\end{align}
Therefore, by demonstrating that $
\begin{tikzpicture}
	\begin{pgfonlayer}{nodelayer}
		\node [style=none] (0) at (0.5, 0.5) {};
		\node [style=none] (1) at (0.5, -0.25) {};
		\node [style=none] (2) at (1, 0) {$rhs$};
		\node [style=none] (3) at (1.5, -0.25) {};
		\node [style=none] (4) at (1.5, 0.5) {};
		\node [style=none] (5) at (0.75, -0.25) {};
		\node [style=none] (6) at (1.25, -0.25) {};
		\node [style=none] (7) at (0.75, -0.75) {};
		\node [style=none] (8) at (1.25, -0.75) {};
		\node [style=none] (9) at (1, -0.5) {$\dots$};
		\node [style={x_dot}] (10) at (1, 1) {};
		\node [style=none] (11) at (1, 0.5) {};
		\node [style=none] (12) at (-1.5, 0.5) {};
		\node [style=none] (13) at (-1.5, -0.25) {};
		\node [style=none] (14) at (-1, 0) {$lhs$};
		\node [style=none] (15) at (-0.5, -0.25) {};
		\node [style=none] (16) at (-0.5, 0.5) {};
		\node [style=none] (17) at (-1.25, -0.25) {};
		\node [style=none] (18) at (-0.75, -0.25) {};
		\node [style=none] (19) at (-1.25, -0.75) {};
		\node [style=none] (20) at (-0.75, -0.75) {};
		\node [style=none] (21) at (-1, -0.5) {$\dots$};
		\node [style={x_dot}] (22) at (-1, 1) {};
		\node [style=none] (23) at (-1, 0.5) {};
		\node [style=none] (24) at (0, 0) {$=$};
	\end{pgfonlayer}
	\begin{pgfonlayer}{edgelayer}
		\draw (5.center) to (7.center);
		\draw (6.center) to (8.center);
		\draw (1.center) to (3.center);
		\draw (4.center) to (3.center);
		\draw (0.center) to (1.center);
		\draw (0.center) to (4.center);
		\draw (10) to (11.center);
		\draw (17.center) to (19.center);
		\draw (18.center) to (20.center);
		\draw (13.center) to (15.center);
		\draw (16.center) to (15.center);
		\draw (12.center) to (13.center);
		\draw (12.center) to (16.center);
		\draw (22) to (23.center);
	\end{pgfonlayer}
\end{tikzpicture}

$ and $
\begin{tikzpicture}
	\begin{pgfonlayer}{nodelayer}
		\node [style=none] (0) at (0.5, 0.5) {};
		\node [style=none] (1) at (0.5, -0.25) {};
		\node [style=none] (2) at (1, 0) {$rhs$};
		\node [style=none] (3) at (1.5, -0.25) {};
		\node [style=none] (4) at (1.5, 0.5) {};
		\node [style=none] (5) at (0.75, -0.25) {};
		\node [style=none] (6) at (1.25, -0.25) {};
		\node [style=none] (7) at (0.75, -0.75) {};
		\node [style=none] (8) at (1.25, -0.75) {};
		\node [style=none] (9) at (1, -0.5) {$\dots$};
		\node [style={x_dot}] (10) at (1, 1) {$\pi$};
		\node [style=none] (11) at (1, 0.5) {};
		\node [style=none] (12) at (-1.5, 0.5) {};
		\node [style=none] (13) at (-1.5, -0.25) {};
		\node [style=none] (14) at (-1, 0) {$lhs$};
		\node [style=none] (15) at (-0.5, -0.25) {};
		\node [style=none] (16) at (-0.5, 0.5) {};
		\node [style=none] (17) at (-1.25, -0.25) {};
		\node [style=none] (18) at (-0.75, -0.25) {};
		\node [style=none] (19) at (-1.25, -0.75) {};
		\node [style=none] (20) at (-0.75, -0.75) {};
		\node [style=none] (21) at (-1, -0.5) {$\dots$};
		\node [style={x_dot}] (22) at (-1, 1) {$\pi$};
		\node [style=none] (23) at (-1, 0.5) {};
		\node [style=none] (24) at (0, 0) {$=$};
	\end{pgfonlayer}
	\begin{pgfonlayer}{edgelayer}
		\draw (5.center) to (7.center);
		\draw (6.center) to (8.center);
		\draw (1.center) to (3.center);
		\draw (4.center) to (3.center);
		\draw (0.center) to (1.center);
		\draw (0.center) to (4.center);
		\draw (10) to (11.center);
		\draw (17.center) to (19.center);
		\draw (18.center) to (20.center);
		\draw (13.center) to (15.center);
		\draw (16.center) to (15.center);
		\draw (12.center) to (13.center);
		\draw (12.center) to (16.center);
		\draw (22) to (23.center);
	\end{pgfonlayer}
\end{tikzpicture}

$ we actually establish that $\interp{
\begin{tikzpicture}
	\begin{pgfonlayer}{nodelayer}
		\node [style=none] (0) at (-0.5, 0.5) {};
		\node [style=none] (1) at (-0.5, -0.25) {};
		\node [style=none] (2) at (0, 0) {$lhs$};
		\node [style=none] (3) at (0.5, -0.25) {};
		\node [style=none] (4) at (0.5, 0.5) {};
		\node [style=none] (5) at (-0.25, -0.25) {};
		\node [style=none] (6) at (0.25, -0.25) {};
		\node [style=none] (7) at (-0.25, -0.5) {};
		\node [style=none] (8) at (0.25, -0.5) {};
		\node [style=none] (9) at (0, -0.5) {$\dots$};
		\node [style=none] (10) at (0, 0.75) {};
		\node [style=none] (11) at (0, 0.5) {};
	\end{pgfonlayer}
	\begin{pgfonlayer}{edgelayer}
		\draw (5.center) to (7.center);
		\draw (6.center) to (8.center);
		\draw (1.center) to (3.center);
		\draw (4.center) to (3.center);
		\draw (0.center) to (1.center);
		\draw (0.center) to (4.center);
		\draw (10.center) to (11.center);
	\end{pgfonlayer}
\end{tikzpicture}

} = \interp{
\begin{tikzpicture}
	\begin{pgfonlayer}{nodelayer}
		\node [style=none] (0) at (-0.5, 0.5) {};
		\node [style=none] (1) at (-0.5, -0.25) {};
		\node [style=none] (2) at (0, 0) {$rhs$};
		\node [style=none] (3) at (0.5, -0.25) {};
		\node [style=none] (4) at (0.5, 0.5) {};
		\node [style=none] (5) at (-0.25, -0.25) {};
		\node [style=none] (6) at (0.25, -0.25) {};
		\node [style=none] (7) at (-0.25, -0.5) {};
		\node [style=none] (8) at (0.25, -0.5) {};
		\node [style=none] (9) at (0, -0.5) {$\dots$};
		\node [style=none] (10) at (0, 0.75) {};
		\node [style=none] (11) at (0, 0.5) {};
	\end{pgfonlayer}
	\begin{pgfonlayer}{edgelayer}
		\draw (5.center) to (7.center);
		\draw (6.center) to (8.center);
		\draw (1.center) to (3.center);
		\draw (4.center) to (3.center);
		\draw (0.center) to (1.center);
		\draw (0.center) to (4.center);
		\draw (10.center) to (11.center);
	\end{pgfonlayer}
\end{tikzpicture}

} $. 

\begin{align}

\input{./ZX_diagrams/new_lemmas/factor_beta_proof/lhs_zero_0.tikz}
 
\equal{ \bo \\ \ref{lemma:zero-triangle-down} \\ \ref{lemma:zero-triangle-up}}

\input{./ZX_diagrams/new_lemmas/factor_beta_proof/lhs_zero_1.tikz}

\equal{\bo}

\input{./ZX_diagrams/new_lemmas/factor_beta_proof/lhs_zero_2.tikz}
 
\equal{\bo}

\input{./ZX_diagrams/new_lemmas/factor_beta_proof/rhs_zero_1.tikz}
 
\equal{\bo \\ \ref{lemma:zero-triangle-down} \\ \ref{lemma:zero-triangle-up}}

\input{./ZX_diagrams/new_lemmas/factor_beta_proof/rhs_zero_0.tikz}

\end{align}
\begin{align}

\input{./ZX_diagrams/new_lemmas/factor_beta_proof/lhs_pi_0.tikz}
 
\equal{ \bo \\ \ref{lemma:pi-triangle-down} \\ \ref{lemma:pi-triangle-up}}

\input{./ZX_diagrams/new_lemmas/factor_beta_proof/lhs_pi_1.tikz}

\equal{\soo}

\begin{tikzpicture}
	\begin{pgfonlayer}{nodelayer}
		\node [style=none] (135) at (-0.25, -0.25) {};
		\node [style=none] (172) at (0.25, -0.25) {};
		\node [style={x_dot}] (177) at (0, 0.25) {$\beta$};
	\end{pgfonlayer}
	\begin{pgfonlayer}{edgelayer}
		\draw (177) to (172.center);
		\draw (177) to (135.center);
	\end{pgfonlayer}
\end{tikzpicture}

\equal{\stt}

\input{./ZX_diagrams/new_lemmas/factor_beta_proof/rhs_pi_1.tikz}
 
\equal{ \bo \\ \ref{lemma:pi-triangle-down} \\ \ref{lemma:pi-triangle-up}}

\input{./ZX_diagrams/new_lemmas/factor_beta_proof/rhs_pi_0.tikz}

\end{align}

The lemma condition follows from the completeness of $\ZX$-calculus.
\qedhere
\end{proof}

\subsection{Linear diagrams}

In ZX-diagrams the angles inside spiders are real numbers or elements of some group $\mathcal{G}$. However, in many equations such as \soo, \kt, \h, and \supp we used \textit{symbols} $\alpha$ and $\beta$ instead of numbers. The symbolic notation was extremely handy to define \textit{families of equalities}, i.e. equalities that hold for any real values of angles. By explicitly keeping symbols inside spiders we can draw \textit{parameterized diagrams}. A \textit{parameterized diagram} with $n$ different symbols can be interpreted as a function that associates to each element of the space $\mathbb{R}^n$ a well-defined ZX-diagram.

More precisely, we say that a diagram $D$ is parametrized by $\beta_1, \dots, \beta_k$ if its angles are some functions on $\beta_1, \dots, \beta_k$. We denote such a diagram by $D(\beta_1, \dots, \beta_k)$. It is possible to \textit{evaluate} a parametrized diagram $D(\bm{\beta})$, $\bm{\beta} \in \mathbb{R}^k$ in a point $\bm{\tilde{\beta}} \in \mathbb{R}^k$ by replacing every occurrence of $\beta_i$ with the respective value $\tilde{\beta}_i$. The result of the evaluation is a diagram $D\left(\bm{\tilde{\beta}}\right)$ from $\ZX_\mathbb{R}$.

Among parameterized diagrams, we distinguish a family of \textit{linear diagrams} denoted $\ZX(\bm{\beta})$:

\begin{defi}[Linear diagrams \cite{completeness_nancy}]\label{def:linear_diagrams}
A ZX-diagram is linear in $\beta_1, \dots, \beta_k$ with constants in $L \subset \mathbb{R}$ if it is generated by $
\input{./ZX_diagrams/generators/gdot-s.tikz}
$, $
\input{./ZX_diagrams/generators/rdot-s.tikz}
$, $
\begin{tikzpicture}
	\begin{pgfonlayer}{nodelayer}
		\node [style=none] (0) at (0, 0.25) {};
		\node [style=none] (1) at (0, -0.25) {};
		\node [style={h_box}] (2) at (0, 0) {};
	\end{pgfonlayer}
	\begin{pgfonlayer}{edgelayer}
		\draw (0.center) to (2);
		\draw (2) to (1.center);
	\end{pgfonlayer}
\end{tikzpicture}

$, $
\begin{tikzpicture}[scale=0.2]
	\begin{pgfonlayer}{nodelayer}
		\node [style=identity] (0) at (0, 0) {}; 
	\end{pgfonlayer}
\end{tikzpicture}

$, $
\begin{tikzpicture}
	\begin{pgfonlayer}{nodelayer}
		\node [style=none] (0) at (0, 0.25) {};
		\node [style=none] (1) at (0, -0.25) {};
	\end{pgfonlayer}
	\begin{pgfonlayer}{edgelayer}
		\draw (0.center) to (1.center);
	\end{pgfonlayer}
\end{tikzpicture}

$, $\!\!\tikzformula{generators/swap-s}\!\!$, $
\begin{tikzpicture}
	\begin{pgfonlayer}{nodelayer}
		\node [style=none] (1) at (-0.25, 0.175) {};
		\node [style=none] (3) at (0.25, 0.175) {};
	\end{pgfonlayer}
	\begin{pgfonlayer}{edgelayer}
		\draw [in=270, out=-90, looseness=2.00] (1.center) to (3.center);
	\end{pgfonlayer}
\end{tikzpicture}

$, $
\begin{tikzpicture}
	\begin{pgfonlayer}{nodelayer}
		\node [style=none] (1) at (-0.25, -0.1) {};
		\node [style=none] (3) at (0.25, -0.1) {};
	\end{pgfonlayer}
	\begin{pgfonlayer}{edgelayer}
		\draw [in=-270, out=90, looseness=2.00] (1.center) to (3.center);
	\end{pgfonlayer}
\end{tikzpicture}

$ combined by tensor product and composition such that $\alpha$ is of the form $\sum_{i}n_i \beta_i + c$ with $n_i \in \mathbb{Z}$ and $c \in L$.
\end{defi}
In other words, in a parameterized linear diagram $D = D(\beta_1, \dots, \beta_k)$ each angle $\alpha$ of a (red/green) spider is of the form  $\sum_{i}n_i \beta_i + c$ for some integer $n_i$ and $c \in L$

It was shown in \cite{completeness_nancy} that for $L = \{\frac{n\pi}{4}\}_{n \in \mathbb{Z}}$ the Clifford+T axiomatization (\cref{fig:ZX_rules}) is complete for linear diagrams.

The family of linear diagrams may appear restricted compared to $\ZX$-diagrams that allow angles from a more general class of functions. It is, however, sufficient for applications in variational quantum algorithms as they use circuits where parameters appear in a linear fashion \cite{Preskill_NISQ}. More importantly, for this family we are able to demonstrate simple formulas for the derivative. We believe that such formulas are not obtainable even for a slightly more general fragment $\ZX_{\mathcal{A}_n}$ where angles belong to the group of affine functions $\mathcal{A}_n = \{(\beta) \rightarrow c^T\beta + c_0|\xspace c \in \mathbb{R}^n, c_0 \in \mathbb{R}\}$ (i.e. where coefficients are \textit{real numbers}). Intuitively, the difficulty comes from the absence of a simple representation of a general matrix over real numbers in terms of spiders. This restriction is removed in the algebraic $\ZX$-calculus \cite{normal_form_algebraic} at the cost of an extended set of generators.

\section{Related work}\label{section:related_work}
In this section we provide a broader context for addition and differentiation in quantum computing. In particular, we discuss how the addition of two unitary matrices may appear in quantum applications and how it can be expressed in the circuit notation \cite{linear_combination}. We also review other attempts to manipulate sums of ZX-diagrams.

The differentiation in quantum computing appears in the context of variational algorithms. We review how derivatives of ZX-diagrams may be helpful for the analysis of these algorithms, in particular in the detection of \textit{barren plateaus} and in the design of \textit{parameter shift rules}. These ideas point out possible directions for future work. 

\subsection{Addition in quantum computing}

\paragraph{Addition with circuits}
In pure qubit quantum mechanics computations are performed by \textit{unitary transformations}. Unitary transformations form a group with respect to multiplication, but the sum of two unitary transformations $U_a$ and $U_b$ is not unitary. 

However, in some context it is meaningful to consider transformations of the form $\frac{1}{c}\left(U_a + U_b\right)$. For instance, the work \cite{linear_combination} suggests to approximate the evolution of a quantum state under some Hamiltonian by a linear combination of unitary operators. It was shown that such an approach typically outperforms the traditional technique for simulation based on \emph{Trotterization}.

The protocol for the addition of the unitaries $U_a$ and $U_b$ in \cite{linear_combination} is:

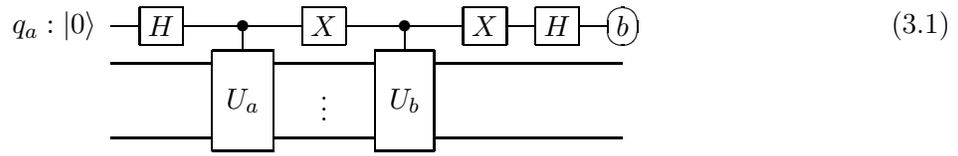
\begin{align}
\Qcircuit @C=1.0em @R=0.0em @!R {
 \lstick{q_a: |0\rangle } & \gate{H} & \ctrl{1} & \gate{X} & \ctrl{1} & \gate{X} & \gate{H} & \measure{b}\\
 & \qw & \multigate{2}{U_a} & \qw &  \multigate{2}{U_b} & \qw & \qw & \qw \\
  & & & \vdots & & & & \\
 & \qw & \ghost{U_a} & \qw & \ghost{U_b} & \qw & \qw & \qw 
}\label{circuit:unitary_sum}
\end{align}
where $q_a$ is an ancilla qubit initialized in a zero-state, and we have access to the controlled-version of the unitary gates $U_a$ and $U_b$.

The protocol (\ref{circuit:unitary_sum}) can be easily
verified. Starting in the state $|0\rangle|\psi\rangle$ we obtain:
\begin{align}
|0\rangle |\psi\rangle \rightarrow \frac{|0\rangle + |1\rangle}{\sqrt{2}} |\psi\rangle & \rightarrow  \frac{1}{\sqrt{2}} \left( |0\rangle \otimes U_b |\psi\rangle + |1\rangle \otimes U_a |\psi\rangle \right) \\
& \rightarrow \frac{1}{2} \left( |0\rangle \otimes ( U_b + U_a)|\psi\rangle + |1\rangle \otimes (U_b - U_a) |\psi\rangle \right)
\end{align}
If the ancilla qubit is measured in the $|0\rangle$ state, the input state $|\psi\rangle$ was successfully transformed to $\frac{1}{c}(U_a + U_b)|\psi\rangle$ ($c$ is the normalization constant). 

We remark that protocol (\ref{circuit:unitary_sum}) uses an ancilla qubit and post-selection on the measurement output and, therefore, exceeds the framework of pure qubit computations. We also emphasize that the circuit involves \textit{controlled versions} of unitary transformations. In order to be executed on a quantum computer, the controlled unitaries have to be compiled into a sequence of elementary gates. An efficient compilation is non-trivial even for a simple controlled-controlled-not gate usually called \textit{Toffoli gate}. Interestingly, the use of graphical representations may enhance the compilation of controlled gates \cite{Toffoli_t_count}.

\paragraph{Addition with the ZX-Calculus}

In the context of the ZX-calculus, there is no natural way to perform a linear combination of arbitrary ZX-diagrams. Fundamentally, this deficiency is due to the absence of a proper physical interpretation for a sum of diagrams. Indeed, ZX-calculus is a language for a process theory representation of quantum computing - i.e. each diagram is a \textit{process} \cite{dodo_book}. The composition of processes can be interpreted as their sequential application, the tensor product corresponds to the parallel application, but the addition doesn't have a meaningful physical interpretation.

Several attempts were made to address this issue. The first option selected in \cite{zx_differentation_toumi, tobias} consists in an extension of the formalism from diagrams to \textit{bags of diagrams}, i.e. formal sums of diagrams. The use of formal sums means computations and diagram simplifications cannot be done purely in the context of diagrams, but also require basic calculus.

Independently from our result,  \cite{zx_addition_oxford} suggested an alternative protocol for the addition of two ZX-diagrams.
This article uses the algebraic ZXW-calculus instead of vanilla ZX-calculus.

There are some similarities with our work, in the sense that their result can also be seen as based on a translation to a controlled form. However, their definition of a controlled form is not equivalent to ours. In addition, instead of inductive procedure, they rely on a decomposition of a diagram on elementary matrices that have an exponential complexity in the worst-case. A more detailed comparison of the two approaches is presented in \cref{subsection:addition}.

\subsection{Differentiation in variational algorithms}

In quantum computing a \textit{variational algorithm (VQA)} is defined by a \textit{parameterized quantum circuit (PQC)} (also called \textit{ansatz}) and a \textit{loss function} that helps to discriminate between the different values of the parameters. Usually, the \textit{loss function} is given as an expectation of an energy operator in the state prepared by PQC. Such expectation can be represented as a scalar $\ZX$-diagram with some angles depending on the parameters. 

Many state-of-the-art methods for the optimization of the loss function such as \textit{Quantum Analytic Descent} \cite{QAD} and \textit{meta-learning optimizers} \cite{Meta-learning-for-parameters} use derivatives \cite{parameter_opt_compare}. For efficient training it is important to detect deficiencies in the parameter landscape. In particular, gradient-based methods fail to coverge to an optimal solution in the presence of so-called \textit{barren plateaus} - zones with exponentially vanishing gradient values. It was shown that some specific structure of the \textit{cost function} \cite{Barren_plateau_nonlocal_cost} or, alternatively, the structure of the \textit{ansatz itself} \cite{Barren_plateaus_random} may lead to a landscape containing such zones. 

As was demonstrated in \cite{barren_plateau_zx}, the diagrammatic representation is beneficial in the analysis of the barren plateau phenomena. Yet the work \cite{barren_plateau_zx} pointed out a crucial obstacle limiting the application of ZX-calculus to variational algorithms - notably the absence of convenient tools for the \textit{addition} and \textit{differentiation} of parametrized diagrams. Due to this obstacle, in \cite{barren_plateau_zx} the analysis was limited to diagrams with only two occurrences of the parameter. In \cite{zx_differentation_toumi} an arbitrary number of parameter occurrences is tackled with explicitly represented product rules leading to \textit{bags of diagrams}. 

Other than in the analysis of the barren plateau, diagrammatic representation can be used to design \textit{parameter shift rules}. 

\paragraph{Parameter shift rules}

\textit{Parameter shift rules} appeared as a solution to the problem of gradient evaluation. Indeed, the gradient of the loss function $\frac{\partial \langle \psi_0 | U^\dagger(\bm{\alpha}) O U(\bm{\alpha})|\psi_0 \rangle }{\partial \alpha_i}$ (where $O$ is the energy operator) in general can't be directly evaluated on quantum hardware. However, in some cases the gradient can be expressed as a linear combination of the loss function values at different points:

\begin{align}
\frac{\partial L}{\partial \alpha_{i^*}}(\bm{\hat{\alpha}}) = \sum_{k=1}^m  \epsilon_k L(\bm{\alpha^k}) \label{eq:phase_shift_rappel}
\end{align}
where $ \alpha^k_i = \begin{cases} \hat{\alpha}_i , \quad & \forall i \neq {i^*} \\ \hat{\alpha}_i + \phi_i, \quad &i = i^*\end{cases}$. The values $L(\bm{\alpha^k})$ can be computed with stochastic approximation.

Decompositions of the form (\ref{eq:phase_shift_rappel}) are called \textit{parameter shift rules}. We remark that the parameter shift rules lead to an unbiased estimator for the gradient compared to the \textit{finite difference approximation} $\frac{\partial L}{\partial \alpha} (\alpha_0) = \frac{L(\alpha_0 + \delta) - L(\alpha_0)}{\delta}$ which, in turn, enhances the parameter optimization process \cite{finite_difference_vs_phase_shift}.

The first shift rule was derived for the case $U(\alpha)=W e^{i\alpha H} V$ where $W$ and $V$ don't depend on $\alpha$, while $H$ (called \textit{generator}) has two eigenvalues $+1$ and $-1$ \cite{phase_shift_original}. It was further extended to generators with arbitrary symmetric eigenvalues $+r$ and $-r$ \cite{phase_shift_basic}:

\begin{align}
\frac{\partial L}{\partial \alpha} (\alpha) = \frac{r}{4} \left( L\left(\alpha + \frac{\pi}{4r}\right)- L\left(\alpha - \frac{\pi}{4r}\right) \right)
\end{align}

For more complex $U(\alpha)$ we can decompose $U(\alpha)$ on a product of elementary matrices $ U(\alpha) = E_1(\alpha)\times \dots \times E_M(\alpha)$ and use product rules. Alternatively, more sophisticated techniques \cite{generalized_phase_shift_pennylane, generalized_phase_shift_kyriienko, generalized_phase_shift_algebraic} involve a \textit{spectral decomposition} of the generator $H$. These advanced rules imply a smaller number of elements in the linear combination (\ref{eq:phase_shift_rappel}) compared to the decomposition, leading to an economy in the number of calls made to the quantum computer.

We remark that simply having a ZX-diagram for the derivative of the loss function is not enough to design new shift rules. We need to further extend the toolbox of ZX-calculus, for instance, we may need a procedure that compiles a diagram into a linear combination of other diagrams corresponding to valid quantum expectations. This extension constitutes one of the possible directions for future work. 

\section{Addition of ZX-diagrams}\label{section:addition}

The addition is a natural operation on matrices from a Hilbert space. However, it doesn't correspond to a physical process, hence it is not reflected in the standard $\ZX$-calculus \cite{dodo_book}. 

It is not hard to show that it is not possible directly, from two diagrams $D_1$ and $D_2$ corresponding to two matrices $M_1$ and $M_2$, to obtain a diagram corresponding to the sum $M_1 + M_2$ by just plugging $D_1$ and $D_2$ into a bigger diagram. Indeed, any diagram that contains a subdiagram representing the $0$ matrix also represents a $0$ matrix.
So if $D_1$ represents a $0$ matrix and $D_2$ represents a nonzero matrix, then plugging $D_1$ and $D_2$ into a bigger diagram will give a zero matrix, and therefore will not represent the sum of the two matrices.

On the other hand, for any two diagrams $D_1,D_2 : n\to m$, the universality of the $\ZX$-calculus guarantees that there exists a diagram $D:n\to m$ such that $\interp D = \interp{D_1}+\interp {D_2}$. 

We provide in this section a general construction for such a diagram. 

The key ingredient is to use controlled versions of the original diagrams, rather than the original diagrams. As pointed out in \cite{normal_form_nancy}, one can inductively define the addition on 'controlled' versions of the diagrams. A controlled version of a diagram $D_0$ is roughly speaking a diagram with an extra input such that when this extra input is set to $|1\rangle$ the diagram behaves as $D_0$ and when it is set to $|0\rangle$ the diagram behaves as a neutral diagram. Roughly speaking, the addition procedure puts controlled versions of two diagrams aside and plugs a diagram corresponding to $const(|10\rangle + |01\rangle)$ to the inputs.

To build a controlled version of a diagram our construction uses \textit{controlled states} originally introduced in \cite{normal_form_nancy}.

\subsection{Controlled states}

\begin{defi}[Controlled state \cite{normal_form_nancy}]\label{def:controlled_state}
A $\ZX$-diagram $C:1 \rightarrow n$ is a controlled state if $\interp{C}|0\rangle = \sum_{x \in \{0,1\}^n}|x\rangle = \interp{\underbrace{\gket \dots \gket}_n}$.
\end{defi}

Intuitively, a controlled state is a way to encode the state $\interp{C}|1\rangle$. A controlled state with no outputs is called \textit{controlled scalar}. For instance, the scalar $0$ could be encoded by the controlled scalar $C_0 = 
\input{./ZX_diagrams/controlled_states/scalar_zero.tikz}
$. Indeed, we can explicitly check that $\interp{C_0}|0\rangle = 1$  and $\interp{C_0}|1\rangle = 0$:
\begin{align}
\interp{
\input{./ZX_diagrams/controlled_states/zero_lhs.tikz}
} \equal{\ref{lemma:sqrt-2-sqrt-1-over-2}} \interp{
\begin{tikzpicture}
	\begin{pgfonlayer}{nodelayer}
		\node [style=identity] (0) at (0, 0) {};
	\end{pgfonlayer}
\end{tikzpicture}

} = 1, \qquad
\interp{
\input{./ZX_diagrams/controlled_states/one_lhs.tikz}
} = \interp{\normalizer^{\otimes 2}} \times \interp{
\begin{tikzpicture}
	\begin{pgfonlayer}{nodelayer}
		\node [style={x_dot}] (3) at (0, 0) {$\pi$};
	\end{pgfonlayer}
\end{tikzpicture}

} \equal{\ref{eq:scalar_diagrams}} 0
\end{align}

It will be more useful in what follows to perform \textit{graphical proofs} instead of matrix computation to manipulate controlled states. With these rules, we verify the condition $\interp{C}|0\rangle = \sum_{x \in \{0,1\}^n}|x\rangle$ by checking if $\ZX \vdash 
\input{./ZX_diagrams/controlled_states/definition_zero.tikz}
$. The diagram corresponding to the encoded state is simply  $
\input{./ZX_diagrams/controlled_states/definition_one.tikz}
$.

\begin{exa}[Proven in \cite{normal_form_nancy}]\label{lemma:controlled_scalars}
Diagrams $C_{1/\sqrt{2}} = 
\input{./ZX_diagrams/controlled_states/scalar_one_over_sqrt_two.tikz}
$ and $C_2 = 
\begin{tikzpicture}
	\begin{pgfonlayer}{nodelayer}
		\node [style={z_dot}] (0) at (0, -0.25) {};
		\node [style=none] (1) at (0, 0.5) {};
	\end{pgfonlayer}
	\begin{pgfonlayer}{edgelayer}
		\draw [style=plus-edge] (0) to (1.center);
	\end{pgfonlayer}
\end{tikzpicture}

$ are controlled scalars: 
\begin{align}

\input{./ZX_diagrams/controlled_states/diag_pi_one_over_sqrt_two.tikz}
 = 
\begin{tikzpicture}
	\begin{pgfonlayer}{nodelayer}
		\node [style={x_dot}] (0) at (0, 0.25) {};
		\node [style={z_dot}] (1) at (-0.5, -0.25) {$\frac{\pi}{4}$};
		\node [style={z_dot}] (2) at (0.5, -0.25) {$-\frac{\pi}{4}$};
		\node [style={x_dot}] (3) at (-1, 0.25) {};
		\node [style={z_dot}] (4) at (-1, -0.25) {};
		\node [style={x_dot}] (5) at (0, 0.75) {$\pi$};
		\node [style={x_dot}] (6) at (-0.5, 1) {};
		\node [style={z_dot}] (7) at (-0.5, 0.5) {};
	\end{pgfonlayer}
	\begin{pgfonlayer}{edgelayer}
		\draw (3) to (4);
		\draw [bend right=45] (3) to (4);
		\draw [bend left=45] (3) to (4);
		\draw (5) to (0);
		\draw (0) to (2);
		\draw (0) to (1);
		\draw (6) to (7);
		\draw [bend right=45] (6) to (7);
		\draw [bend left=45] (6) to (7);
	\end{pgfonlayer}
\end{tikzpicture}

 = \normalizer
\quad \text{ and } \quad 
\input{./ZX_diagrams/controlled_states/diag_pi_two.tikz}
 = 
\begin{tikzpicture}
	\begin{pgfonlayer}{nodelayer}
		\node [style={z_dot}] (0) at (0, -0.25) {};
		\node [style=none] (1) at (0, 0.5) {};
		\node [style={x_dot}] (2) at (0, 0.5) {$\pi$};
		\node [style={x_dot}] (3) at (-0.5, 0.75) {};
		\node [style={z_dot}] (4) at (-0.5, 0.25) {};
	\end{pgfonlayer}
	\begin{pgfonlayer}{edgelayer}
		\draw [style=plus-edge] (0) to (1.center);
		\draw (3) to (4);
		\draw [bend right=45] (3) to (4);
		\draw [bend left=45] (3) to (4);
	\end{pgfonlayer}
\end{tikzpicture}

 = \gdot
\end{align}
representing respectively $\frac{1}{\sqrt{2}}$ and $2$.
\end{exa}

Controlled states have nice properties that allows to perform element-wise addition and tensor product of corresponding vectors:

\begin{lem}[Sum and tensor product \cite{normal_form_nancy}]\label{lemma:sum_and_product_of_cs}
For any controlled states $C_1, C_2: 1 \rightarrow n$ and $C_3: 1 \rightarrow m$ the diagrams:
\begin{align}
C_+ = 
\begin{tikzpicture}
	\begin{pgfonlayer}{nodelayer}
		\node [style={z_dot}] (0) at (-1, 1) {};
		\node [style={z_dot}] (1) at (0, 1.5) {};
		\node [style={x_dot}] (2) at (1, 1) {};
		\node [style=none] (3) at (-1.5, 0.5) {};
		\node [style=none] (4) at (1.5, 0.5) {};
		\node [style={z_dot}] (7) at (-1.5, -1.25) {};
		\node [style={z_dot}] (8) at (1.5, -1.25) {};
		\node [style=none] (9) at (-1.75, -0.5) {};
		\node [style=none] (10) at (-1.25, -0.5) {};
		\node [style=none] (11) at (-1.25, -0.75) {$\dots$};
		\node [style=none] (12) at (1.25, -0.5) {};
		\node [style=none] (13) at (1.75, -0.5) {};
		\node [style=none] (14) at (1.25, -0.75) {$\dots$};
		\node [style=none] (15) at (-1.5, -1.75) {};
		\node [style=none] (16) at (1.5, -1.75) {};
		\node [style=none] (17) at (0, -1.25) {$\dots$};
		\node [style=none] (18) at (0, 2) {};
		\node [style={z_dot}] (19) at (-1, 1.5) {};
		\node [style={x_dot}] (20) at (-1, 2) {};
		\node [style=none] (21) at (-1.5, 0.5) {};
		\node [style=none] (22) at (-1.5, 0) {$C_1$};
		\node [style=none] (25) at (-2, 0.5) {};
		\node [style=none] (26) at (-2, -0.5) {};
		\node [style=none] (27) at (-1, -0.5) {};
		\node [style=none] (28) at (-1, 0.5) {};
		\node [style=none] (29) at (1.5, 0.5) {};
		\node [style=none] (30) at (1.5, 0) {$C_2$};
		\node [style=none] (33) at (1, 0.5) {};
		\node [style=none] (34) at (1, -0.5) {};
		\node [style=none] (35) at (2, -0.5) {};
		\node [style=none] (36) at (2, 0.5) {};
	\end{pgfonlayer}
	\begin{pgfonlayer}{edgelayer}
		\draw [style=plus-edge] (0) to (1);
		\draw (0) to (2);
		\draw (1) to (2);
		\draw (2) to (4.center);
		\draw (0) to (3.center);
		\draw (9.center) to (7);
		\draw (10.center) to (8);
		\draw (12.center) to (7);
		\draw (13.center) to (8);
		\draw (7) to (15.center);
		\draw (8) to (16.center);
		\draw (18.center) to (1);
		\draw (20) to (19);
		\draw (25.center) to (28.center);
		\draw (28.center) to (27.center);
		\draw (25.center) to (26.center);
		\draw (26.center) to (27.center);
		\draw (33.center) to (36.center);
		\draw (36.center) to (35.center);
		\draw (33.center) to (34.center);
		\draw (34.center) to (35.center);
	\end{pgfonlayer}
\end{tikzpicture}

 \quad \text{and} \quad C_\otimes = 
\begin{tikzpicture}
	\begin{pgfonlayer}{nodelayer}
		\node [style={z_dot}] (1) at (0, 0.5) {};
		\node [style=none] (3) at (-1, 0) {};
		\node [style=none] (4) at (1, 0) {};
		\node [style=none] (5) at (-1, -0.5) {$C_1$};
		\node [style=none] (6) at (1, -0.5) {$C_3$};
		\node [style=none] (9) at (-1.4, -1) {};
		\node [style=none] (10) at (-0.6, -1) {};
		\node [style=none] (11) at (-1, -1.5) {$\dots$};
		\node [style=none] (12) at (0.6, -1) {};
		\node [style=none] (13) at (1.4, -1) {};
		\node [style=none] (14) at (1, -1.5) {$\dots$};
		\node [style=none] (15) at (-1.4, -1.625) {};
		\node [style=none] (16) at (1.4, -1.65) {};
		\node [style=none] (18) at (0, 1) {};
		\node [style=none] (20) at (0.6, -1.65) {};
		\node [style=none] (21) at (-0.6, -1.625) {};
		\node [style=none] (22) at (0.5, 0) {};
		\node [style=none] (23) at (1.5, -1) {};
		\node [style=none] (24) at (1.5, 0) {};
		\node [style=none] (25) at (0.5, -1) {};
		\node [style=none] (26) at (-1.5, 0) {};
		\node [style=none] (27) at (-1.5, -1) {};
		\node [style=none] (28) at (-0.5, -1) {};
		\node [style=none] (29) at (-0.5, 0) {};
	\end{pgfonlayer}
	\begin{pgfonlayer}{edgelayer}
		\draw [bend right=15] (1) to (3.center);
		\draw [bend left=15] (1) to (4.center);
		\draw (9.center) to (15.center);
		\draw (10.center) to (21.center);
		\draw (12.center) to (20.center);
		\draw (13.center) to (16.center);
		\draw (18.center) to (1);
		\draw (26.center) to (29.center);
		\draw (29.center) to (28.center);
		\draw (26.center) to (27.center);
		\draw (27.center) to (28.center);
		\draw (22.center) to (25.center);
		\draw (25.center) to (23.center);
		\draw (23.center) to (24.center);
		\draw (22.center) to (24.center);
	\end{pgfonlayer}
\end{tikzpicture}

\end{align}
are controlled states such that $\interp{C_+}|1\rangle = \interp{C_1}|1\rangle + \interp{C_2}|1\rangle$ and $\interp{C_\otimes}|1\rangle = \interp{C_1}|1\rangle \otimes \interp{C_3}|1\rangle$.
\end{lem}

The previous lemma can be used to add two controlled states. More precisely, if one of the controlled states encodes a state diagram $D_1: 0\rightarrow n$ and the second controlled state encodes $D_2: 0\rightarrow n$ the diagram $
\begin{tikzpicture}[scale=0.4]
	\begin{pgfonlayer}{nodelayer}
		\node [style=none] (0) at (2.5, 0.5) {};
		\node [style=none] (1) at (2.5, -0.5) {};
		\node [style=none] (2) at (1, 0.5) {};
		\node [style=none] (3) at (1, -0.5) {};
		\node [style=none] (4) at (2.25, -0.5) {};
		\node [style=none] (5) at (1.25, -0.5) {};
		\node [style=none] (6) at (1.75, 0.5) {};
		\node [style={x_dot}] (8) at (1.75, 1.5) {$\pi$};
		\node [style=none] (10) at (2.25, -1) {};
		\node [style=none] (11) at (1.25, -1) {};
		\node [style=none] (14) at (1.75, 0) {\footnotesize $C_+$};
		\node [style=none] (15) at (1.75, -0.75) {\tiny $\ldots$};
		\node [style={z_dot}] (16) at (0.75, 2) {};
		\node [style={x_dot}] (17) at (0.75, 1) {};
		\node [style=none] (18) at (-0.5, 0.5) {};
		\node [style=none] (19) at (-0.5, -0.5) {};
		\node [style=none] (20) at (-2, 0.5) {};
		\node [style=none] (21) at (-2, -0.5) {};
		\node [style=none] (22) at (-0.75, -0.5) {};
		\node [style=none] (23) at (-1.75, -0.5) {};
		\node [style=none] (24) at (-1.25, 0.5) {};
		\node [style=none] (25) at (-0.75, -1) {};
		\node [style=none] (26) at (-1.75, -1) {};
		\node [style=none] (27) at (-1.25, 0) {\footnotesize $D_+$};
		\node [style=none] (28) at (-1.25, -0.75) {\tiny $\ldots$};
		\node [style=none] (29) at (0.25, 0) {$=$};
	\end{pgfonlayer}
	\begin{pgfonlayer}{edgelayer}
		\draw (0.center) to (1.center);
		\draw (1.center) to (3.center);
		\draw (3.center) to (2.center);
		\draw (2.center) to (0.center);
		\draw (6.center) to (8);
		\draw (4.center) to (10.center);
		\draw (11.center) to (5.center);
		\draw [bend left=45, looseness=1.25] (16) to (17);
		\draw [bend right=45, looseness=1.25] (16) to (17);
		\draw (16) to (17);
		\draw (18.center) to (19.center);
		\draw (19.center) to (21.center);
		\draw (21.center) to (20.center);
		\draw (20.center) to (18.center);
		\draw (22.center) to (25.center);
		\draw (26.center) to (23.center);
	\end{pgfonlayer}
\end{tikzpicture}

$ correspond to the vector $\interp{D_1} + \interp{D_2}$. Moreover, as the construction returns a controlled state we can directly proceed to the addition of new terms: for $N$ controlled versions $C_1, \dots, C_N$ of $D_1, \dots, D_N$ the diagram:
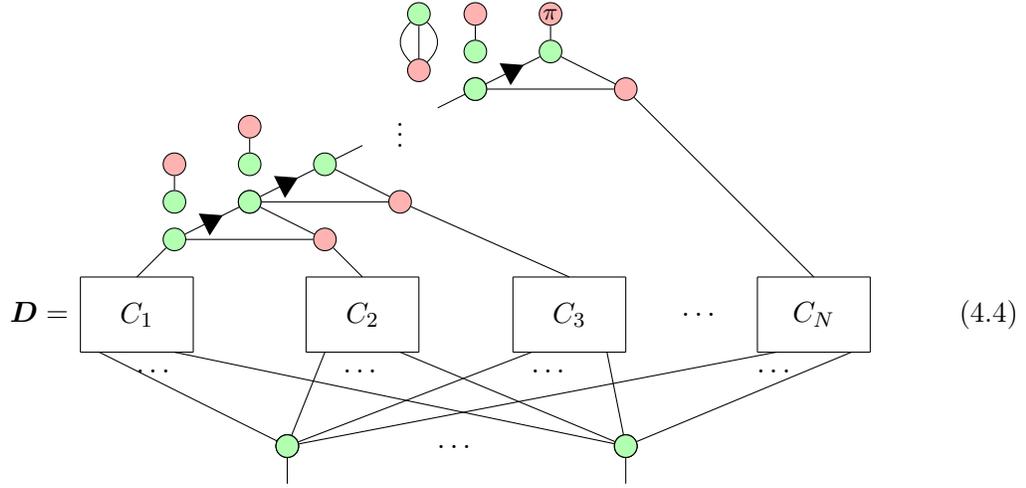
\begin{align}
\bm{D} = 
\begin{tikzpicture}
	\begin{pgfonlayer}{nodelayer}
		\node [style={z_dot}] (0) at (-1, 1) {};
		\node [style={z_dot}] (1) at (0, 1.5) {};
		\node [style={x_dot}] (2) at (1, 1) {};
		\node [style=none] (3) at (-1.5, 0.5) {};
		\node [style=none] (4) at (1.5, 0.5) {};
		\node [style={z_dot}] (7) at (0.5, -1.75) {};
		\node [style={z_dot}] (8) at (5, -1.75) {};
		\node [style=none] (9) at (-2, -0.5) {};
		\node [style=none] (10) at (-1, -0.5) {};
		\node [style=none] (11) at (-1.25, -0.75) {$\dots$};
		\node [style=none] (12) at (1, -0.5) {};
		\node [style=none] (13) at (2, -0.5) {};
		\node [style=none] (14) at (1.5, -0.75) {$\dots$};
		\node [style=none] (15) at (0.5, -2.25) {};
		\node [style=none] (16) at (5, -2.25) {};
		\node [style=none] (17) at (2.75, -1.75) {$\dots$};
		\node [style={z_dot}] (19) at (-1, 1.5) {};
		\node [style={x_dot}] (20) at (-1, 2) {};
		\node [style=none] (21) at (-1.5, 0.5) {};
		\node [style=none] (22) at (-1.5, 0) {$C_1$};
		\node [style=none] (25) at (-2.25, 0.5) {};
		\node [style=none] (26) at (-2.25, -0.5) {};
		\node [style=none] (27) at (-0.75, -0.5) {};
		\node [style=none] (28) at (-0.75, 0.5) {};
		\node [style=none] (29) at (1.5, 0.5) {};
		\node [style=none] (30) at (1.5, 0) {$C_2$};
		\node [style=none] (33) at (0.75, 0.5) {};
		\node [style=none] (34) at (0.75, -0.5) {};
		\node [style=none] (35) at (2.25, -0.5) {};
		\node [style=none] (36) at (2.25, 0.5) {};
		\node [style={z_dot}] (37) at (0, 1.5) {};
		\node [style={z_dot}] (38) at (1, 2) {};
		\node [style={x_dot}] (39) at (2, 1.5) {};
		\node [style=none] (40) at (1.5, 2.25) {};
		\node [style={z_dot}] (41) at (0, 2) {};
		\node [style={x_dot}] (42) at (0, 2.5) {};
		\node [style=none] (43) at (4.25, 0.5) {};
		\node [style={z_dot}] (44) at (0.5, -1.75) {};
		\node [style={z_dot}] (45) at (5, -1.75) {};
		\node [style=none] (46) at (3.75, -0.5) {};
		\node [style=none] (47) at (4.75, -0.5) {};
		\node [style=none] (48) at (4, -0.75) {$\dots$};
		\node [style=none] (50) at (4.25, 0.5) {};
		\node [style=none] (51) at (4.25, 0) {$C_3$};
		\node [style=none] (52) at (3.5, 0.5) {};
		\node [style=none] (53) at (3.5, -0.5) {};
		\node [style=none] (54) at (5, -0.5) {};
		\node [style=none] (55) at (5, 0.5) {};
		\node [style={z_dot}] (56) at (3, 3) {};
		\node [style={z_dot}] (57) at (3, 3) {};
		\node [style={z_dot}] (58) at (4, 3.5) {};
		\node [style={x_dot}] (59) at (5, 3) {};
		\node [style={x_dot}] (60) at (4, 4) {$\pi$};
		\node [style={z_dot}] (61) at (3, 3.5) {};
		\node [style={x_dot}] (62) at (3, 4) {};
		\node [style=none] (63) at (2, 2.5) {$\vdots$};
		\node [style=none] (65) at (2.5, 2.75) {};
		\node [style=none] (66) at (7.5, 0.5) {};
		\node [style={z_dot}] (67) at (0.5, -1.75) {};
		\node [style={z_dot}] (68) at (5, -1.75) {};
		\node [style=none] (69) at (7, -0.5) {};
		\node [style=none] (70) at (8, -0.5) {};
		\node [style=none] (71) at (7, -0.75) {$\dots$};
		\node [style=none] (72) at (6, 0) {$\dots$};
		\node [style=none] (73) at (7.5, 0.5) {};
		\node [style=none] (74) at (7.5, 0) {$C_N$};
		\node [style=none] (75) at (6.75, 0.5) {};
		\node [style=none] (76) at (6.75, -0.5) {};
		\node [style=none] (77) at (8.25, -0.5) {};
		\node [style=none] (78) at (8.25, 0.5) {};
		\node [style={z_dot}] (79) at (2.25, 4) {};
		\node [style={x_dot}] (80) at (2.25, 3.25) {};
	\end{pgfonlayer}
	\begin{pgfonlayer}{edgelayer}
		\draw [style=plus-edge] (0) to (1);
		\draw (0) to (2);
		\draw (1) to (2);
		\draw (2) to (4.center);
		\draw (0) to (3.center);
		\draw (9.center) to (7);
		\draw (10.center) to (8);
		\draw (12.center) to (7);
		\draw (13.center) to (8);
		\draw (7) to (15.center);
		\draw (8) to (16.center);
		\draw (20) to (19);
		\draw (25.center) to (28.center);
		\draw (28.center) to (27.center);
		\draw (25.center) to (26.center);
		\draw (26.center) to (27.center);
		\draw (33.center) to (36.center);
		\draw (36.center) to (35.center);
		\draw (33.center) to (34.center);
		\draw (34.center) to (35.center);
		\draw [style=plus-edge] (37) to (38);
		\draw (37) to (39);
		\draw (38) to (39);
		\draw (40.center) to (38);
		\draw (42) to (41);
		\draw (46.center) to (44);
		\draw (47.center) to (45);
		\draw (52.center) to (55.center);
		\draw (55.center) to (54.center);
		\draw (52.center) to (53.center);
		\draw (53.center) to (54.center);
		\draw (39) to (50.center);
		\draw [style=plus-edge] (57) to (58);
		\draw (57) to (59);
		\draw (58) to (59);
		\draw (60) to (58);
		\draw (62) to (61);
		\draw (65.center) to (57);
		\draw (69.center) to (67);
		\draw (70.center) to (68);
		\draw (75.center) to (78.center);
		\draw (78.center) to (77.center);
		\draw (75.center) to (76.center);
		\draw (76.center) to (77.center);
		\draw (59) to (73.center);
		\draw [bend left=45, looseness=1.25] (79) to (80);
		\draw [bend right=45, looseness=1.25] (79) to (80);
		\draw (79) to (80);
	\end{pgfonlayer}
\end{tikzpicture}

\end{align}
correspond to the state $\interp{\bm{D}} = \interp{D_1} + \dots + \interp{D_N}$. 

We remark that for each number of output wires $n$ the zero-element $C_0^n:1 \rightarrow n$ is a controlled state $
\input{./ZX_diagrams/controlled_states/zero_n.tikz}
$. Indeed, we can prove the following lemma:
\begin{lem}\label{lemma:controlled_zero_add_and_mult}
For any controlled state $C:1 \rightarrow n$
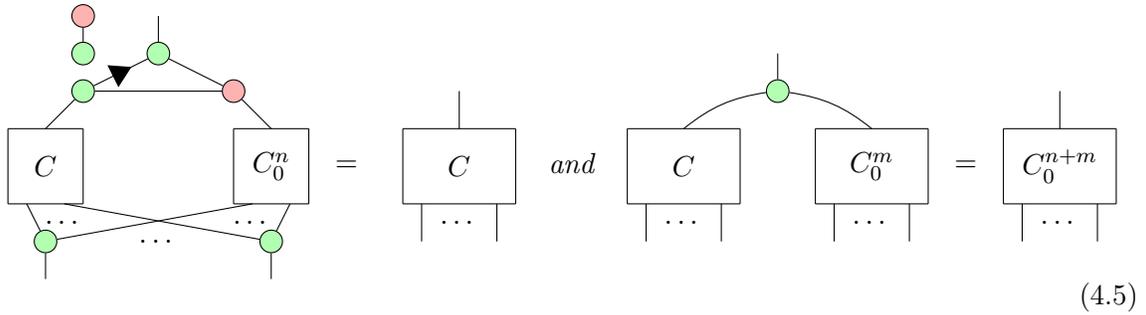
\begin{align}

\input{./ZX_diagrams/old_lemmas/sum_and_product_of_cs/lemma/s0.tikz}
 \quad \text{and} \quad 
\input{./ZX_diagrams/old_lemmas/sum_and_product_of_cs/lemma/p0.tikz}

\end{align}
\end{lem}
\begin{proof}
We start by proving the first equality. The left-hand side is:
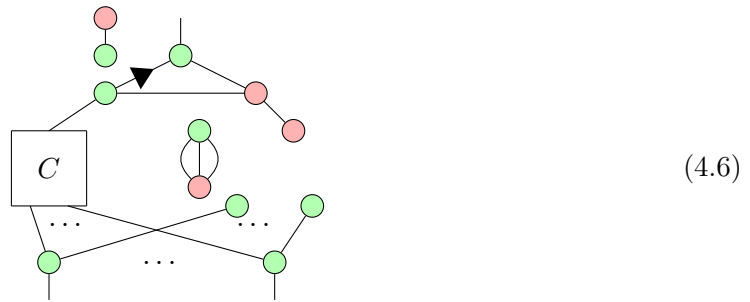
\begin{align}

\input{./ZX_diagrams/old_lemmas/sum_and_product_of_cs/lemma/s1.tikz}

\end{align}
The top part of the diagram can be transformed as follows:
\begin{align}

\input{./ZX_diagrams/old_lemmas/sum_and_product_of_cs/lemma/s2.tikz}

\equal{ \ref{lemma:sqrt-2-sqrt-1-over-2} \\ \ref{lemma:controlled_zero_add_and_mult} }

\input{./ZX_diagrams/old_lemmas/sum_and_product_of_cs/lemma/s3.tikz}

\equal{\bo}

\input{./ZX_diagrams/old_lemmas/sum_and_product_of_cs/lemma/s4.tikz}

\equal{\ref{lemma:sqrt-2-sqrt-1-over-2} \\ \ref{lemma:zero-triangle-down}}

\begin{tikzpicture}
	\begin{pgfonlayer}{nodelayer}
		\node [style={z_dot}] (1) at (0, 0) {};
		\node [style=none] (2) at (0, -0.75) {};
		\node [style=none] (18) at (0, 0.75) {};
	\end{pgfonlayer}
	\begin{pgfonlayer}{edgelayer}
		\draw (1) to (2.center);
		\draw (18.center) to (1);
	\end{pgfonlayer}
\end{tikzpicture}

\end{align} 
The statement follows from \stt and \soo. 

For the second equality we have:
\begin{align}

\input{./ZX_diagrams/old_lemmas/sum_and_product_of_cs/lemma/p1.tikz}
 \equal{\bo}
\input{./ZX_diagrams/old_lemmas/sum_and_product_of_cs/lemma/p2.tikz}

\end{align}
The lemma follows from the definition of the controlled state.
\qedhere
\end{proof}

We follow  in this work the definition of a controlled state in~\cite{normal_form_nancy}. There exist other definitions, see e.g. \cite{zx_addition_oxford}, which essentially differ in the output of the zero-case.
%In their definition a controlled state $\overtilde{C}:1 \rightarrow n$, when applied to the zero-input is transformed to $ \normalizer^{\otimes n}\ketzero \dots \ketzero$ instead of the uniform superposition $\underbrace{\gket \dots \gket}_n$.

\subsection{Controlizer}

The lemma \ref{lemma:sum_and_product_of_cs} provides a way to obtain the sum of two diagrams in a controlled state form.

The definition and the lemma technically allow only to solve the problem for states rather than arbitrary matrices, but notice that going from a matrix $n\rightarrow m$ to a state $0\rightarrow n+m$ is just a matter of bending wires. We remark that bending wires alone can't lead to a procedure for addition, as there is no obvious way to directly sum state diagrams without controlled versions.  

What remains to be done is to explain how to compute algorithmically, from a state (diagram)  $D$, a controlled version of the same state (diagram).

We introduce for this the notion of a \emph{controlizer} - a map that associates diagrams with the corresponding controlled states.

\begin{nota}
$\ZX[n,m]$ denotes the family of all ZX-diagrams with $n$ inputs and $m$ outputs.
\end{nota}
The formal definition of the controlizer is:
\begin{defi}[Controlizer]\label{def:controlizer}
We say that a map $C: \ZX[n, m] \rightarrow \ZX[1, n+m]$ that associates to every diagram $D:n \rightarrow m$ a diagram $C(D): 1 \rightarrow n + m$ is a controlizer if the following conditions hold for any $\ZX$-diagram $D$:
\begin{itemize}
\item $C(D)$ is a controlled state
\item The state represents $D$ up to some scalars in the sense that:
\begin{align}
\interp{D} = \interp{
\input{./ZX_diagrams/controlizer/definition.tikz}
}\label{eq:controlizer_property}
\end{align}
\end{itemize} 
\end{defi}

Several maps satisfy the definition of the controlizer. For instance, the map presented in \cite{normal_form_nancy} that associates to each diagram its \textit{normal form} is in fact a controlizer. In \cite{zx_addition_oxford} an alternative controlled state form is constructed from the $2^n$ coefficients of the algebraic normal form. 

In our approach the controlizer is defined by induction over the composition and the tensor product and doesn't rely on normal forms.
It means that the shape of the controlizer will follow closely the shape of the original diagram. In terms of complexity, this also implies that the size of the controlizer output is polynomial in the size of the diagram. 

\begin{defi}[Inductive controlizer]\label{example:controlizer}
We define the map $C: \ZX[n,m] \rightarrow \ZX[1, n+m]$ that associates to each diagram $D: n\rightarrow m$ a diagram $C(D): 1\rightarrow n+m$ as follows:
\begin{itemize}
\item \label{controlizer:base} For the generators $\rbeta$ , $
\input{./ZX_diagrams/generators/tree_rdot-.tikz}
$ , $
\begin{tikzpicture}
	\begin{pgfonlayer}{nodelayer}
		\node [style=none] (0) at (0, 0.25) {};
		\node [style=none] (1) at (0, -0.25) {};
		\node [style={h_box}] (2) at (0, 0) {};
	\end{pgfonlayer}
	\begin{pgfonlayer}{edgelayer}
		\draw (0.center) to (2);
		\draw (2) to (1.center);
	\end{pgfonlayer}
\end{tikzpicture}

$ , $
\begin{tikzpicture}
	\begin{pgfonlayer}{nodelayer}
		\node [style=none] (1) at (-0.25, -0.1) {};
		\node [style=none] (3) at (0.25, -0.1) {};
	\end{pgfonlayer}
	\begin{pgfonlayer}{edgelayer}
		\draw [in=-270, out=90, looseness=2.00] (1.center) to (3.center);
	\end{pgfonlayer}
\end{tikzpicture}

$ , $
\begin{tikzpicture}
	\begin{pgfonlayer}{nodelayer}
		\node [style=none] (1) at (-0.25, 0.175) {};
		\node [style=none] (3) at (0.25, 0.175) {};
	\end{pgfonlayer}
	\begin{pgfonlayer}{edgelayer}
		\draw [in=270, out=-90, looseness=2.00] (1.center) to (3.center);
	\end{pgfonlayer}
\end{tikzpicture}

$ , $
\input{./ZX_diagrams/generators/swap-s.tikz}
$ :
\begin{align}
&\controlizer{\rbeta} = 
\input{./ZX_diagrams/controlizer/rbeta.tikz}
, \quad
\controlizer{
\input{./ZX_diagrams/generators/tree_rdot-.tikz}
} = 
\input{./ZX_diagrams/controlizer/tree_rdot.tikz}
, \quad
\controlizer{
\begin{tikzpicture}
	\begin{pgfonlayer}{nodelayer}
		\node [style=none] (0) at (0, 0.25) {};
		\node [style=none] (1) at (0, -0.25) {};
		\node [style={h_box}] (2) at (0, 0) {};
	\end{pgfonlayer}
	\begin{pgfonlayer}{edgelayer}
		\draw (0.center) to (2);
		\draw (2) to (1.center);
	\end{pgfonlayer}
\end{tikzpicture}

} =
\input{./ZX_diagrams/controlizer/hadamard.tikz}
 \nonumber \\
&
\controlizer{
\begin{tikzpicture}
	\begin{pgfonlayer}{nodelayer}
		\node [style=none] (1) at (-0.25, 0.175) {};
		\node [style=none] (3) at (0.25, 0.175) {};
	\end{pgfonlayer}
	\begin{pgfonlayer}{edgelayer}
		\draw [in=270, out=-90, looseness=2.00] (1.center) to (3.center);
	\end{pgfonlayer}
\end{tikzpicture}

} = \controlizer{
\begin{tikzpicture}
	\begin{pgfonlayer}{nodelayer}
		\node [style=none] (1) at (-0.25, -0.1) {};
		\node [style=none] (3) at (0.25, -0.1) {};
	\end{pgfonlayer}
	\begin{pgfonlayer}{edgelayer}
		\draw [in=-270, out=90, looseness=2.00] (1.center) to (3.center);
	\end{pgfonlayer}
\end{tikzpicture}

} = 
\input{./ZX_diagrams/controlizer/cap.tikz}
, \quad
\controlizer{
\input{./ZX_diagrams/generators/swap-s.tikz}
} = 
\input{./ZX_diagrams/controlizer/swap.tikz}

\end{align}
\item \label{controlizer:base_spiders} Generators $
\input{./ZX_diagrams/generators/rdot-s.tikz}
$ and $
\input{./ZX_diagrams/generators/gdot-s.tikz}
$ can be decomposed as follows using the above generators:
\begin{align}
&\controlizer{
\input{./ZX_diagrams/generators/rdot.tikz}
} = \controlizer{
\input{./ZX_diagrams/controlizer/rdot.tikz}
}, \quad
\controlizer{
\input{./ZX_diagrams/generators/gdot.tikz}
} = \controlizer{
\input{./ZX_diagrams/controlizer/gdot.tikz}
} \nonumber 
\end{align}
\item  For tensor product $D_\otimes = D_2 \otimes D_1$ and composition $D_{\circ} = D_3 \circ D_1$ where $D_1: n \rightarrow m$, $D_2: k \rightarrow l$ and $D_3: m \rightarrow k$:
\begin{align}
C(D_\otimes) = \tikzformula{controlizer/tensor_product}, \quad C(D_\circ) = \tikzformula{controlizer/composition}\label{controlizer:step}
\end{align}
\end{itemize}

\end{defi}

We remark that the output of the inductive controlizer is not unique as it depends on the decomposition order. For instance, two diagrams that are equivalent up to deformation (the \emph{topology matters} rule) will produce different outputs, that are not equivalent up to deformation. However, the semantics is preserved: both representations will of course represent the same matrix.

We now verify that our inductive controlizer \ref{example:controlizer} satisfies the definition \ref{def:controlizer}:

\begin{proof}
We prove the claim for the generators $\rbeta$ , $
\input{./ZX_diagrams/generators/tree_rdot-.tikz}
$ , $
\begin{tikzpicture}
	\begin{pgfonlayer}{nodelayer}
		\node [style=none] (0) at (0, 0.25) {};
		\node [style=none] (1) at (0, -0.25) {};
		\node [style={h_box}] (2) at (0, 0) {};
	\end{pgfonlayer}
	\begin{pgfonlayer}{edgelayer}
		\draw (0.center) to (2);
		\draw (2) to (1.center);
	\end{pgfonlayer}
\end{tikzpicture}

$ , $
\begin{tikzpicture}
	\begin{pgfonlayer}{nodelayer}
		\node [style=none] (1) at (-0.25, -0.1) {};
		\node [style=none] (3) at (0.25, -0.1) {};
	\end{pgfonlayer}
	\begin{pgfonlayer}{edgelayer}
		\draw [in=-270, out=90, looseness=2.00] (1.center) to (3.center);
	\end{pgfonlayer}
\end{tikzpicture}

$ , $
\begin{tikzpicture}
	\begin{pgfonlayer}{nodelayer}
		\node [style=none] (1) at (-0.25, 0.175) {};
		\node [style=none] (3) at (0.25, 0.175) {};
	\end{pgfonlayer}
	\begin{pgfonlayer}{edgelayer}
		\draw [in=270, out=-90, looseness=2.00] (1.center) to (3.center);
	\end{pgfonlayer}
\end{tikzpicture}

$ , $
\input{./ZX_diagrams/generators/swap-s.tikz}
$  and both compositions. The case of $\tikzformula{generators/rdot-s}$ and $\tikzformula{generators/gdot-s}$ is a direct consequence of the axiom $\soo$.

First, we inductively prove that for all diagrams $D$ the diagram $C(D)$ is controlled state.
\begin{itemize}
\item \textbf{Generators:}
\begin{align}
\controlizer{\rbeta} \circ \ketzero = 
\begin{tikzpicture}
	\begin{pgfonlayer}{nodelayer}
		\node [style={x_dot}] (0) at (0, 0.25) {$\pi$};
		\node [style={x_dot}] (1) at (0, -0.75) {$\beta$};
		\node [style={z_dot}] (2) at (0, 0.75) {};
		\node [style=none] (3) at (0, -1) {};
		\node [style={z_dot}] (4) at (0.75, 0.75) {};
		\node [style={x_dot}] (5) at (0, 1.25) {};
		\node [style={x_dot}] (6) at (-0.5, 1.25) {};
		\node [style={z_dot}] (7) at (-0.5, 0.75) {};
	\end{pgfonlayer}
	\begin{pgfonlayer}{edgelayer}
		\draw (3.center) to (1);
		\draw [style=plus-edge] (1) to (0);
		\draw [style=plus-edge] (4) to (2);
		\draw (6) to (7);
		\draw [bend right=45] (6) to (7);
		\draw [bend left=45] (6) to (7);
		\draw (5) to (2);
		\draw (2) to (0);
	\end{pgfonlayer}
\end{tikzpicture}

\equal{\bo}

\begin{tikzpicture}
	\begin{pgfonlayer}{nodelayer}
		\node [style={x_dot}] (0) at (0, 0.25) {$\pi$};
		\node [style={x_dot}] (1) at (0, -0.75) {$\beta$};
		\node [style={x_dot}] (2) at (0.5, 0) {};
		\node [style=none] (3) at (0, -1) {};
		\node [style={z_dot}] (4) at (1.25, 0) {};
		\node [style={x_dot}] (5) at (0, 0.75) {};
		\node [style={x_dot}] (6) at (-0.5, 1) {};
		\node [style={z_dot}] (7) at (-0.5, 0.5) {};
		\node [style={x_dot}] (8) at (0.5, 1) {};
		\node [style={z_dot}] (9) at (0.5, 0.5) {};
	\end{pgfonlayer}
	\begin{pgfonlayer}{edgelayer}
		\draw (3.center) to (1);
		\draw [style=plus-edge] (1) to (0);
		\draw [style=plus-edge] (4) to (2);
		\draw (6) to (7);
		\draw [bend right=45] (6) to (7);
		\draw [bend left=45] (6) to (7);
		\draw (5) to (0);
		\draw (8) to (9);
		\draw [bend right=45] (8) to (9);
		\draw [bend left=45] (8) to (9);
	\end{pgfonlayer}
\end{tikzpicture}

\equal{\ref{lemma:controlled_scalars} \\ \ref{lemma:pi-triangle-up}} 
\begin{tikzpicture}
	\begin{pgfonlayer}{nodelayer}
		\node [style={z_dot}] (0) at (0.25, 0.25) {};
		\node [style={x_dot}] (1) at (0.25, -0.5) {$\beta$};
		\node [style=none] (3) at (0.25, -1) {};
		\node [style={x_dot}] (6) at (-0.25, 0.5) {};
		\node [style={z_dot}] (7) at (-0.25, 0) {};
		\node [style={x_dot}] (10) at (-0.25, -0.5) {};
		\node [style={z_dot}] (11) at (-0.25, -1) {};
	\end{pgfonlayer}
	\begin{pgfonlayer}{edgelayer}
		\draw (3.center) to (1);
		\draw (1) to (0);
		\draw (6) to (7);
		\draw [bend right=45] (6) to (7);
		\draw [bend left=45] (6) to (7);
		\draw (10) to (11);
	\end{pgfonlayer}
\end{tikzpicture}

\equal{\ref{lemma:sqrt-2-sqrt-1-over-2}}

\begin{tikzpicture}
	\begin{pgfonlayer}{nodelayer}
		\node [style=none] (2) at (0, -0.25) {};
		\node [style={z_dot}] (12) at (0, 0.25) {};
	\end{pgfonlayer}
	\begin{pgfonlayer}{edgelayer}
		\draw (12) to (2.center);
	\end{pgfonlayer}
\end{tikzpicture}

\end{align}

\begin{align}
\controlizer{
\begin{tikzpicture}
	\begin{pgfonlayer}{nodelayer}
		\node [style={x_dot}] (0) at (0, 0) {};
		\node [style=none] (1) at (0, -0.25) {};
		\node [style=none] (2) at (0.25, 0) {};
		\node [style=none] (3) at (0, 0.25) {}; 
	\end{pgfonlayer}
	\begin{pgfonlayer}{edgelayer}
		\draw (0) to (1.center);
		\draw (3.center) to (0);
		\draw (0) to (2.center);
	\end{pgfonlayer}
\end{tikzpicture}

} \circ \ketzero &= 
\begin{tikzpicture}
	\begin{pgfonlayer}{nodelayer}
		\node [style={x_dot}] (0) at (0, -0.25) {$\pi$};
		\node [style={z_dot}] (1) at (0, 0.5) {};
		\node [style=none] (2) at (-0.5, -0.75) {};
		\node [style=none] (3) at (0, -0.75) {};
		\node [style=none] (4) at (0.5, -0.75) {};
		\node [style={z_dot}] (5) at (-0.5, -0.25) {};
		\node [style={x_dot}] (7) at (-0.5, 1) {};
		\node [style={z_dot}] (8) at (-0.5, 0.5) {};
		\node [style={x_dot}] (9) at (0, 1) {};
		\node [style={z_dot}] (10) at (0.75, 0.5) {};
	\end{pgfonlayer}
	\begin{pgfonlayer}{edgelayer}
		\draw [style=plus-edge] (1) to (0);
		\draw (0) to (2.center);
		\draw (0) to (3.center);
		\draw (0) to (4.center);
		\draw (7) to (8);
		\draw [bend right=45] (7) to (8);
		\draw [bend left=45] (7) to (8);
		\draw [style=plus-edge] (10) to (1);
		\draw (9) to (1);
	\end{pgfonlayer}
\end{tikzpicture}

\equal{\bo}

\begin{tikzpicture}
	\begin{pgfonlayer}{nodelayer}
		\node [style={x_dot}] (0) at (0, -0.25) {$\pi$};
		\node [style={x_dot}] (1) at (0, 0.5) {};
		\node [style=none] (2) at (-0.5, -0.75) {};
		\node [style=none] (3) at (0, -0.75) {};
		\node [style=none] (4) at (0.5, -0.75) {};
		\node [style={z_dot}] (5) at (-0.5, -0.25) {};
		\node [style={x_dot}] (7) at (-0.5, 0.75) {};
		\node [style={z_dot}] (8) at (-0.5, 0.25) {};
		\node [style={x_dot}] (11) at (0.5, -0.25) {};
		\node [style={x_dot}] (12) at (0.5, 0.75) {};
		\node [style={z_dot}] (13) at (0.5, 0.25) {};
		\node [style={z_dot}] (15) at (1.25, -0.25) {};
	\end{pgfonlayer}
	\begin{pgfonlayer}{edgelayer}
		\draw [style=plus-edge] (1) to (0);
		\draw (0) to (2.center);
		\draw (0) to (3.center);
		\draw (0) to (4.center);
		\draw (7) to (8);
		\draw [bend right=45] (7) to (8);
		\draw [bend left=45] (7) to (8);
		\draw (12) to (13);
		\draw [bend right=45] (12) to (13);
		\draw [bend left=45] (12) to (13);
		\draw [style=plus-edge] (15) to (11);
	\end{pgfonlayer}
\end{tikzpicture}

\equal{\ref{lemma:controlled_scalars} \\ \ref{lemma:zero-triangle-down}}

\begin{tikzpicture}
	\begin{pgfonlayer}{nodelayer}
		\node [style={x_dot}] (0) at (0, -0.25) {$\pi$};
		\node [style={z_dot}] (1) at (0, 0.5) {};
		\node [style=none] (2) at (-0.5, -0.75) {};
		\node [style=none] (3) at (0, -0.75) {};
		\node [style=none] (4) at (0.5, -0.75) {};
		\node [style={z_dot}] (5) at (-0.5, -0.25) {};
		\node [style={x_dot}] (7) at (-0.5, 0.75) {};
		\node [style={z_dot}] (8) at (-0.5, 0.25) {};
		\node [style={x_dot}] (9) at (0.5, 0.75) {};
		\node [style={z_dot}] (10) at (0.5, 0.25) {};
	\end{pgfonlayer}
	\begin{pgfonlayer}{edgelayer}
		\draw (1) to (0);
		\draw (0) to (2.center);
		\draw (0) to (3.center);
		\draw (0) to (4.center);
		\draw (7) to (8);
		\draw [bend right=45] (7) to (8);
		\draw [bend left=45] (7) to (8);
		\draw (9) to (10);
	\end{pgfonlayer}
\end{tikzpicture}

\equal{\bo \\ \ref{lemma:sqrt-2-sqrt-1-over-2}} 

\begin{tikzpicture}
	\begin{pgfonlayer}{nodelayer}
		\node [style=none] (2) at (-0.5, -0.25) {};
		\node [style=none] (3) at (0, -0.25) {};
		\node [style=none] (4) at (0.5, -0.25) {};
		\node [style={z_dot}] (12) at (-0.5, 0.25) {};
		\node [style={z_dot}] (13) at (0, 0.25) {};
		\node [style={z_dot}] (14) at (0.5, 0.25) {};
	\end{pgfonlayer}
	\begin{pgfonlayer}{edgelayer}
		\draw (12) to (2.center);
		\draw (13) to (3.center);
		\draw (14) to (4.center);
	\end{pgfonlayer}
\end{tikzpicture}

\end{align}

\begin{align}
\controlizer{
\begin{tikzpicture}
	\begin{pgfonlayer}{nodelayer}
		\node [style=none] (0) at (0, 0.25) {};
		\node [style=none] (1) at (0, -0.25) {};
		\node [style={h_box}] (2) at (0, 0) {};
	\end{pgfonlayer}
	\begin{pgfonlayer}{edgelayer}
		\draw (0.center) to (2);
		\draw (2) to (1.center);
	\end{pgfonlayer}
\end{tikzpicture}

} \circ \ketzero = 
\begin{tikzpicture}
	\begin{pgfonlayer}{nodelayer}
		\node [style={z_dot}] (0) at (-0.5, 0.5) {};
		\node [style={z_dot}] (1) at (-0.5, -0.25) {};
		\node [style={z_dot}] (2) at (0.5, -0.25) {};
		\node [style={x_dot}] (3) at (0, 0.5) {$\pi$};
		\node [style=none] (5) at (-0.5, -1) {};
		\node [style=none] (6) at (0.5, -1) {};
		\node [style={h_box}] (7) at (-0.5, -0.75) {};
		\node [style={x_dot}] (8) at (0.5, 1) {$\pi$};
		\node [style={z_dot}] (9) at (0.5, 0.5) {$\frac{\pi}{4}$};
		\node [style={z_dot}] (10) at (0.5, 1.5) {$-\frac{\pi}{4}$};
		\node [style={z_dot}] (13) at (0, 1) {};
		\node [style={x_dot}] (14) at (0, 1.5) {};
		\node [style={x_dot}] (15) at (-0.5, 1.5) {};
		\node [style={z_dot}] (16) at (-0.5, 1) {};
	\end{pgfonlayer}
	\begin{pgfonlayer}{edgelayer}
		\draw (7) to (1);
		\draw (5.center) to (7);
		\draw (6.center) to (2);
		\draw (1) to (3);
		\draw [style=plus-edge] (2) to (3);
		\draw [style=plus-edge] (1) to (2);
		\draw (8) to (13);
		\draw (10) to (8);
		\draw (9) to (8);
		\draw (3) to (13);
		\draw (13) to (14);
		\draw (16) to (15);
		\draw [bend left=45] (16) to (15);
		\draw [bend right=45] (16) to (15);
	\end{pgfonlayer}
\end{tikzpicture}

\equal{\bo \\ \ref{lemma:sqrt-2-sqrt-1-over-2} }

\begin{tikzpicture}
	\begin{pgfonlayer}{nodelayer}
		\node [style={z_dot}] (0) at (-1, 0.25) {};
		\node [style={z_dot}] (1) at (-1, -0.25) {};
		\node [style={z_dot}] (2) at (0, -0.25) {};
		\node [style={x_dot}] (3) at (-0.5, 0.5) {$\pi$};
		\node [style=none] (5) at (-1, -1) {};
		\node [style=none] (6) at (0, -1) {};
		\node [style={h_box}] (7) at (-1, -0.75) {};
		\node [style={x_dot}] (14) at (-0.5, 1.25) {};
		\node [style={x_dot}] (15) at (-1, 1.25) {};
		\node [style={z_dot}] (16) at (-1, 0.75) {};
		\node [style={x_dot}] (17) at (1, 1) {};
		\node [style={z_dot}] (18) at (1, 0.5) {$\frac{\pi}{4}$};
		\node [style={z_dot}] (19) at (1, 1.5) {$-\frac{\pi}{4}$};
		\node [style={x_dot}] (20) at (0.5, 1) {$\pi$};
		\node [style={x_dot}] (22) at (0, 1.25) {};
		\node [style={z_dot}] (23) at (0, 0.75) {};
		\node [style={z_dot}] (24) at (0.5, 0) {};
		\node [style={x_dot}] (25) at (0.5, -0.5) {};
		\node [style={x_dot}] (26) at (1, 0) {};
		\node [style={z_dot}] (27) at (1, -0.5) {};
	\end{pgfonlayer}
	\begin{pgfonlayer}{edgelayer}
		\draw (7) to (1);
		\draw (5.center) to (7);
		\draw (6.center) to (2);
		\draw (1) to (3);
		\draw [style=plus-edge] (2) to (3);
		\draw [style=plus-edge] (1) to (2);
		\draw (16) to (15);
		\draw [bend left=45] (16) to (15);
		\draw [bend right=45] (16) to (15);
		\draw (17) to (20);
		\draw (19) to (17);
		\draw (18) to (17);
		\draw (23) to (22);
		\draw [bend left=45] (23) to (22);
		\draw [bend right=45] (23) to (22);
		\draw (25) to (24);
		\draw (27) to (26);
		\draw [bend left=45] (27) to (26);
		\draw [bend right=45] (27) to (26);
		\draw (3) to (14);
	\end{pgfonlayer}
\end{tikzpicture}

\equal{ \ref{lemma:controlled_scalars} \\ \ref{lemma:opposite-triangles-pi}}

\begin{tikzpicture}
	\begin{pgfonlayer}{nodelayer}
		\node [style={z_dot}] (0) at (-0.75, -0.5) {};
		\node [style={x_dot}] (1) at (-0.25, 0.25) {};
		\node [style={z_dot}] (2) at (0.25, 0.25) {};
		\node [style=none] (5) at (-0.25, -0.5) {};
		\node [style=none] (6) at (0.25, -0.5) {};
		\node [style={h_box}] (7) at (-0.25, -0.25) {};
		\node [style={x_dot}] (15) at (-0.75, 0.5) {};
		\node [style={z_dot}] (16) at (-0.75, 0) {};
		\node [style=none] (17) at (-0.5, 0.75) {$\otimes 2$};
	\end{pgfonlayer}
	\begin{pgfonlayer}{edgelayer}
		\draw (1) to (7);
		\draw (7) to (5.center);
		\draw (2) to (6.center);
		\draw (15) to (16);
		\draw [bend right=45] (15) to (16);
		\draw [bend left=45] (15) to (16);
	\end{pgfonlayer}
\end{tikzpicture}

\equal{\h \\ \ref{lemma:sqrt-2-sqrt-1-over-2}}

\begin{tikzpicture}
	\begin{pgfonlayer}{nodelayer}
		\node [style={z_dot}] (1) at (-0.25, 0.25) {};
		\node [style={z_dot}] (2) at (0.25, 0.25) {};
		\node [style=none] (5) at (-0.25, -0.5) {};
		\node [style=none] (6) at (0.25, -0.5) {};
	\end{pgfonlayer}
	\begin{pgfonlayer}{edgelayer}
		\draw (2) to (6.center);
		\draw (1) to (5.center);
	\end{pgfonlayer}
\end{tikzpicture}

\end{align}

\begin{align}
\controlizer{
\begin{tikzpicture}
	\begin{pgfonlayer}{nodelayer}
		\node [style=none] (1) at (-0.25, -0.1) {};
		\node [style=none] (3) at (0.25, -0.1) {};
	\end{pgfonlayer}
	\begin{pgfonlayer}{edgelayer}
		\draw [in=-270, out=90, looseness=2.00] (1.center) to (3.center);
	\end{pgfonlayer}
\end{tikzpicture}

} \circ \ketzero &= 
\begin{tikzpicture}
	\begin{pgfonlayer}{nodelayer}
		\node [style={x_dot}] (0) at (0, -0.25) {$\pi$};
		\node [style={z_dot}] (1) at (0, 0.5) {};
		\node [style=none] (2) at (-0.5, -0.75) {};
		\node [style=none] (4) at (0.5, -0.75) {};
		\node [style={z_dot}] (5) at (-0.75, -0.25) {};
		\node [style={x_dot}] (6) at (-0.75, 0.25) {};
		\node [style={x_dot}] (7) at (-0.5, 1) {};
		\node [style={z_dot}] (8) at (-0.5, 0.5) {};
		\node [style={x_dot}] (9) at (0, 1) {};
		\node [style={z_dot}] (10) at (0.75, 0.5) {};
	\end{pgfonlayer}
	\begin{pgfonlayer}{edgelayer}
		\draw [style=plus-edge] (1) to (0);
		\draw (0) to (2.center);
		\draw (0) to (4.center);
		\draw (6) to (5);
		\draw (7) to (8);
		\draw [bend right=45] (7) to (8);
		\draw [bend left=45] (7) to (8);
		\draw [style=plus-edge] (10) to (1);
		\draw (9) to (1);
	\end{pgfonlayer}
\end{tikzpicture}

\equal{\bo}

\begin{tikzpicture}
	\begin{pgfonlayer}{nodelayer}
		\node [style={x_dot}] (0) at (0, 0) {$\pi$};
		\node [style={x_dot}] (1) at (0, 0.75) {};
		\node [style=none] (2) at (-0.5, -0.5) {};
		\node [style=none] (4) at (0.5, -0.5) {};
		\node [style={z_dot}] (5) at (-0.75, -0.5) {};
		\node [style={x_dot}] (6) at (-0.75, 0) {};
		\node [style={x_dot}] (7) at (-0.75, 1) {};
		\node [style={z_dot}] (8) at (-0.75, 0.5) {};
		\node [style={x_dot}] (11) at (0.5, 0) {};
		\node [style={x_dot}] (12) at (0.5, 1) {};
		\node [style={z_dot}] (13) at (0.5, 0.5) {};
		\node [style={z_dot}] (15) at (1.25, 0) {};
	\end{pgfonlayer}
	\begin{pgfonlayer}{edgelayer}
		\draw [style=plus-edge] (1) to (0);
		\draw (0) to (2.center);
		\draw (0) to (4.center);
		\draw (6) to (5);
		\draw (7) to (8);
		\draw [bend right=45] (7) to (8);
		\draw [bend left=45] (7) to (8);
		\draw (12) to (13);
		\draw [bend right=45] (12) to (13);
		\draw [bend left=45] (12) to (13);
		\draw [style=plus-edge] (15) to (11);
	\end{pgfonlayer}
\end{tikzpicture}

\equal{\ref{lemma:controlled_scalars} \\ \ref{lemma:zero-triangle-down}}

\begin{tikzpicture}
	\begin{pgfonlayer}{nodelayer}
		\node [style={x_dot}] (0) at (0, 0) {$\pi$};
		\node [style={z_dot}] (1) at (0, 0.75) {};
		\node [style=none] (2) at (-0.25, -0.5) {};
		\node [style=none] (4) at (0.25, -0.5) {};
		\node [style={z_dot}] (5) at (-0.5, 0.25) {};
		\node [style={x_dot}] (6) at (-0.5, 0.75) {};
	\end{pgfonlayer}
	\begin{pgfonlayer}{edgelayer}
		\draw (1) to (0);
		\draw (0) to (2.center);
		\draw (0) to (4.center);
		\draw (6) to (5);
	\end{pgfonlayer}
\end{tikzpicture}

\equal{\bo \\ \ref{lemma:sqrt-2-sqrt-1-over-2}}

\begin{tikzpicture}
	\begin{pgfonlayer}{nodelayer}
		\node [style={z_dot}] (1) at (0.25, 0.25) {};
		\node [style=none] (2) at (-0.25, -0.25) {};
		\node [style=none] (4) at (0.25, -0.25) {};
		\node [style={z_dot}] (11) at (-0.25, 0.25) {};
	\end{pgfonlayer}
	\begin{pgfonlayer}{edgelayer}
		\draw (11) to (2.center);
		\draw (1) to (4.center);
	\end{pgfonlayer}
\end{tikzpicture}

\end{align}

\begin{align}
\controlizer{
\input{./ZX_diagrams/generators/swap.tikz}
} & \circ \ketzero = 
\input{./ZX_diagrams/controlizer/proof/zero/swap/step1.tikz}

\equal{\bo}

\input{./ZX_diagrams/controlizer/proof/zero/swap/step1bis.tikz}
 \nonumber \\
&\equal{\ref{lemma:controlled_scalars} \\ \ref{lemma:pi-push} \\ \bo}

\input{./ZX_diagrams/controlizer/proof/zero/swap/step2.tikz}

\equal{\ref{lemma:pi-triangle-up}}

\input{./ZX_diagrams/controlizer/proof/zero/swap/step3.tikz}
 
\equal{\bo \\ \ref{lemma:sqrt-2-sqrt-1-over-2}}

\input{./ZX_diagrams/controlizer/proof/zero/swap/step5.tikz}

\end{align}
\item \textbf{Tensor product:} it directly follows from lemma (\ref{lemma:sum_and_product_of_cs}) that if $C(D_1)$ and $C(D_2)$ are controlled states, then the diagram $C(D_1 \otimes D_2)$ from (\ref{controlizer:step}) is controlled state.

\item \textbf{Composition:} let's assume that $C(D_1)$ and $C(D_2)$ are controlled states. We show that $C(D_1 \circ D_2)$ is also a controlled state:
\begin{align}
C(D_1 \circ D_2) & \circ \ketzero  = 


\end{align}

\end{itemize}

Next we show that the map $C$ satisfy the property (\ref{eq:controlizer_property}), i.e. it correctly encodes the input diagram.
\begin{itemize}
\item \textbf{Generators:}
\begin{align}
\controlizer{\rbeta} \circ \ketone = 
%

\end{align}
\item \textbf{Composition:} Let's assume that the property (\ref{eq:controlizer_property}) holds for diagrams $D_1: n \rightarrow m$ and $D_2: m \rightarrow k$.The controller for their composition $D_1\circ D_2$ is given by:
\begin{align}
&
%

 = D_2 \circ D_1
\end{align}
\item \textbf{Tensor product:} Let's assume that the property (\ref{eq:controlizer_property}) holds for diagrams $D_1: n \rightarrow m$ and $D_2: k \rightarrow l$.  The controller for their tensor $D_1\circ D_2$ is given by:

\begin{align}
&
%

 \label{eq:controlizer_tp_diagram}
\end{align}
The diagram (\ref{eq:controlizer_tp_diagram}) is equal to $
%

$ by the induction assumption.
\end{itemize}
\qedhere
\end{proof}

To give a flavor of our computation process we show how to obtain a controlled version of $
%

\label{example:controlled_pi}
\end{align}

\paragraph{Addition of arbitrary diagrams}

The controlizer allows to map every two ZX-diagrams $D_1:n \rightarrow m$ and $D_2:n \rightarrow m$ to controlled states $C_1:1 \rightarrow n+m$ and $C_2:1 \rightarrow n+m$. The lemma \ref{lemma:sum_and_product_of_cs} provides a way to obtain a sum of resulting controlled states. By combining this two results we can recover the diagram for addition of $D_1$ and $D_2$. The addition protocol is resumed in the following theorem:

\begin{thm}\label{theorem:sum}
For diagrams $D_1: n \rightarrow m$ and $D_2: n \rightarrow m$ the diagram

$D_+ = 
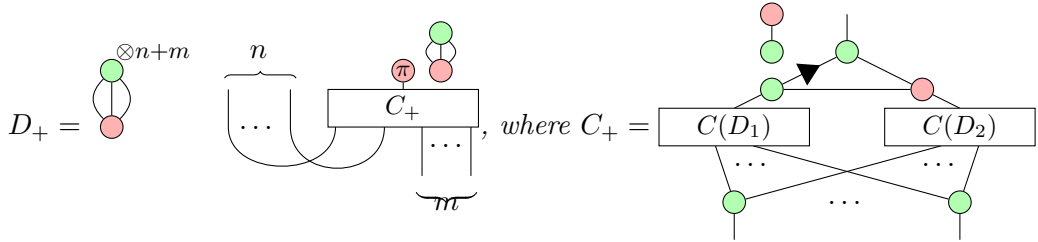
\begin{tikzpicture}
	\begin{pgfonlayer}{nodelayer}
		\node [style=none] (3) at (1, 0) {};
		\node [style=none] (4) at (1.35, -0.25) {$\dots$};
		\node [style=none] (5) at (1.65, 0) {};
		\node [style=none] (6) at (-0.15, 0) {};
		\node [style=none] (8) at (0.5, 0) {};
		\node [style=none] (9) at (-1.575, 0) {};
		\node [style=none] (10) at (-1.175, 0) {$\dots$};
		\node [style=none] (11) at (-0.75, 0) {};
		\node [style={big_diag}] (12) at (0.75, 0.25) {$C_+$};
		\node [style=none] (13) at (1, -0.65) {};
		\node [style=none] (14) at (1.65, -0.65) {};
		\node [style=none] (15) at (-1.575, 0.5) {};
		\node [style=none] (16) at (-0.75, 0.5) {};
		\node [style={x_dot}] (17) at (0.75, 0.75) {$\pi$};
		\node [style={x_dot}] (18) at (1.25, 0.75) {};
		\node [style={z_dot}] (19) at (1.25, 1.25) {};
		\node [style=none] (20) at (-1.625, 0.65) {};
		\node [style=none] (21) at (-0.675, 0.65) {};
		\node [style=none] (22) at (0.925, -0.8) {};
		\node [style=none] (23) at (1.725, -0.8) {};
		\node [style=none] (24) at (-1.175, 1) {$n$};
		\node [style=none] (25) at (1.325, -1.025) {$m$};
		\node [style={z_dot}] (26) at (-3.125, 0.75) {};
		\node [style={x_dot}] (27) at (-3.125, 0) {};
		\node [style=none] (28) at (-2.575, 1) {\footnotesize $\otimes n{+}m$};
	\end{pgfonlayer}
	\begin{pgfonlayer}{edgelayer}
		\draw (18) to (19);
		\draw [bend left=45, looseness=1.50] (18) to (19);
		\draw [bend right=45, looseness=1.50] (18) to (19);
		\draw (12) to (17);
		\draw (13.center) to (3.center);
		\draw (14.center) to (5.center);
		\draw [in=-90, out=-90, looseness=1.25] (9.center) to (6.center);
		\draw [in=270, out=-90, looseness=1.50] (11.center) to (8.center);
		\draw (9.center) to (15.center);
		\draw (11.center) to (16.center);
		\draw [style=underbrace] (21.center) to (20.center);
		\draw [style=brace] (23.center) to (22.center);
		\draw (26) to (27);
		\draw [bend right=45, looseness=1.25] (26) to (27);
		\draw [bend left=45, looseness=1.25] (26) to (27);
	\end{pgfonlayer}
\end{tikzpicture}

$, where $C_+ = 
\begin{tikzpicture}
	\begin{pgfonlayer}{nodelayer}
		\node [style={z_dot}] (0) at (-1, 0.5) {};
		\node [style={z_dot}] (1) at (0, 1) {};
		\node [style={x_dot}] (2) at (1, 0.5) {};
		\node [style=none] (3) at (-1.5, 0.25) {};
		\node [style=none] (4) at (1.5, 0.25) {};
		\node [style={big_diag}] (5) at (-1.5, 0) {$C(D_1)$};
		\node [style={big_diag}] (6) at (1.5, 0) {$C(D_2)$};
		\node [style={z_dot}] (7) at (-1.5, -1) {};
		\node [style={z_dot}] (8) at (1.5, -1) {};
		\node [style=none] (9) at (-1.75, -0.25) {};
		\node [style=none] (10) at (-1.25, -0.25) {};
		\node [style=none] (11) at (-1.25, -0.5) {$\dots$};
		\node [style=none] (12) at (1.25, -0.25) {};
		\node [style=none] (13) at (1.75, -0.25) {};
		\node [style=none] (14) at (1.25, -0.5) {$\dots$};
		\node [style=none] (15) at (-1.5, -1.5) {};
		\node [style=none] (16) at (1.5, -1.5) {};
		\node [style=none] (17) at (0, -1) {$\dots$};
		\node [style=none] (18) at (0, 1.5) {};
		\node [style={z_dot}] (19) at (-1, 1) {};
		\node [style={x_dot}] (20) at (-1, 1.5) {};
	\end{pgfonlayer}
	\begin{pgfonlayer}{edgelayer}
		\draw [style=plus-edge] (0) to (1);
		\draw (0) to (2);
		\draw (1) to (2);
		\draw (2) to (4.center);
		\draw (0) to (3.center);
		\draw (9.center) to (7);
		\draw (10.center) to (8);
		\draw (12.center) to (7);
		\draw (13.center) to (8);
		\draw (7) to (15.center);
		\draw (8) to (16.center);
		\draw (18.center) to (1);
		\draw (20) to (19);
	\end{pgfonlayer}
\end{tikzpicture}

$

is such that $\interp{D_+} = \interp{D_1} + \interp{D_2}$.
\end{thm}

\begin{proof}[Proof (theorem \ref{theorem:sum})]
The theorem follows from the definition of the controlizer and the lemma \ref{lemma:sum_and_product_of_cs}.
\qedhere
\end{proof}

We illustrate the  diagrammatic addition with a simple example.

\begin{exa}\label{example:sum}Using \ref{theorem:sum}, we construct a diagram $D$ corresponding to the addition of $
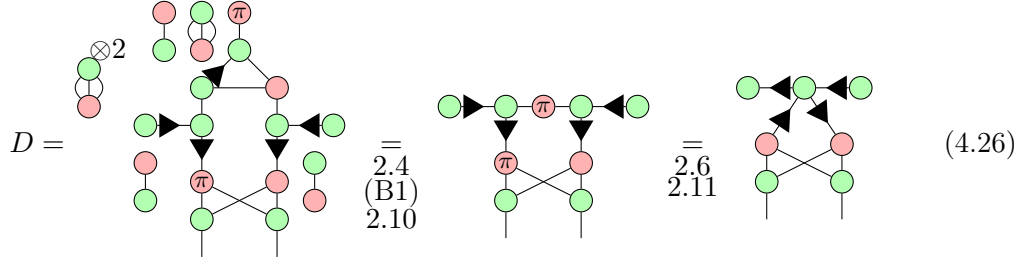
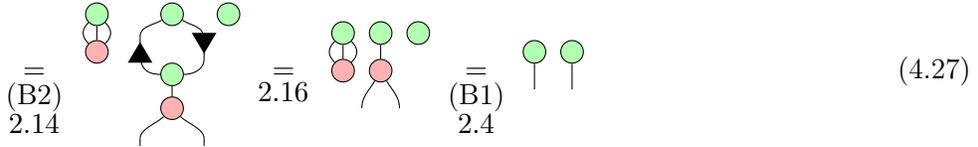
\begin{tikzpicture}
	\begin{pgfonlayer}{nodelayer}
		\node [style=none] (1) at (-0.25, -0.1) {};
		\node [style=none] (3) at (0.25, -0.1) {};
	\end{pgfonlayer}
	\begin{pgfonlayer}{edgelayer}
		\draw [in=-270, out=90, looseness=2.00] (1.center) to (3.center);
	\end{pgfonlayer}
\end{tikzpicture}
$ and $
\begin{tikzpicture}
	\begin{pgfonlayer}{nodelayer}
		\node [style={x_dot}] (0) at (0, 0.2) {$\pi$};
		\node [style=none] (1) at (-0.25, -0.225) {};
		\node [style=none] (2) at (0.25, -0.225) {};
	\end{pgfonlayer}
	\begin{pgfonlayer}{edgelayer}
		\draw [in=90, out=-135, looseness=1.25] (0) to (1.center);
		\draw [in=90, out=-45, looseness=1.25] (0) to (2.center);
	\end{pgfonlayer}
\end{tikzpicture}

$.

\begin{align}
D &= 
\begin{tikzpicture}
	\begin{pgfonlayer}{nodelayer}
		\node [style={z_dot}] (0) at (0.5, 0.25) {};
		\node [style={z_dot}] (1) at (1.25, 0.25) {};
		\node [style={x_dot}] (2) at (0.5, -0.5) {};
		\node [style=none] (3) at (0.5, 0.75) {};
		\node [style={z_dot}] (6) at (1, -0.25) {};
		\node [style={x_dot}] (7) at (1, -0.75) {};
		\node [style={x_dot}] (8) at (-0.5, -0.5) {$\pi$};
		\node [style={z_dot}] (9) at (0.5, -1) {};
		\node [style={z_dot}] (10) at (-0.5, -1) {};
		\node [style={z_dot}] (11) at (-1.25, -0.75) {};
		\node [style={x_dot}] (12) at (-1.25, -0.25) {};
		\node [style={z_dot}] (13) at (-0.5, 0.25) {};
		\node [style={z_dot}] (14) at (-1.25, 0.25) {};
		\node [style=none] (15) at (-0.5, 0.75) {};
		\node [style={z_dot}] (16) at (-0.5, 0.75) {};
		\node [style={z_dot}] (17) at (0, 1.25) {};
		\node [style={x_dot}] (18) at (0.5, 0.75) {};
		\node [style={x_dot}] (19) at (0, 1.75) {$\pi$};
		\node [style={z_dot}] (20) at (-1, 1.25) {};
		\node [style={x_dot}] (21) at (-1, 1.75) {};
		\node [style={z_dot}] (22) at (-0.5, 1.75) {};
		\node [style={x_dot}] (23) at (-0.5, 1.25) {};
		\node [style={z_dot}] (24) at (-2, 1) {};
		\node [style={x_dot}] (25) at (-2, 0.5) {};
		\node [style=none] (26) at (-1.75, 1.25) {$\otimes 2$};
		\node [style=none] (27) at (-0.5, -1.5) {};
		\node [style=none] (28) at (0.5, -1.5) {};
	\end{pgfonlayer}
	\begin{pgfonlayer}{edgelayer}
		\draw [style=plus-edge] (1) to (0);
		\draw [style=plus-edge] (0) to (2);
		\draw (3.center) to (0);
		\draw (6) to (7);
		\draw (8) to (9);
		\draw (8) to (10);
		\draw (12) to (11);
		\draw [style=plus-edge] (14) to (13);
		\draw (15.center) to (13);
		\draw [style=plus-edge] (13) to (8);
		\draw [style=plus-edge] (16) to (17);
		\draw (16) to (18);
		\draw (17) to (18);
		\draw (19) to (17);
		\draw (21) to (20);
		\draw (22) to (23);
		\draw [bend right=45, looseness=1.25] (22) to (23);
		\draw [bend left=45, looseness=1.25] (22) to (23);
		\draw (24) to (25);
		\draw [bend right=45, looseness=1.25] (24) to (25);
		\draw [bend left=45, looseness=1.25] (24) to (25);
		\draw (2) to (9);
		\draw (2) to (10);
		\draw (10) to (27.center);
		\draw (9) to (28.center);
	\end{pgfonlayer}
\end{tikzpicture}

\equal{ \ref{lemma:sqrt-2-sqrt-1-over-2} \\ \bo \\ \ref{lemma:pi-triangle-up}}

\begin{tikzpicture}
	\begin{pgfonlayer}{nodelayer}
		\node [style={z_dot}] (0) at (0.5, 0.5) {};
		\node [style={z_dot}] (1) at (1.25, 0.5) {};
		\node [style={x_dot}] (2) at (0.5, -0.25) {};
		\node [style={x_dot}] (8) at (-0.5, -0.25) {$\pi$};
		\node [style={z_dot}] (9) at (0.5, -0.75) {};
		\node [style={z_dot}] (10) at (-0.5, -0.75) {};
		\node [style={z_dot}] (13) at (-0.5, 0.5) {};
		\node [style={z_dot}] (14) at (-1.25, 0.5) {};
		\node [style={x_dot}] (19) at (0, 0.5) {$\pi$};
		\node [style=none] (27) at (-0.5, -1.25) {};
		\node [style=none] (28) at (0.5, -1.25) {};
	\end{pgfonlayer}
	\begin{pgfonlayer}{edgelayer}
		\draw [style=plus-edge] (1) to (0);
		\draw [style=plus-edge] (0) to (2);
		\draw (8) to (9);
		\draw (8) to (10);
		\draw [style=plus-edge] (14) to (13);
		\draw [style=plus-edge] (13) to (8);
		\draw (2) to (9);
		\draw (2) to (10);
		\draw (10) to (27.center);
		\draw (9) to (28.center);
		\draw (13) to (19);
		\draw (19) to (0);
	\end{pgfonlayer}
\end{tikzpicture}

\equal{\ref{lemma:pi-push} \\ \ref{lemma:pi-triangle-push}}

\begin{tikzpicture}
	\begin{pgfonlayer}{nodelayer}
		\node [style={z_dot}] (1) at (0.75, 0.75) {};
		\node [style={x_dot}] (2) at (0.5, 0) {};
		\node [style={x_dot}] (8) at (-0.5, 0) {};
		\node [style={z_dot}] (9) at (0.5, -0.5) {};
		\node [style={z_dot}] (10) at (-0.5, -0.5) {};
		\node [style={z_dot}] (13) at (0, 0.75) {};
		\node [style={z_dot}] (14) at (-0.75, 0.75) {};
		\node [style=none] (27) at (-0.5, -1) {};
		\node [style=none] (28) at (0.5, -1) {};
	\end{pgfonlayer}
	\begin{pgfonlayer}{edgelayer}
		\draw (8) to (9);
		\draw (8) to (10);
		\draw [style=plus-edge] (13) to (14);
		\draw [style=plus-edge] (8) to (13);
		\draw (2) to (9);
		\draw (2) to (10);
		\draw (10) to (27.center);
		\draw (9) to (28.center);
		\draw [style=plus-edge] (1) to (13);
		\draw [style=plus-edge] (13) to (2);
	\end{pgfonlayer}
\end{tikzpicture}

 \\
& \equal{\bt \\ \ref{lemma:green-dots-with-opposed-triangles}}

\begin{tikzpicture}
	\begin{pgfonlayer}{nodelayer}
		\node [style={z_dot}] (1) at (0.75, 0.75) {};
		\node [style=none] (2) at (0.425, 0.15) {};
		\node [style=none] (8) at (-0.425, 0.225) {};
		\node [style=none] (9) at (0.425, -0.9) {};
		\node [style=none] (10) at (-0.425, -0.9) {};
		\node [style={z_dot}] (13) at (0, 0.75) {};
		\node [style=none] (27) at (-0.425, -1) {};
		\node [style=none] (28) at (0.425, -1) {};
		\node [style={z_dot}] (29) at (-1, 0.75) {};
		\node [style={x_dot}] (30) at (-1, 0.25) {};
		\node [style={z_dot}] (31) at (0, -0.05) {};
		\node [style={x_dot}] (32) at (0, -0.5) {};
		\node [style=none] (33) at (-0.425, 0.45) {};
		\node [style=none] (34) at (0.425, 0.4) {};
	\end{pgfonlayer}
	\begin{pgfonlayer}{edgelayer}
		\draw [style=plus-edge] (8.center) to (33.center);
		\draw (10.center) to (27.center);
		\draw (9.center) to (28.center);
		\draw [style=plus-edge] (34.center) to (2.center);
		\draw (29) to (30);
		\draw [bend right=45, looseness=1.25] (29) to (30);
		\draw [bend left=45, looseness=1.25] (29) to (30);
		\draw (31) to (32);
		\draw [in=-180, out=-90, looseness=1.25] (8.center) to (31);
		\draw [in=-90, out=0] (31) to (2.center);
		\draw [in=225, out=90, looseness=0.75] (10.center) to (32);
		\draw [in=90, out=-45, looseness=0.75] (32) to (9.center);
		\draw [in=210, out=90] (33.center) to (13);
		\draw [in=90, out=-30, looseness=1.25] (13) to (34.center);
	\end{pgfonlayer}
\end{tikzpicture}

\equal{\ref{lemma:opposite-triangles}}

\begin{tikzpicture}
	\begin{pgfonlayer}{nodelayer}
		\node [style={z_dot}] (1) at (0.5, 0.5) {};
		\node [style=none] (27) at (-0.25, -0.5) {};
		\node [style=none] (28) at (0.25, -0.5) {};
		\node [style={z_dot}] (29) at (-0.5, 0.5) {};
		\node [style={x_dot}] (30) at (-0.5, 0) {};
		\node [style={z_dot}] (31) at (0, 0.5) {};
		\node [style={x_dot}] (32) at (0, 0) {};
	\end{pgfonlayer}
	\begin{pgfonlayer}{edgelayer}
		\draw (29) to (30);
		\draw [bend right=45, looseness=1.25] (29) to (30);
		\draw [bend left=45, looseness=1.25] (29) to (30);
		\draw (31) to (32);
		\draw [in=90, out=-60] (32) to (28.center);
		\draw [in=90, out=-120] (32) to (27.center);
	\end{pgfonlayer}
\end{tikzpicture}

\equal{\bo \\ \ref{lemma:sqrt-2-sqrt-1-over-2}}

\begin{tikzpicture}
	\begin{pgfonlayer}{nodelayer}
		\node [style={z_dot}] (1) at (0.25, 0.25) {};
		\node [style=none] (27) at (-0.25, -0.25) {};
		\node [style=none] (28) at (0.25, -0.25) {};
		\node [style={z_dot}] (31) at (-0.25, 0.25) {};
	\end{pgfonlayer}
	\begin{pgfonlayer}{edgelayer}
		\draw (31) to (27.center);
		\draw (1) to (28.center);
	\end{pgfonlayer}
\end{tikzpicture}

\label{eq:sum_example}
\end{align}
\end{exa}

In this example we simplified the diagram with graphical rewriting to obtain a known result:
\begin{align}
\interp{
\begin{tikzpicture}
	\begin{pgfonlayer}{nodelayer}
		\node [style=none] (1) at (-0.25, -0.1) {};
		\node [style=none] (3) at (0.25, -0.1) {};
	\end{pgfonlayer}
	\begin{pgfonlayer}{edgelayer}
		\draw [in=-270, out=90, looseness=2.00] (1.center) to (3.center);
	\end{pgfonlayer}
\end{tikzpicture}
}+\interp{
\begin{tikzpicture}
	\begin{pgfonlayer}{nodelayer}
		\node [style={x_dot}] (0) at (0, 0.2) {$\pi$};
		\node [style=none] (1) at (-0.25, -0.225) {};
		\node [style=none] (2) at (0.25, -0.225) {};
	\end{pgfonlayer}
	\begin{pgfonlayer}{edgelayer}
		\draw [in=90, out=-135, looseness=1.25] (0) to (1.center);
		\draw [in=90, out=-45, looseness=1.25] (0) to (2.center);
	\end{pgfonlayer}
\end{tikzpicture}

}=(|00\rangle + |11\rangle)+ (|01\rangle + |10\rangle) =2| ++ \rangle= \interp {
\begin{tikzpicture}
	\begin{pgfonlayer}{nodelayer}
		\node [style={z_dot}] (1) at (0.25, 0.25) {};
		\node [style=none] (27) at (-0.25, -0.25) {};
		\node [style=none] (28) at (0.25, -0.25) {};
		\node [style={z_dot}] (31) at (-0.25, 0.25) {};
	\end{pgfonlayer}
	\begin{pgfonlayer}{edgelayer}
		\draw (31) to (27.center);
		\draw (1) to (28.center);
	\end{pgfonlayer}
\end{tikzpicture}

}
\end{align}

%Relation to coherent unitary control

\subsection{Discussion}\label{subsection:addition}

The previous example shows that our inductive procedure leads to large diagrams, even when the initial diagrams are fairly simple. The resulting diagrams roughly follow the shape of the initial diagrams, but are still quite different. This defect is however consistent with intuition. Indeed, contrary to sequential and parallel compositions, the addition is not a physical operation, hence it is not surprising that can't be easily incorporated into the framework of ZX-diagrams. On the positive side, we can rely on the powerful equation theory of the ZX-calculus to simplify, when it is possible, the diagrams representing the sum of diagrams. 

An alternative approach for the addition in the algebraic ZXW-calculus was presented in \cite{zx_addition_oxford}. It also relies on the controlled forms of the diagram, but their controlled forms are different from ours. Notably, contrary to our approach that accepts diagrams with an arbitrary number of input and output wires, the controlled forms in \cite{zx_addition_oxford} are defined only for two families of diagrams: state diagrams $\ZX[0, n]$ and diagrams $\ZX[m, m]$ that correspond to \textit{square matrices}. 
Controlled matrices \cite{zx_addition_oxford} are built from the decomposition of the initial diagram $D$ on a product of diagrams for so-called \textit{elementary matrices} \cite{elementary_matrix}. In the general case, the decomposition on elementary matrices seems to require an explicit computation of the matrix corresponding to the diagram. Therefore, although for given controlled forms the addition from \cite{zx_addition_oxford} is done diagrammatically with the help of \textit{W-spider}, the computation of the controlled forms uses the semantics. In contrast, in our inductive procedure the entire process is maintained in the graphical framework.

However, in some special cases of practical importance the approach suggested in \cite{zx_addition_oxford} is beneficial. For instance, as was shown in \cite{zx_addition_oxford} that for local Hamiltonians made out of Pauli tensors the decomposition on elementary matrices is straightforward. As a consequence, the controlled versions are easy to compute and the overall procedure leads to better-looking diagrams. 

\section{Differentiation of ZX-diagrams}\label{section:differentiation}

A basic definition of the \textit{derivative} is the derivative for a smooth function $f:\mathbb{R} \rightarrow \mathbb{R}$:
\begin{align}
\frac{\partial f}{\partial x}(x_0) = \lim_{\delta \rightarrow 0} \frac{f(x_0 + \delta) - f(x_0)}{\delta}\label{calculus_derivative}
\end{align}

In traditional calculus the derivative usually is computed from the decomposition of the function $f:\mathbb{R} \rightarrow \mathbb{R}$ on a sum, product, and composition of elementary functions. Derivatives for compositions of function functions are computed using linearity and the product rules: $\partial (f \circ g) = f \circ \partial g + \partial f \circ g $ and similarly for the tensor product.

A parameterized ZX-diagram $D(\beta)$ can be understood as a function $f: \mathbb{R} \rightarrow \ZX$ that associates to each real value $\beta \in \mathbb{R}$ a ZX-diagram. In the context of variational algorithms, we consider the derivatives over the \textit{parameters}. 

In order to keep the notations simple in what follows we restrict our attention to the set of linear diagrams over one variable, denoted by $\ZX(\beta)$. We denote the corresponding matrices by $\Mat(\beta)$. The elements of matrices in $\Mat(\beta)$ belong to the ring generated by complex numbers and the smooth functions $e^{i \mathcal{A}}$ where $\mathcal{A} = \{k\beta \;| \; k \in \mathbb{Z}\}$ is the group of affine functions with integer coefficients.
We remark that even if we restrict the consideration to diagrams over one variable, all results are easily extended to the case of partial derivatives $\partial_{\beta_i}$ for linear diagrams with an arbitrary number of variables mutatis mutandis.

The sum for the matrices in $\Mat(\beta)$ is naturally defined by entry-wise addition. Therefore, we can define the derivative in $\Mat(\beta)$ by specifying the general definition for monoidal categories with sums given in \cite{zx_differentation_toumi}:

\begin{defi}[Derivative in $\Mat(\beta)$]\label{def:lm_derivative}
A derivative $\partial_M$ in $\Mat(\beta)$ is a linear unary operator on $\Mat(\beta)$ that associates to a matrix $L \in \Mat(\beta), \; L:n \rightarrow m$ another matrix $L' \in \Mat(\beta), \; L':n \rightarrow m$ of the same dimensions. It satisfies the following axioms:
\item $\circ$-product rule :
$\partial_M[A \circ B] = \partial_M[A] \circ B + A \circ \partial_M[B]$
\item $\otimes$-product rule : 
$\partial_M[A \otimes B] = \partial_M[A] \otimes B + A \otimes \partial_M[B]$
\end{defi}

Rather than defining the derivative as a limit of finite difference and deriving the product rules, the work \cite{zx_differentation_toumi} suggests an axiomatic definition that proclaims \textit{product rules} as the fundamental property of the derivative operator. We remark that the entry-wise differentiation of matrix elements satisfies this definition. 

Although our theorem \ref{theorem:sum} provides a fully diagrammatic way for the addition of diagrams, we avoid extending the language with a formal introduction of a sum operator in order to circumvent unnecessary complications. Therefore, the general definition from \cite{zx_differentation_toumi} doesn't apply. Instead, we suggest an alternative semantics for the derivative in the $\ZX$-calculus requiring the coherence between the derivative of a diagram and the derivative of the corresponding linear map:

\begin{defi}[$\ZX$-derivative]
\label{def:diagrammatic_derivative}
A derivative $\partial_{\ZX}: \ZX[n, m] \rightarrow \ZX[n, m]$ is a unary operator that commutes with the standard interpretation:
\begin{align}
\interp{\partial_{ZX} D} = \partial_{M}\interp{D} \label{eq:derivative_definition}
\end{align}
where $\partial_M$ is an (arbitrary) fixed derivative in $\Mat(\beta)$ satisfying the definition (\ref{def:lm_derivative}).
\end{defi}

The property (\ref{eq:derivative_definition}) is called \textit{diagrammatic differentiation}.

\subsection{Diagrammatic differentiation with controlizers}\label{section:inductive_differentiation}

The derivative in $\Mat(\beta)$ is defined through product rules that involve sums that will be translated to the diagrammatic framework using \textit{controlizers}. 

\begin{nota}
In what follows we denote by $C$ any map that satisfies the definition \ref{def:controlizer} of controlizer, for instance the one of \ref{example:controlizer}.
\end{nota}

\begin{defi}\label{def:c-derivative}
  Given a controlizer $C$, the  $C$-derivative is the map $\Delta: \ZX(\beta) \rightarrow \ZX(\beta)$ that associates to a parameterized diagram $D:n \rightarrow m$ another diagram $\Delta\left(D\right): 1 \rightarrow n + m$ defined as follows:
\begin{itemize}
\item \textbf{Generators:}
For parametrized spiders:
\begin{align}
&\Delta\left(\rbeta\right) = 
\input{./ZX_diagrams/c-derivative/ketbeta.tikz}
, \quad \Delta\left(\rminusbeta\right) = 
\input{./ZX_diagrams/c-derivative/ketminusbeta.tikz}
 \\ &\Delta\left(
\input{./ZX_diagrams/c-derivative/rdot.tikz}
\right) = \Delta\left(
\input{./ZX_diagrams/c-derivative/rdot_delta.tikz}
\right), 
\quad \Delta\left(
\input{./ZX_diagrams/c-derivative/gdot.tikz}
\right) = 
\Delta\left(
\input{./ZX_diagrams/c-derivative/gdot_delta.tikz}
\right)
\end{align}

For all other generators $g:n \rightarrow m$ that do not contain an instance of the variable $\beta$, we define $\Delta(g) = 
\input{./ZX_diagrams/controlled_states/scalar_zero.tikz}
 \underbrace{\gket \dots \gket}_{n+m}$. 

\item \textbf{Tensor product:} for $D_1:n \rightarrow m$ and $D_2: k \rightarrow l$ the diagram $\Delta(D_1 \otimes D_2)$ is:
\begin{align}
\Delta(D_1 \otimes D_2) = \tikzformula{c-derivative/tensor_product}
\end{align}
\item \textbf{Composition:} for $D_1:n \rightarrow m$ and $D_2: m \rightarrow k$ the diagram $\Delta(D_2 \circ D_1)$ is:
\begin{align}
\Delta(D_2 \circ D_1) = \tikzformula{c-derivative/composition}
\end{align}
\end{itemize}
\end{defi}

As should be clear from the definition, the $C$-derivative of a diagram is not derivative of a diagram. Rather, it corresponds to a controlled-version of the derivative.

A simple verification shows that for any generator $G$ the $C$-derivative $\Delta\left(G\right)$ is a controlled state. From lemma \ref{lemma:sum_and_product_of_cs} it follows that for every diagram $D: n \rightarrow m$ the $C$-derivative $\Delta\left(D\right): 1 \rightarrow n+m$ is a controlled state. Moreover, as was proven in lemma \ref{lemma:controlled_zero_add_and_mult} the $C$-derivative $
\input{./ZX_diagrams/controlled_states/scalar_zero.tikz}
 \underbrace{\gket \dots \gket}_{n+m}$ of generators that are independent on $\beta$ is precisely the zero-element $C_0^{n+m}$ for the addition of controlled states. The same lemma implies, in particular, that the $C$-derivative of any diagram $D:n \rightarrow m$ that is independent on $\beta$ is $
\input{./ZX_diagrams/controlled_states/scalar_zero.tikz}
 \underbrace{\gket \dots \gket}_{n+m}$.

Similarly to controlizers, a step-by-step application of the map $\Delta$ may lead to different diagrams depending on the order of decomposition on tensor products and compositions. However, all possible outputs are semantically equivalent and, by the completeness of $\ZX$-calculus, are equivalent as diagrams. 

By combining $C$-derivatives and controlizers we can reproduce the semantics of product rules in a graphical representation:

\begin{defi}\label{def:partial_c}
Given a $C$-derivative $\Delta$, let $\partial_C:\ZX(\beta) \rightarrow \ZX(\beta)$ be the  unary operator such that for any diagram $D: n \rightarrow m$
\begin{align}
\partial_C[D] = 
\input{./ZX_diagrams/c-derivative/partial_c.tikz}

\end{align} 
\end{defi}

\begin{thm}\label{theorem:partial_c}
The operator $\partial_C$ satisfies the definition (\ref{eq:derivative_definition}) of diagrammatic differentiation.
\end{thm}

\begin{proof}[Proof (theorem \ref{theorem:partial_c})]
We give the proof by induction over tensor product and composition.
\begin{itemize}
\item \textbf{Generators}: First, we show that $\partial_M\left[ \interp{\rbeta}\right] = i e^{i\beta}|-\rangle$ and $\partial_M\left[ \interp{\rminusbeta}\right] = -i e^{-i\beta}|-\rangle$. Indeed, we have:

\begin{align}
\partial_M\interp{\rbeta} = \partial_M\left(|+\rangle + e^{i\beta}|-\rangle\right) = ie^{i\beta}|-\rangle \\
\partial_M\interp{\rminusbeta} = \partial_M\left(|+\rangle + e^{-i\beta}|-\rangle\right) = -ie^{-i\beta}|-\rangle
\end{align}

For the diagrams $\partial_C\left(\rbeta\right)$ and $\partial_C\left(\rminusbeta\right)$ we verify the property of diagrammatic differentiation:
\begin{align}
\interp{\diagi \normalizer \normalizer 
\input{./ZX_diagrams/c-derivative/ketbeta1.tikz}
} =\interp{ 
\begin{tikzpicture}
	\begin{pgfonlayer}{nodelayer}
		\node [style={x_dot}] (1) at (0.75, 0) {$\beta$};
		\node [style={z_dot}] (4) at (0.75, 0.5) {$\pi$};
		\node [style=none] (5) at (0.75, -0.5) {};
		\node [style={x_dot}] (6) at (-0.25, 0.5) {};
		\node [style={z_dot}] (7) at (-0.25, 0) {};
		\node [style={z_dot}] (8) at (-0.75, 0.5) {\small$\pi$};
		\node [style={x_dot}] (9) at (-0.75, 0) {\tiny$\frac{\pi}{2}$};
		\node [style={x_dot}] (10) at (0.25, 0.5) {};
		\node [style={z_dot}] (11) at (0.25, 0) {};
	\end{pgfonlayer}
	\begin{pgfonlayer}{edgelayer}
		\draw (1) to (5.center);
		\draw (4) to (1);
		\draw (6) to (7);
		\draw [bend right=45] (6) to (7);
		\draw [bend left=45] (6) to (7);
		\draw (8) to (9);
		\draw (10) to (11);
		\draw [bend right=45] (10) to (11);
		\draw [bend left=45] (10) to (11);
	\end{pgfonlayer}
\end{tikzpicture}

} = ie^{i\beta}|-\rangle
\end{align}

\begin{align}
\interp{\diagi \normalizer \normalizer 
\input{./ZX_diagrams/c-derivative/ketminusbeta1.tikz}
} =\interp{ 
\begin{tikzpicture}
	\begin{pgfonlayer}{nodelayer}
		\node [style={x_dot}] (1) at (0, -0.25) {$-\beta$};
		\node [style={x_dot}] (2) at (-1.5, -0.25) {};
		\node [style={z_dot}] (3) at (-1.5, 0.25) {};
		\node [style={z_dot}] (4) at (0, 0.5) {$\pi$};
		\node [style=none] (5) at (0, -0.75) {};
		\node [style={z_dot}] (7) at (-1, -0.25) {$\pi$};
		\node [style={x_dot}] (8) at (-1, 0.25) {$\pi$};
		\node [style={x_dot}] (9) at (-2.5, 0.25) {};
		\node [style={z_dot}] (10) at (-2.5, -0.25) {};
		\node [style={z_dot}] (11) at (-3, 0.25) {\small$\pi$};
		\node [style={x_dot}] (12) at (-3, -0.25) {\tiny$\frac{\pi}{2}$};
		\node [style={x_dot}] (13) at (-2, 0.25) {};
		\node [style={z_dot}] (14) at (-2, -0.25) {};
		\node [style={x_dot}] (15) at (-0.5, 0.25) {};
		\node [style={z_dot}] (16) at (-0.5, -0.25) {};
	\end{pgfonlayer}
	\begin{pgfonlayer}{edgelayer}
		\draw (1) to (5.center);
		\draw (3) to (2);
		\draw (8) to (7);
		\draw (4) to (1);
		\draw (9) to (10);
		\draw [bend right=45] (9) to (10);
		\draw [bend left=45] (9) to (10);
		\draw (11) to (12);
		\draw (13) to (14);
		\draw [bend right=45] (13) to (14);
		\draw [bend left=45] (13) to (14);
		\draw (15) to (16);
		\draw [bend right=45] (15) to (16);
		\draw [bend left=45] (15) to (16);
	\end{pgfonlayer}
\end{tikzpicture}

} = -ie^{-i\beta}|-\rangle
\end{align}

For a generator $g:n\rightarrow m$ that doesn't depend on $\beta$:
\begin{align}
\interp{\partial_C(g)} =\interp{
\begin{tikzpicture}
	\begin{pgfonlayer}{nodelayer}
		\node [style={x_dot}] (3) at (0, 0.25) {$\pi$};
		\node [style={x_dot}] (8) at (0, -0.25) {};
		\node [style={z_dot}] (9) at (0.75, -0.25) {};
		\node [style={z_dot}] (10) at (1.75, -0.25) {};
		\node [style=none] (11) at (1.25, -0.5) {$\dots$};
		\node [style=none] (12) at (0.75, -0.75) {};
		\node [style=none] (13) at (1.75, -0.75) {};
		\node [style={x_dot}] (14) at (-2.25, 0.25) {};
		\node [style={z_dot}] (15) at (-2.25, -0.25) {};
		\node [style={z_dot}] (16) at (-2.75, 0.25) {\small$\pi$};
		\node [style={x_dot}] (17) at (-2.75, -0.25) {\tiny$\frac{\pi}{2}$};
		\node [style={x_dot}] (18) at (-1.75, 0.25) {};
		\node [style={z_dot}] (19) at (-1.75, -0.25) {};
		\node [style=none] (20) at (-1, 0.5) {\tiny$n+m+2$};
		\node [style={z_dot}] (21) at (0.75, 0.25) {};
		\node [style={z_dot}] (22) at (1.75, 0.25) {};
		\node [style=none] (23) at (1.25, 0.5) {$\dots$};
		\node [style=none] (24) at (0.75, 0.75) {};
		\node [style=none] (25) at (1.75, 0.75) {};
		\node [style=none] (26) at (0.5, 0.75) {};
		\node [style=none] (27) at (2, 0.75) {};
		\node [style=none] (28) at (1.25, 1.25) {$n$};
		\node [style=none] (29) at (0.5, -0.75) {};
		\node [style=none] (30) at (2, -0.75) {};
		\node [style=none] (31) at (1.25, -1.25) {$m$};
	\end{pgfonlayer}
	\begin{pgfonlayer}{edgelayer}
		\draw (3) to (8);
		\draw (9) to (12.center);
		\draw (10) to (13.center);
		\draw (14) to (15);
		\draw [bend right=45] (14) to (15);
		\draw [bend left=45] (14) to (15);
		\draw (16) to (17);
		\draw (18) to (19);
		\draw [bend right=45] (18) to (19);
		\draw [bend left=45] (18) to (19);
		\draw (21) to (24.center);
		\draw (22) to (25.center);
		\draw [style=brace] (26.center) to (27.center);
		\draw [style=brace] (30.center) to (29.center);
	\end{pgfonlayer}
\end{tikzpicture}

} \equal{\soo} \interp{
\begin{tikzpicture}
	\begin{pgfonlayer}{nodelayer}
		\node [style={x_dot}] (3) at (0, 0) {$\pi$};
	\end{pgfonlayer}
\end{tikzpicture}

 \otimes \dots} = \begin{pmatrix}0\end{pmatrix}_{n \times m}
\end{align}

\item \textbf{Tensor product}: Let's assume that $\interp{\partial_C(D_1)} = \partial_M\interp{D_1}$ and $\interp{\partial_C(D_2)} = \partial_M \interp{D_2}$ for $D_1: n \rightarrow m$ and $D_2: l\rightarrow k$. For $D = D_1\otimes D_2$:
\begin{align}
\interp{\partial_C(D)} = \interp{\tikzformula{c-derivative/proof/partial_c/tensor_product/step1}}
\end{align} 

By applying the definition \ref{def:c-derivative} of the $C$-derivative for the tensor product we obtain the diagram:
\begin{align}
\interp{\tikzformula{c-derivative/proof/partial_c/tensor_product/step2}}\label{eq:c-derivative-tensor-1}
\end{align}

We denote by $\tikzformula{c-derivative/proof/partial_c/tensor_product/t}$ the part of the diagram surrounded by the red frame. According to the definition, $C$-derivatives $\Delta(D_1)$, $\Delta(D_2)$ and controlizer outputs $\controlizer{D_1}$, $\controlizer{D_2}$ are controlled states. Therefore, from lemma \ref{lemma:sum_and_product_of_cs} we get:

\begin{align}
\interp{\tikzformula{c-derivative/proof/partial_c/tensor_product/t}} & = \interp{\tikzformula{c-derivative/proof/partial_c/tensor_product/step3_d2}} +  \interp{\tikzformula{c-derivative/proof/partial_c/tensor_product/step3_d1}} \label{zx_diag:t}
\end{align}

Meanwhile, by bending wires we see that the resting part of the diagram (\ref{eq:c-derivative-tensor-1}) is:
\begin{align}
B = \tikzformula{c-derivative/proof/partial_c/tensor_product/step3}
\end{align}

The sum in $\Mat(\beta)$ is distributive over the composition. Therefore,
\begin{align}
& \interp{\tikzformula{c-derivative/proof/partial_c/tensor_product/step4_0}}  = \\ 
&\qquad \interp{\tikzformula{c-derivative/proof/partial_c/tensor_product/step4}} \\
+ &\qquad \interp{\tikzformula{c-derivative/proof/partial_c/tensor_product/step4_1}}
\end{align} 
The claim follows from the induction assumption and the definition of controlizer.
\item \textbf{Composition:} Following the same reasoning as for the tensor product we obtain for $D_1:n\rightarrow m$ and $D_2:m \rightarrow l$
\begin{align}
&\partial_C[ D_2 \circ D_1] = \tikzformula{c-derivative/proof/partial_c/composition/s1}
\end{align}
where the diagram $T$ is exactly the same as in (\ref{zx_diag:t}).
On more time, by distributivity of the sum in $\Mat(\beta)$ over the composition we obtain
\begin{align}
\interp{\partial_C[ D_2 \circ D_1]} & = \tikzformula{c-derivative/proof/partial_c/composition/s2} \\
&\qquad + \tikzformula{c-derivative/proof/partial_c/composition/s3}
\end{align}
The claim follows from the definition of controlizer an the induction assumption. \qedhere
\end{itemize}
\end{proof}

We consider a simple example to illustrate how to perform the inductive procedure.

\begin{exa}\label{example:cos_bad}
We apply the definition \ref{def:partial_c} to the simple scalar diagram $
\begin{tikzpicture}
	\begin{pgfonlayer}{nodelayer}
		\node [style={x_dot}] (17) at (-0.5, 0.25) {\small$-\beta$};
		\node [style={x_dot}] (18) at (0.5, 0.25) {\small$\beta$};
		\node [style={z_dot}] (22) at (0, -0.25) {$\pi$};
	\end{pgfonlayer}
	\begin{pgfonlayer}{edgelayer}
		\draw (17) to (22);
		\draw (22) to (18);
	\end{pgfonlayer}
\end{tikzpicture}

 = 
\begin{tikzpicture}
	\begin{pgfonlayer}{nodelayer}
		\node [style={x_dot}] (17) at (-0.5, 0.5) {\tiny$-\beta$};
		\node [style={x_dot}] (18) at (0.5, 0.5) {\small$\beta$};
		\node [style={h_box}] (19) at (-0.5, 0) {};
		\node [style={h_box}] (20) at (0.5, 0) {};
		\node [style={x_dot}] (21) at (0, -0.25) {};
		\node [style={x_dot}] (22) at (0, -0.25) {$\pi$};
	\end{pgfonlayer}
	\begin{pgfonlayer}{edgelayer}
		\draw (17) to (19);
		\draw (19) to (21);
		\draw (21) to (20);
		\draw (18) to (20);
		\draw (21) to (22);
	\end{pgfonlayer}
\end{tikzpicture}

 $.
Notice that $C\left(
\begin{tikzpicture}
	\begin{pgfonlayer}{nodelayer}
		\node [style=none] (0) at (0, 0.25) {};
		\node [style=none] (1) at (0, -0.25) {};
		\node [style={h_box}] (2) at (0, 0) {};
	\end{pgfonlayer}
	\begin{pgfonlayer}{edgelayer}
		\draw (0.center) to (2);
		\draw (2) to (1.center);
	\end{pgfonlayer}
\end{tikzpicture}

 \otimes 
\begin{tikzpicture}
	\begin{pgfonlayer}{nodelayer}
		\node [style=none] (0) at (0, 0.25) {};
		\node [style=none] (1) at (0, -0.25) {};
		\node [style={h_box}] (2) at (0, 0) {};
	\end{pgfonlayer}
	\begin{pgfonlayer}{edgelayer}
		\draw (0.center) to (2);
		\draw (2) to (1.center);
	\end{pgfonlayer}
\end{tikzpicture}

\right) =
\input{./ZX_diagrams/cos_example/controlizer/2_hadamards.tikz}
 \equal{\ref{lemma:cs-plus-commutation} \\ (\ref{d:triangle}) \\ \ref{lemma:pi-push}} 
\input{./ZX_diagrams/cos_example/controlizer/2_hadamards_simplified.tikz}
$, moreover $C(
\begin{tikzpicture}
	\begin{pgfonlayer}{nodelayer}
		\node [style={x_dot}] (0) at (0, 0.2) {$\pi$};
		\node [style=none] (1) at (-0.25, -0.225) {};
		\node [style=none] (2) at (0.25, -0.225) {};
	\end{pgfonlayer}
	\begin{pgfonlayer}{edgelayer}
		\draw [in=90, out=-135, looseness=1.25] (0) to (1.center);
		\draw [in=90, out=-45, looseness=1.25] (0) to (2.center);
	\end{pgfonlayer}
\end{tikzpicture}

)$ was already found in the example \ref{example:controlled_pi}. We obtain a diagram for $C\left(
\input{./ZX_diagrams/cos_example/controlizer/bottom.tikz}
\right)$:
\begin{align}

\input{./ZX_diagrams/cos_example/controlizer/bottom_1.tikz}
 
\equal{\ref{lemma:same-direction-triangles-link-removal} }

\input{./ZX_diagrams/cos_example/controlizer/bottom_2.tikz}
 
\equal{\ref{lemma:same-direction-triangles-link-removal}}

\input{./ZX_diagrams/cos_example/controlizer/bottom_3.tikz}
 
\equal{\ref{lemma:opposite-triangles}}

\input{./ZX_diagrams/cos_example/controlizer/bottom_final.tikz}

\end{align}
We know from \ref{lemma:controlled_zero_add_and_mult} that $\Delta\left(
\input{./ZX_diagrams/cos_example/controlizer/bottom.tikz}
\right) = \tikzformula{controlled_states/scalar_zero} 
\begin{tikzpicture}
	\begin{pgfonlayer}{nodelayer}
		\node [style={z_dot}] (1) at (0.25, 0.25) {};
		\node [style=none] (27) at (-0.25, -0.25) {};
		\node [style=none] (28) at (0.25, -0.25) {};
		\node [style={z_dot}] (31) at (-0.25, 0.25) {};
	\end{pgfonlayer}
	\begin{pgfonlayer}{edgelayer}
		\draw (31) to (27.center);
		\draw (1) to (28.center);
	\end{pgfonlayer}
\end{tikzpicture}

$. By definition, $\Delta\left(\rminusbeta \rbeta\right) = 
\input{./ZX_diagrams/cos_example/controlizer/betas.tikz}
$ and $\partial_C\left(
\begin{tikzpicture}
	\begin{pgfonlayer}{nodelayer}
		\node [style={x_dot}] (17) at (-0.5, 0.5) {\tiny$-\beta$};
		\node [style={x_dot}] (18) at (0.5, 0.5) {\small$\beta$};
		\node [style={h_box}] (19) at (-0.5, 0) {};
		\node [style={h_box}] (20) at (0.5, 0) {};
		\node [style={x_dot}] (21) at (0, -0.25) {};
		\node [style={x_dot}] (22) at (0, -0.25) {$\pi$};
	\end{pgfonlayer}
	\begin{pgfonlayer}{edgelayer}
		\draw (17) to (19);
		\draw (19) to (21);
		\draw (21) to (20);
		\draw (18) to (20);
		\draw (21) to (22);
	\end{pgfonlayer}
\end{tikzpicture}

\right) = 
\input{./ZX_diagrams/cos_example/controlizer/final.tikz}
$
\end{exa}

\subsection{Formula for derivatives in \texorpdfstring{$\ZX(\beta)$}{ZX(β)}}\label{section:diff_with_formulas}

Although perfectly correct, the differentiation procedure described above leads to very puzzling output even for small diagrams (see example \ref{example:cos_bad}). In this section we provide a simpler approach to obtain the derivative of a diagram in $\ZX(\beta)$ written in a special form. We formalize it in definitions $\partial_{\ZX}$ and $\partial_P$ of unary operators that satisfy the property of diagrammatic differentiation (\ref{eq:derivative_definition}). 

First, we remark that even in the inductive procedure the effort can be significantly reduced if we separate the parts that depend on the parameter from the rest of the diagram that is constant on $\beta$. Indeed, for constant diagrams $A:k \rightarrow l$ and $B:m \rightarrow k$ that do not depend on $\beta$, the $C$-derivative is the controlled version of the zero diagrams $C_0^{k+l}$ and $C_0^{m+k}$. Therefore, for a parameterized diagram $D(\beta): n \rightarrow m$ the lemma \ref{lemma:controlled_zero_add_and_mult} implies :
\begin{align}
\Delta(D(\beta) \otimes A) = \tikzformula{c-derivative/constant/tensor_product}, \quad \Delta(D(\beta) \circ A) = \tikzformula{c-derivative/constant/composition} 
\end{align}
These expressions are significantly simpler that the general case (\ref{def:c-derivative}) where both parts of the compositions depend on $\beta$.

We want to further explore this property to find simplified formulas for the derivative.
\bigskip
Let's denote by $X_\beta(n,m)$ the family of diagrams 
$\underbrace{\rbeta \dots \rbeta}_{n}\underbrace{\rminusbeta \dots \rminusbeta}_{m}$ from $\ZX(\beta)$.

\begin {clm}\label{claim:beta_factored}
Using the rules of ZX-calculus, each diagram $D(\beta)$ from $ZX(\beta)$ may be transformed into the form 
\begin{align}
D(\beta) = 
\begin{tikzpicture}
	\begin{pgfonlayer}{nodelayer}
		\node [style=none] (0) at (0, 0) {$D_2$};
		\node [style=none] (1) at (-1.75, 0.5) {};
		\node [style=none] (2) at (-1.75, -0.5) {};
		\node [style=none] (3) at (1.75, -0.5) {};
		\node [style=none] (4) at (1.75, 0.5) {};
		\node [style=none] (5) at (-1, 1.25) {$D_1$};
		\node [style=none] (6) at (-1.75, 1.5) {};
		\node [style=none] (7) at (-1.75, 1) {};
		\node [style=none] (8) at (-0.25, 1) {};
		\node [style=none] (9) at (-0.25, 1.5) {};
		\node [style=none] (10) at (1, 1.5) {$X_\beta(n, m)$};
		\node [style=none] (11) at (0, 2) {};
		\node [style=none] (12) at (0, 1) {};
		\node [style=none] (13) at (2, 1) {};
		\node [style=none] (14) at (2, 2) {};
		\node [style=none] (15) at (-1.5, 1) {};
		\node [style=none] (16) at (-0.5, 1) {};
		\node [style=none] (17) at (-1.5, 0.5) {};
		\node [style=none] (18) at (-0.5, 0.5) {};
		\node [style=none] (19) at (0.5, 1) {};
		\node [style=none] (20) at (1.5, 1) {};
		\node [style=none] (21) at (0.5, 0.5) {};
		\node [style=none] (22) at (1.5, 0.5) {};
		\node [style=none] (23) at (1, 0.75) {$\dots$};
		\node [style=none] (24) at (-1, 0.75) {$\dots$};
		\node [style=none] (25) at (-1.5, 1.5) {};
		\node [style=none] (26) at (-0.5, 1.5) {};
		\node [style=none] (27) at (-0.5, 2) {};
		\node [style=none] (28) at (-1.5, 2) {};
		\node [style=none] (29) at (-1, 1.75) {$\dots$};
		\node [style=none] (30) at (-1.25, -1) {};
		\node [style=none] (31) at (1.25, -1) {};
		\node [style=none] (32) at (1.25, -0.5) {};
		\node [style=none] (33) at (-1.25, -0.5) {};
		\node [style=none] (34) at (0, -0.75) {$\dots$};
	\end{pgfonlayer}
	\begin{pgfonlayer}{edgelayer}
		\draw (1.center) to (4.center);
		\draw (4.center) to (3.center);
		\draw (1.center) to (2.center);
		\draw (2.center) to (3.center);
		\draw (6.center) to (9.center);
		\draw (9.center) to (8.center);
		\draw (6.center) to (7.center);
		\draw (7.center) to (8.center);
		\draw (11.center) to (14.center);
		\draw (14.center) to (13.center);
		\draw (11.center) to (12.center);
		\draw (12.center) to (13.center);
		\draw (28.center) to (25.center);
		\draw (15.center) to (17.center);
		\draw (27.center) to (26.center);
		\draw (16.center) to (18.center);
		\draw (19.center) to (21.center);
		\draw (20.center) to (22.center);
		\draw (33.center) to (30.center);
		\draw (32.center) to (31.center);
	\end{pgfonlayer}
\end{tikzpicture}

 = 
\begin{tikzpicture}
	\begin{pgfonlayer}{nodelayer}
		\node [style=none] (0) at (0.5, 0) {$D_2$};
		\node [style=none] (1) at (-2, 0.5) {};
		\node [style=none] (2) at (-2, -0.5) {};
		\node [style=none] (3) at (2.75, -0.5) {};
		\node [style=none] (4) at (2.75, 0.5) {};
		\node [style=none] (5) at (-1.25, 1) {$D_1$};
		\node [style=none] (6) at (-2, 1.25) {};
		\node [style=none] (7) at (-2, 0.75) {};
		\node [style=none] (8) at (-0.5, 0.75) {};
		\node [style=none] (9) at (-0.5, 1.25) {};
		\node [style=none] (15) at (-1.75, 0.75) {};
		\node [style=none] (16) at (-0.75, 0.75) {};
		\node [style=none] (17) at (-1.75, 0.5) {};
		\node [style=none] (18) at (-0.75, 0.5) {};
		\node [style={x_dot}] (19) at (0, 1.25) {$\beta$};
		\node [style={x_dot}] (20) at (2.5, 1.25) {$-\beta$};
		\node [style=none] (21) at (0, 0.5) {};
		\node [style=none] (22) at (2.5, 0.5) {};
		\node [style=none] (23) at (0.5, 1) {$\dots$};
		\node [style=none] (24) at (-1.25, 0.75) {$\dots$};
		\node [style=none] (25) at (-1.75, 1.25) {};
		\node [style=none] (26) at (-0.75, 1.25) {};
		\node [style=none] (27) at (-0.75, 1.5) {};
		\node [style=none] (28) at (-1.75, 1.5) {};
		\node [style=none] (29) at (-1.25, 1.5) {$\dots$};
		\node [style=none] (30) at (-1.5, -0.75) {};
		\node [style=none] (31) at (2.25, -0.75) {};
		\node [style=none] (32) at (2.25, -0.5) {};
		\node [style=none] (33) at (-1.5, -0.5) {};
		\node [style=none] (34) at (0.5, -0.75) {$\dots$};
		\node [style={x_dot}] (35) at (1, 1.25) {$\beta$};
		\node [style=none] (36) at (1, 0.5) {};
		\node [style={x_dot}] (37) at (1.5, 1.25) {$-\beta$};
		\node [style=none] (38) at (1.5, 0.5) {};
		\node [style=none] (39) at (2, 1) {$\dots$};
	\end{pgfonlayer}
	\begin{pgfonlayer}{edgelayer}
		\draw (1.center) to (4.center);
		\draw (4.center) to (3.center);
		\draw (1.center) to (2.center);
		\draw (2.center) to (3.center);
		\draw (6.center) to (9.center);
		\draw (9.center) to (8.center);
		\draw (6.center) to (7.center);
		\draw (7.center) to (8.center);
		\draw (28.center) to (25.center);
		\draw (15.center) to (17.center);
		\draw (27.center) to (26.center);
		\draw (16.center) to (18.center);
		\draw (19) to (21.center);
		\draw (20) to (22.center);
		\draw (33.center) to (30.center);
		\draw (32.center) to (31.center);
		\draw (35) to (36.center);
		\draw (37) to (38.center);
	\end{pgfonlayer}
\end{tikzpicture}

 \label{def:factored_form}
\end{align}
where $n$, $m$ are some integer numbers and $D_1$, $D_2$ are constant with respect to $\beta$. We call diagrams in this form $\beta$-factored. 
\end {clm}

A rigorous demonstration of claim \ref{claim:beta_factored} may be found in \cite{completeness_nancy}. It essentially follows from the \h rule to transform any green spider into a red spider, the \soo rule to separate the $\beta$ part from the rest of the diagram, and the paradigm \textit{Only topology matters} to rearrange the diagram.

The diagram $X_\beta(n,m)$ has a particularly regular form, so we expect it to have a nice-looking derivative. We would like to develop the intuition from the semantics $\partial_M\interp{X_\beta(n,m)}$.

\subsubsection{Intuition}

Due to the form obtained in the previous claim, it remains to find the derivative of $X_\beta(n,m)$ to obtain the desired results.

For reasons that will become more clear later, we first focus on a variant of 
$X_\beta(n,m)$ with green spiders instead of red spiders.

We denote by $|\beta\rangle = |0\rangle + e^{i\beta}|1\rangle$ the state $\interp{\gbeta}$ and by $|-\beta\rangle = |0\rangle + e^{-i\beta}|1\rangle$ the state $\interp{\gminusbeta}$. 

We also introduce the notation:
\begin{align}
|\beta(n, m)\rangle = |\beta\rangle^{\otimes n} \otimes |-\beta\rangle^{\otimes m} = \sum_{x \in \{0,1\}^{n+m}} e^{i\beta(|x^+| - |x^-|)} |x^+\rangle\otimes|x^-\rangle
\end{align} 

where $x^+ \in \{0, 1\}^n$ corresponds to the first $n$ qubits that are in the state $|\beta\rangle^{\otimes n}$ and $x^- \in \{0, 1\}^m$ corresponds to the last $m$ qubits that are in the state $|-\beta\rangle^{\otimes m}$. The notation $|x|: \{0, 1\}^k \rightarrow \mathbb{Z}_+$ stands for the Hamming weight of the binary string $x$.    

By applying entrywise differentiation to the vector $|\beta(n, m)\rangle$, we obtain the following result:
\begin{align}
\partial_M\left(|\beta(n, m)\rangle\right) = \sum_{x \in \{0,1\}^{n+m}}i(|x^+| - |x^-|)e^{i\beta(|x^+| - |x^-|)}|x^+\rangle\otimes|x^-\rangle
\end{align}

We observe that $\partial_M\left(|\beta(n, m)\rangle\right) = i M_\Delta|\beta(n, m)\rangle$ where $M_\Delta:2^{n+m} \rightarrow 2^{n+m}$ is a linear map such that $M_\Delta: |x\rangle \rightarrow (|x^+| - |x^-|) |x\rangle$.

Therefore, obtaining the derivative of $\left(|\beta(n, m)\rangle\right)$ requires the ability to produce the matrix $M_\Delta$.

The first step is finding a diagram for the controlled-triangle $D_\lambda: 2 \rightarrow 2$. Semantically, we want the diagram $D_\lambda$ to apply the triangle $\dtriangle$ to the second input if the first one is in the $|1\rangle$ state and do nothing otherwise. We denote the matrix corresponding to $D_\lambda$ by $M_\lambda$. Crucially, when the matrix $M_\lambda$ is applied to a vector $|b\rangle \otimes \left(k|0\rangle + |1\rangle\right)$ for $b \in \{0, 1\}$ the coefficient $k$ is modified depending on the value of $b$:
\begin{align}
|b\rangle \otimes \left(k|0\rangle + |1\rangle\right) \xrightarrow{M_\lambda} |b\rangle \otimes \left((k+b)|0\rangle + |1\rangle\right)
\end{align}
Similarly, when the input vector is $|b\rangle \otimes \left(k|0\rangle - |1\rangle\right)$ the value of $b$ is subtracted from $k$:
\begin{align}
|b\rangle \otimes \left(k|0\rangle + |1\rangle\right) \xrightarrow{M_\lambda} |b\rangle \otimes \left((k-b)|0\rangle - |1\rangle\right)
\end{align}

We remark the matrix $M_\lambda$ is not norm-preserving, so it doesn't correspond to a valid quantum gate. However, it doesn't matter as ZX-diagrams are universal for general linear maps of dimensions $2^n \times 2^m$.

We prove that $D_\lambda = \tikzformula{intuition/c_triangle_definition} =  
\begin{tikzpicture}
	\begin{pgfonlayer}{nodelayer}
		\node [style={z_dot}] (0) at (-0.25, -0.25) {};
		\node [style={z_dot}] (1) at (0.75, 0.25) {$\pi$};
		\node [style={z_dot}] (2) at (0.75, -0.75) {$\pi$};
		\node [style={z_dot}] (3) at (0.75, 1) {};
		\node [style={z_dot}] (4) at (0.75, -1.25) {};
		\node [style={z_dot}] (5) at (1.5, 1) {};
		\node [style={z_dot}] (6) at (1.5, -1.25) {};
		\node [style={h_box}] (7) at (0.25, 0) {};
		\node [style={h_box}] (8) at (0.25, -0.5) {};
		\node [style=none] (9) at (-0.25, 1.5) {};
		\node [style=none] (10) at (0.75, 1.5) {};
		\node [style=none] (11) at (0.75, -1.75) {};
		\node [style=none] (12) at (-0.25, -1.75) {};
	\end{pgfonlayer}
	\begin{pgfonlayer}{edgelayer}
		\draw [style=plus-edge] (3) to (5);
		\draw [style=plus-edge] (1) to (3);
		\draw [style=plus-edge] (2) to (1);
		\draw [style=plus-edge] (6) to (4);
		\draw (10.center) to (3);
		\draw (2) to (4);
		\draw (4) to (11.center);
		\draw (9.center) to (0);
		\draw (0) to (12.center);
		\draw (0) to (7);
		\draw (7) to (1);
		\draw (0) to (8);
		\draw (8) to (2);
	\end{pgfonlayer}
\end{tikzpicture}

$ by explicitly checking cases when the first input is in $|0\rangle$ and in $|1\rangle$ states.

For the $|0\rangle$-case we have:
\begin{align}
\tikzformula{intuition/triangle_zero/step1} 
\equal{\bo \\ \h } 
\tikzformula{intuition/triangle_zero/step2} 
\equal{ \ref{lemma:two-pi-triangle-is-identity}}
 \tikzformula{intuition/triangle_zero/step3}
 \equal{\ref{lemma:green-dots-with-opposed-triangles} \\ \ref{lemma:sqrt-2-sqrt-1-over-2}}
 \tikzformula{intuition/triangle_zero/step5}
\end{align}
When the first input is in the $|1\rangle$ state we get:
\begin{align}
\tikzformula{intuition/triangle_one/step1} 
\equal{ \ref{lemma:pi-push} \\ \bo \\ \h } \tikzformula{intuition/triangle_one/step2} 
\equal{ \ref{lemma:cs-two-push} }
\tikzformula{intuition/triangle_one/step3} 
\equal{\ref{lemma:green-dots-with-opposed-triangles} \\ \ref{lemma:sqrt-2-sqrt-1-over-2}} \tikzformula{intuition/triangle_one/step5}
\end{align}

It turns out that the matrix $M_\Delta$ can be written as an evolution in $n+m+1$ space with one ancilla input followed by a projection on $n+m$ dimensional space. The ancilla input is used to \textit{accumulate} the coefficient $\alpha_x$ of the basis state $|x\rangle$. We suggest a construction in which the coefficient $\alpha_x$ is recovered from the projection $\langle 0| \psi_a\rangle$ of the state $|\psi_a\rangle$ of the ancilla input.

To be exact, we design an iterative procedure that multiplies the input vector \linebreak$|b_0, \dots, b_{m+1}\rangle$ by the matrix $M_\Delta$. Initially, we set $|\psi_a\rangle = |1\rangle$. Then, for $i \in [1, \dots, n]$ we apply the transformation $M_\lambda$ to $|b_i\rangle|\psi_a\rangle$ controlled by the input $i$. After $n$ applications the state $|\psi_a\rangle$ becomes $|\psi_a\rangle = \left(\sum_{i=1}^n b_i\right)|0\rangle + |1\rangle$. To proceed further we apply $Z$ matrix to $|\psi_a\rangle$ and obtain  $|\psi_a\rangle = \left(\sum_{i=1}^n b_i\right)|0\rangle - |1\rangle$. Then the state $|\psi_a\rangle$ is transformed to $|\psi_a\rangle = \left(\sum_{i=1}^n b_i - \sum_{i=n+1}^{m} b_i \right)|0\rangle - |1\rangle$ by iterative application of $M_\lambda$ to $|b_i\rangle|\psi_a\rangle$. Finally, the coefficient $\alpha_x$ is:
\begin{align}
\alpha_x = \langle 0|\psi_a\rangle = \left(\sum_{i=1}^n b_i - \sum_{i=n+1}^{m} b_i \right)
\end{align}
As the original input wires are used only in control, the output of such a procedure is $\alpha_x |b_0,\dots, b_{n+m}\rangle$.
 
 This procedure can be directly translated in a diagram $D_\lambda$;
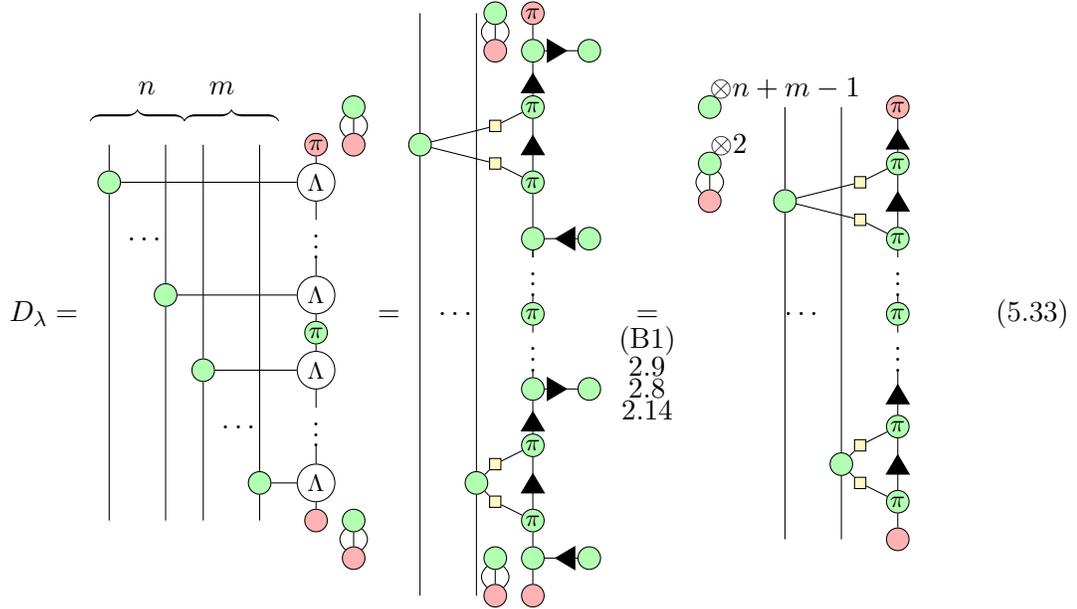
\begin{align}
D_{\lambda} = 
\begin{tikzpicture}
	\begin{pgfonlayer}{nodelayer}
		\node [style=none] (0) at (-1.25, 2.25) {};
		\node [style=none] (1) at (-1.25, -2.75) {};
		\node [style=none] (2) at (-0.75, 1) {$\dots$};
		\node [style=none] (3) at (-0.5, 2.25) {};
		\node [style=none] (4) at (-0.5, -2.75) {};
		\node [style=none] (5) at (0, 2.25) {};
		\node [style=none] (6) at (0, -2.75) {};
		\node [style=none] (7) at (0.5, -1.5) {$\dots$};
		\node [style=none] (8) at (0.75, 2.25) {};
		\node [style=none] (9) at (0.75, -2.75) {};
		\node [style={x_dot}] (10) at (1.5, 2.25) {$\pi$};
		\node [style={x_dot}] (11) at (1.5, -2.75) {};
		\node [style=none] (12) at (-1.5, 2.5) {};
		\node [style=none] (13) at (-0.25, 2.5) {};
		\node [style=none] (14) at (-0.25, 2.5) {};
		\node [style=none] (15) at (1, 2.5) {};
		\node [style=none] (16) at (-0.75, 3) {$n$};
		\node [style=none] (17) at (0.25, 3) {$m$};
		\node [style={z_dot}] (18) at (-1.25, 1.75) {};
		\node [style={z_dot}] (19) at (-0.5, 0.25) {};
		\node [style={z_dot}] (20) at (0, -0.75) {};
		\node [style={z_dot}] (21) at (0.75, -2.25) {};
		\node [style={z_dot}] (22) at (1.5, -0.25) {$\pi$};
		\node [style=diag] (23) at (1.5, 1.75) {$\Lambda$};
		\node [style=diag] (24) at (1.5, 0.25) {$\Lambda$};
		\node [style=diag] (25) at (1.5, -0.75) {$\Lambda$};
		\node [style=diag] (26) at (1.5, -2.25) {$\Lambda$};
		\node [style=none] (27) at (1.5, 1.5) {};
		\node [style=none] (28) at (1.5, 1.25) {};
		\node [style=none] (29) at (1.5, 0.75) {};
		\node [style=none] (30) at (1.5, -1.25) {};
		\node [style=none] (31) at (1.5, -1.5) {$\vdots$};
		\node [style=none] (32) at (1.5, -1.75) {};
		\node [style={z_dot}] (33) at (2, 2.75) {};
		\node [style={x_dot}] (35) at (2, 2.25) {};
		\node [style={z_dot}] (36) at (2, -2.75) {};
		\node [style={x_dot}] (37) at (2, -3.25) {};
		\node [style=none] (38) at (1.5, 1) {$\vdots$};
	\end{pgfonlayer}
	\begin{pgfonlayer}{edgelayer}
		\draw [style=brace] (12.center) to (13.center);
		\draw [style=brace] (14.center) to (15.center);
		\draw (8.center) to (21);
		\draw (21) to (9.center);
		\draw (21) to (26);
		\draw (20) to (25);
		\draw (19) to (24);
		\draw (18) to (23);
		\draw (10) to (23);
		\draw (23) to (28.center);
		\draw (29.center) to (24);
		\draw (24) to (22);
		\draw (22) to (25);
		\draw (25) to (30.center);
		\draw (26) to (11);
		\draw (0.center) to (18);
		\draw (18) to (1.center);
		\draw (3.center) to (19);
		\draw (19) to (4.center);
		\draw (20) to (6.center);
		\draw (33) to (35);
		\draw [bend right=45, looseness=1.25] (33) to (35);
		\draw [bend left=45, looseness=1.25] (33) to (35);
		\draw (36) to (37);
		\draw [bend right=45, looseness=1.25] (36) to (37);
		\draw [bend left=45, looseness=1.25] (36) to (37);
		\draw (20) to (5.center);
		\draw (26) to (32.center);
	\end{pgfonlayer}
\end{tikzpicture}

 = 
\begin{tikzpicture}
	\begin{pgfonlayer}{nodelayer}
		\node [style={z_dot}] (0) at (-0.75, 2.25) {};
		\node [style={z_dot}] (1) at (0.75, 2.75) {$\pi$};
		\node [style={z_dot}] (2) at (0.75, 1.75) {$\pi$};
		\node [style={z_dot}] (3) at (0.75, 3.5) {};
		\node [style={z_dot}] (4) at (0.75, 1) {};
		\node [style={z_dot}] (5) at (1.5, 3.5) {};
		\node [style={z_dot}] (6) at (1.5, 1) {};
		\node [style={h_box}] (7) at (0.25, 2.5) {};
		\node [style={h_box}] (8) at (0.25, 2) {};
		\node [style=none] (9) at (0, 4) {};
		\node [style={x_dot}] (10) at (0.75, 4) {$\pi$};
		\node [style=none] (11) at (0.75, 0.75) {};
		\node [style=none] (12) at (0, -3.75) {};
		\node [style={z_dot}] (13) at (0, -2.25) {};
		\node [style={z_dot}] (14) at (0.75, -1.75) {$\pi$};
		\node [style={z_dot}] (15) at (0.75, -2.75) {$\pi$};
		\node [style={z_dot}] (17) at (0.75, -3.25) {};
		\node [style={z_dot}] (19) at (1.5, -3.25) {};
		\node [style={h_box}] (20) at (0.25, -2) {};
		\node [style={h_box}] (21) at (0.25, -2.5) {};
		\node [style=none] (22) at (-0.75, 4) {};
		\node [style={x_dot}] (24) at (0.75, -3.75) {};
		\node [style=none] (25) at (-0.75, -3.75) {};
		\node [style=none] (26) at (0.75, 0.5) {$\vdots$};
		\node [style=none] (28) at (0.75, -0.5) {$\vdots$};
		\node [style=none] (29) at (0.75, -0.25) {};
		\node [style=none] (30) at (0.75, 0.25) {};
		\node [style={z_dot}] (31) at (0.75, 0) {$\pi$};
		\node [style={z_dot}] (32) at (0.75, -1) {};
		\node [style={z_dot}] (33) at (1.5, -1) {};
		\node [style=none] (34) at (0.75, -0.75) {};
		\node [style={z_dot}] (35) at (0.25, 4) {};
		\node [style={x_dot}] (36) at (0.25, 3.5) {};
		\node [style=none] (38) at (-0.25, 0) {$\dots$};
		\node [style={z_dot}] (39) at (0.25, -3.25) {};
		\node [style={x_dot}] (40) at (0.25, -3.75) {};
	\end{pgfonlayer}
	\begin{pgfonlayer}{edgelayer}
		\draw [style=plus-edge] (3) to (5);
		\draw [style=plus-edge] (1) to (3);
		\draw [style=plus-edge] (2) to (1);
		\draw [style=plus-edge] (6) to (4);
		\draw (10) to (3);
		\draw (2) to (4);
		\draw (4) to (11.center);
		\draw (0) to (7);
		\draw (7) to (1);
		\draw (0) to (8);
		\draw (8) to (2);
		\draw [style=plus-edge] (15) to (14);
		\draw [style=plus-edge] (19) to (17);
		\draw (15) to (17);
		\draw (17) to (24);
		\draw (13) to (20);
		\draw (20) to (14);
		\draw (13) to (21);
		\draw (21) to (15);
		\draw (4) to (11.center);
		\draw (30.center) to (31);
		\draw (31) to (29.center);
		\draw [style=plus-edge] (32) to (33);
		\draw (32) to (34.center);
		\draw [style=plus-edge] (14) to (32);
		\draw (35) to (36);
		\draw [bend right=45, looseness=1.25] (35) to (36);
		\draw [bend left=45, looseness=1.25] (35) to (36);
		\draw (22.center) to (0);
		\draw (0) to (25.center);
		\draw (9.center) to (13);
		\draw (13) to (12.center);
		\draw (39) to (40);
		\draw [bend right=45, looseness=1.25] (39) to (40);
		\draw [bend left=45, looseness=1.25] (39) to (40);
	\end{pgfonlayer}
\end{tikzpicture}

\equal{\bo \\ \ref{lemma:pi-triangle-down} \\ \ref{lemma:zero-triangle-up} \\ \ref{lemma:green-dots-with-opposed-triangles}}

\begin{tikzpicture}
	\begin{pgfonlayer}{nodelayer}
		\node [style={z_dot}] (0) at (-0.75, 1.5) {};
		\node [style={z_dot}] (1) at (0.75, 2) {$\pi$};
		\node [style={z_dot}] (2) at (0.75, 1) {$\pi$};
		\node [style={h_box}] (7) at (0.25, 1.75) {};
		\node [style={h_box}] (8) at (0.25, 1.25) {};
		\node [style=none] (9) at (0, 2.75) {};
		\node [style={x_dot}] (10) at (0.75, 2.75) {$\pi$};
		\node [style=none] (11) at (0.75, 0.75) {};
		\node [style=none] (12) at (0, -3) {};
		\node [style={z_dot}] (13) at (0, -2) {};
		\node [style={z_dot}] (14) at (0.75, -1.5) {$\pi$};
		\node [style={z_dot}] (15) at (0.75, -2.5) {$\pi$};
		\node [style={h_box}] (20) at (0.25, -1.75) {};
		\node [style={h_box}] (21) at (0.25, -2.25) {};
		\node [style=none] (22) at (-0.75, 2.75) {};
		\node [style={x_dot}] (24) at (0.75, -3) {};
		\node [style=none] (25) at (-0.75, -3) {};
		\node [style=none] (26) at (0.75, 0.5) {$\vdots$};
		\node [style=none] (28) at (0.75, -0.5) {$\vdots$};
		\node [style=none] (29) at (0.75, -0.25) {};
		\node [style=none] (30) at (0.75, 0.25) {};
		\node [style={z_dot}] (31) at (0.75, 0) {$\pi$};
		\node [style=none] (34) at (0.75, -0.75) {};
		\node [style={z_dot}] (35) at (-1.75, 2) {};
		\node [style={x_dot}] (36) at (-1.75, 1.5) {};
		\node [style=none] (37) at (-1.5, 2.25) {$\otimes 2$};
		\node [style={z_dot}] (38) at (-1.75, 2.75) {};
		\node [style=none] (39) at (-0.75, 3) {$\otimes n+m-1$};
		\node [style=none] (40) at (-0.5, 0) {$\dots$};
	\end{pgfonlayer}
	\begin{pgfonlayer}{edgelayer}
		\draw [style=plus-edge] (2) to (1);
		\draw (0) to (7);
		\draw (7) to (1);
		\draw (0) to (8);
		\draw (8) to (2);
		\draw [style=plus-edge] (15) to (14);
		\draw (13) to (20);
		\draw (20) to (14);
		\draw (13) to (21);
		\draw (21) to (15);
		\draw (30.center) to (31);
		\draw (31) to (29.center);
		\draw (35) to (36);
		\draw [bend right=45, looseness=1.25] (35) to (36);
		\draw [bend left=45, looseness=1.25] (35) to (36);
		\draw (11.center) to (2);
		\draw [style=plus-edge] (14) to (34.center);
		\draw (15) to (24);
		\draw [style=plus-edge] (1) to (10);
		\draw (22.center) to (0);
		\draw (0) to (25.center);
		\draw (9.center) to (13);
		\draw (13) to (12.center);
	\end{pgfonlayer}
\end{tikzpicture}

\end{align}

The diagram for the derivative of $X_\beta(n, m) = \rbeta \dots \rbeta \rminusbeta \dots \rminusbeta$ can be obtained by conjugating every input with Hadamard boxes:

\begin{align}
\partial_M\interp{\tikzformula{intuition/red}} & 
= \partial_M\interp{\tikzformula{intuition/red/s1}} 
= \interp{\tikzformula{intuition/red/s2_2}} \circ \partial_M\interp{\tikzformula{intuition/red/s2_1}} \raisetag{13pt} \\[10pt]
& = 
\begin{tikzpicture}
	\begin{pgfonlayer}{nodelayer}
		\node [style={z_dot}] (0) at (0, 0.25) {\small$\pi$};
		\node [style={x_dot}] (1) at (0, -0.25) {\tiny$\frac{\pi}{2}$};
	\end{pgfonlayer}
	\begin{pgfonlayer}{edgelayer}
		\draw (0) to (1);
	\end{pgfonlayer}
\end{tikzpicture}

 \interp{\tikzformula{intuition/red/s2_2}} \circ \interp{\tikzformula{intuition/red/s3}} = \interp{ 
\begin{tikzpicture}
	\begin{pgfonlayer}{nodelayer}
		\node [style={z_dot}] (0) at (0, 0.25) {\small$\pi$};
		\node [style={x_dot}] (1) at (0, -0.25) {\tiny$\frac{\pi}{2}$};
	\end{pgfonlayer}
	\begin{pgfonlayer}{edgelayer}
		\draw (0) to (1);
	\end{pgfonlayer}
\end{tikzpicture}

\tikzformula{intuition/red/s4}} \raisetag{-5pt}
\end{align}

By applying the rule \h to the final we obtain 
\begin{align}
\partial_M\interp{X_\beta(n, m)} = \interp{\tikzformula{automatic_differentiation/definition}} \label{eq:formula_intuition}
\end{align}

Notice that the Hadamard boxes from $D_\lambda$ have disappeared in the expression of $\partial_M\interp{X_\beta(n, m)}$, which explains why we chose initially red spiders instead of green spiders.

\subsubsection{Diagrammatic differentiation of $\beta$-factored forms}

Motivated by the simplicity of the result (\ref{eq:formula_intuition}), we suggest an alternative non-inductive definition for derivatives of diagrams in $\beta$-factored forms: 

\begin{defi}\label{def:delta_beta}
Given a diagram $D(\beta) = D_2 \circ (D_1\otimes X_\beta(n, m))$ in $\beta$-factored form, let $\partial_{\ZX}[D] = D_2 \circ (D_1\otimes \partial_{\ZX}[X_\beta(n, m)])$ where 

\begin{align}
\partial_{\ZX}&[X_\beta(n,m)] = \partial_{\ZX}\left[ \underbrace{\rbeta \dots \rbeta}_{n} \underbrace{\rminusbeta \dots \rminusbeta}_{m}
\right] \\
&=\tikzformula{automatic_differentiation/definition}\label{eq:automatic_derivative}
\end{align}
\end{defi}

Although our definition is inspired by the intuitive reflection presented above, we provide a rigorous diagrammatic demonstration of the soundness of our definition. More precisely, we prove the following theorem:

\begin{thm}\label{theorem:automatic_derivative}
The operator $\partial_{\ZX}[-]$ defined at (\ref{def:delta_beta}) satisfies the property of diagrammatic differentiation: for any diagram $D(\beta) \in \ZX(\beta)$ in  $\beta$-factored form,
\begin{align}
%\forall D(\beta) \in \ZX(\beta): 
\interp{\partial_{\ZX} D(\beta)} = \partial_M\interp{ D(\beta)} 
\end{align}
\end{thm}

We remark that according to the definition \ref{def:diagrammatic_derivative} the derivative $D':n\rightarrow m$ of a diagram $D:n\rightarrow m$ that is constant on $\beta$ is such that $\interp{D'} = \partial_M\interp{D} = (0)_{n \times m}$. Therefore, theorem \ref{theorem:automatic_derivative} is a direct consequence of the following lemma:
\begin{lem}\label{lemma:automatic_derivative} For any $n, m$:
\begin{align}
\interp{\partial_{\ZX} X_\beta(n, m)} = \partial_M\interp{X_\beta(n, m)}\label{eq:derivative}
\end{align}
\end{lem}

\begin{proof}\label{proof:automatic_derivative}
We prove the lemma by induction. We provide the demonstration for $n = n+1$, the proof for $m=m+1$ is directly obtainable in the same way.  

\textbf{Base:} We already know that $ \partial_M\interp{\rbeta} =  ie^{i\beta}|-\rangle$. 
Thus, for the case $n=1, m=0$ the lemma statement follows from:
\begin{align}
\interp{\partial_{\ZX} X_{\beta}(1,0)} = \interp{
\begin{tikzpicture}
	\begin{pgfonlayer}{nodelayer}
		\node [style={x_dot}] (24) at (-0.75, 0.5) {\small$\pi$};
		\node [style={z_dot}] (25) at (0.25, 0.5) {\small$\pi$};
		\node [style={z_dot}] (26) at (1.25, 0.5) {\small$\pi$};
		\node [style={x_dot}] (33) at (0.75, 0) {\small $\beta$};
		\node [style=none] (35) at (0.75, -0.5) {};
		\node [style={x_dot}] (55) at (1.75, 0.5) {};
		\node [style={z_dot}] (61) at (-1.25, 0) {};
		\node [style={x_dot}] (62) at (-1.25, -0.5) {};
		\node [style=none] (63) at (-1, 0.25) {$\otimes 3$};
		\node [style={z_dot}] (64) at (-2, 0) {\small$\pi$};
		\node [style={x_dot}] (65) at (-2, -0.5) {\small$\frac{\pi}{2}$};
	\end{pgfonlayer}
	\begin{pgfonlayer}{edgelayer}
		\draw [style=plus-edge] (25) to (24);
		\draw [style=plus-edge] (26) to (25);
		\draw (25) to (33);
		\draw (33) to (26);
		\draw (33) to (35.center);
		\draw (26) to (55);
		\draw (61) to (62);
		\draw [bend right=45] (61) to (62);
		\draw [bend left=45] (61) to (62);
		\draw (64) to (65);
	\end{pgfonlayer}
\end{tikzpicture}

} \equal{\ref{lemma:derivative-base-induction} \\ \ref{lemma:sqrt-2-sqrt-1-over-2}} \interp{
\begin{tikzpicture}
	\begin{pgfonlayer}{nodelayer}
		\node [style={z_dot}] (61) at (0, 0.5) {};
		\node [style={x_dot}] (62) at (0, 0) {};
		\node [style=none] (63) at (0.25, 0.75) {$\otimes 2$};
		\node [style={z_dot}] (64) at (-0.75, 0.5) {\small$\pi$};
		\node [style={x_dot}] (65) at (-0.75, 0) {\small$\frac{\pi}{2}$};
		\node [style={z_dot}] (66) at (1, 0.5) {\small$\pi$};
		\node [style={x_dot}] (67) at (1, 0) {\small $\beta$};
		\node [style=none] (68) at (1, -0.5) {};
	\end{pgfonlayer}
	\begin{pgfonlayer}{edgelayer}
		\draw (61) to (62);
		\draw [bend right=45] (61) to (62);
		\draw [bend left=45] (61) to (62);
		\draw (64) to (65);
		\draw (66) to (67);
		\draw (67) to (68.center);
	\end{pgfonlayer}
\end{tikzpicture}

} = ie^{i\beta}|-\rangle
\end{align} 

\textbf{Step:} By induction, we assume that the equation (\ref{eq:derivative}) holds for some $n$ and $m$. We show that under this assumption $\interp{\partial_{\ZX} X_\beta(n+1, m)} = \partial_M\interp{X_\beta(n+1, m)}$. 

In order to prove the induction we will proceed in three steps. Firstly we apply the product rule to the matrix $\partial_M \interp{X_\beta(n+1, m)}$: 
\begin{align}
\partial_M \interp{X_\beta(n+1,  m)}
& = \partial_M \interp{X_\beta(1, 0)} \otimes \interp{X_\beta(n, m)} +  \interp{ X_\beta(1, 0)} \otimes \partial_{M}\interp{ X_\beta(n, m)} \nonumber \\
& = \interp{\partial_{\ZX} \left[ X_\beta(1, 0) \right] \otimes X_\beta(n, m)} + \interp{ X_\beta(1, 0) \otimes \partial_{\ZX} \left[ X_\beta(n, m)\right]} \label{claim:sum}
\end{align}
where the second equality follows from the inductive assumption.

Secondly, we show the last sum in the equation (\ref{claim:sum}) can be represented diagrammatically with a controlled state, i.e. the following claim holds: 

\begin {clm}\label{claim:rhs_as_controlled_diagram}
We can find a controlled state $\tilde{X} : 1 \rightarrow n+1+m$ and a constant scalar $c \in\ZX_{\frac{\pi}{2}}$ such that
\begin{align}
\interp{
\input{./ZX_diagrams/automatic_differentiation/controlled_states/x_tilde.tikz}
} = \interp{\partial_{\ZX} \left[ X_\beta(1, 0) \right] \otimes X_\beta(n, m)} + \interp{ X_\beta(1, 0) \otimes \partial_{\ZX} \left[ X_\beta(n, m)\right]} 
\end{align}
\end {clm}

Finally, we will show that we can transform the diagram $\partial_{\ZX} X_\beta(n+1, m)$ to \linebreak$\interp{
\input{./ZX_diagrams/automatic_differentiation/controlled_states/x_tilde.tikz}
}$ with a sequence of graphical rewrites: 
 
\begin {clm}\label{claim:lhs_and_rhs_diagrams_are_equal}
\begin{align}
\partial_{\ZX} X_\beta(n+1, m) = 
\input{./ZX_diagrams/automatic_differentiation/controlled_states/x_tilde.tikz}

\end{align}
\end {clm}

As the rewrite rules preserve the semantics, it follows from the claims above that:
\begin{align}
\interp{\partial_{\ZX} X_\beta(n+1, m)} 
& \equal{\ref{claim:lhs_and_rhs_diagrams_are_equal}}
\interp{
\input{./ZX_diagrams/automatic_differentiation/controlled_states/x_tilde.tikz}
} \nonumber \\
& \equal{\ref{claim:rhs_as_controlled_diagram}} 
\interp{\partial_{\ZX} \left[ X_\beta(1, 0) \right] \otimes X_\beta(n, m)} + \interp{ X_\beta(1, 0) \otimes \partial_{\ZX} \left[ X_\beta(n, m)\right]} \nonumber \\
&\equal{\ref{claim:sum}}
\partial_M \interp{X_\beta(n+1,  m)}
\end{align}
\qedhere
\end{proof}

\paragraph{Proofs}
\tikzstyle{diag}=[rectangle,fill=white,draw=black,xscale=1,yscale=1,font=\small,inner sep=0.75pt,minimum width=1cm,minimum height=0.5cm]

While proving the claims \ref{claim:rhs_as_controlled_diagram} and \ref{claim:lhs_and_rhs_diagrams_are_equal} we will repeatedly use the following lemmas:

\begin{lem}\label{lemma:drop_link}
$$
\begin{tikzpicture}
	\begin{pgfonlayer}{nodelayer}
		\node [style={z_dot}] (51) at (-2.25, 0.75) {};
		\node [style={z_dot}] (52) at (-0.75, 0.75) {$\pi$};
		\node [style=none] (56) at (-3.5, 0.75) {$\dots$};
		\node [style={x_dot}] (62) at (-1.25, 0) {$\pi$};
		\node [style={z_dot}] (77) at (-3, 0.75) {};
		\node [style={z_dot}] (86) at (-1.5, -0.5) {};
		\node [style={z_dot}] (94) at (-2.25, 0) {};
		\node [style={z_dot}] (108) at (-5.5, 0.75) {};
		\node [style=none] (134) at (-5.5, 1.25) {};
		\node [style=none] (137) at (-1.5, -1) {};
		\node [style={z_dot}] (138) at (-1.5, 0.75) {$\pi$};
		\node [style=none] (139) at (-2.25, 1.25) {};
		\node [style=none] (140) at (-3, 1.25) {};
		\node [style=none] (141) at (-0.25, 0.75) {};
		\node [style=none] (142) at (-3.25, 0.75) {};
		\node [style=none] (143) at (-3.75, 0.75) {};
		\node [style={z_dot}] (144) at (-4, 0.75) {};
		\node [style={z_dot}] (145) at (-4.75, 0.75) {};
		\node [style=none] (146) at (-4, 1.25) {};
		\node [style=none] (147) at (-4.75, 1.25) {};
		\node [style=none] (148) at (0, 0) {$=$};
		\node [style={z_dot}] (149) at (3.5, 0.75) {};
		\node [style={z_dot}] (150) at (5.25, 0.75) {$\pi$};
		\node [style=none] (151) at (2.25, 0.75) {$\dots$};
		\node [style={x_dot}] (152) at (4.75, 0) {};
		\node [style={z_dot}] (153) at (2.75, 0.75) {};
		\node [style={z_dot}] (156) at (0.25, 0.75) {};
		\node [style=none] (157) at (0.25, 1.25) {};
		\node [style=none] (158) at (4.75, -1) {};
		\node [style={z_dot}] (159) at (4.25, 0.75) {$\pi$};
		\node [style=none] (160) at (3.5, 1.25) {};
		\node [style=none] (161) at (2.75, 1.25) {};
		\node [style=none] (162) at (5.75, 0.75) {};
		\node [style=none] (163) at (2.5, 0.75) {};
		\node [style=none] (164) at (2, 0.75) {};
		\node [style={z_dot}] (165) at (1.75, 0.75) {};
		\node [style={z_dot}] (166) at (1, 0.75) {};
		\node [style=none] (167) at (1.75, 1.25) {};
		\node [style=none] (168) at (1, 1.25) {};
		\node [style={x_dot}] (169) at (4.25, 0) {};
		\node [style={z_dot}] (170) at (4.25, -0.5) {};
	\end{pgfonlayer}
	\begin{pgfonlayer}{edgelayer}
		\draw (62) to (52);
		\draw [style=plus-edge] (51) to (77);
		\draw [style=plus-edge] (94) to (62);
		\draw [bend right=15] (108) to (94);
		\draw [style=plus-edge] (86) to (94);
		\draw (62) to (86);
		\draw (86) to (137.center);
		\draw (108) to (134.center);
		\draw [style=plus-edge] (52) to (138);
		\draw (138) to (62);
		\draw (51) to (139.center);
		\draw (140.center) to (77);
		\draw (52) to (141.center);
		\draw [style=plus-edge] (138) to (51);
		\draw [style=plus-edge] (144) to (145);
		\draw (144) to (146.center);
		\draw (147.center) to (145);
		\draw [style=plus-edge] (145) to (108);
		\draw (144) to (143.center);
		\draw (142.center) to (77);
		\draw (152) to (150);
		\draw [style=plus-edge] (149) to (153);
		\draw (156) to (157.center);
		\draw [style=plus-edge] (150) to (159);
		\draw (159) to (152);
		\draw (149) to (160.center);
		\draw (161.center) to (153);
		\draw (150) to (162.center);
		\draw [style=plus-edge] (159) to (149);
		\draw [style=plus-edge] (165) to (166);
		\draw (165) to (167.center);
		\draw (168.center) to (166);
		\draw [style=plus-edge] (166) to (156);
		\draw (165) to (164.center);
		\draw (163.center) to (153);
		\draw (152) to (158.center);
		\draw (169) to (170);
		\draw [bend right=45] (169) to (170);
		\draw [bend left=45] (169) to (170);
	\end{pgfonlayer}
\end{tikzpicture}

$$
\end{lem}

\begin{proof}
The right hand side transformation:
\begin{align}
&
\input{./ZX_diagrams/new_lemmas/drop_link_proof/step1.tikz}

\equal{\soo \\ \ref{lemma:tranzistor-swap-legs}}

\input{./ZX_diagrams/new_lemmas/drop_link_proof/step2.tikz}
 
\equal{\ref{lemma:pi-push}}

\input{./ZX_diagrams/new_lemmas/drop_link_proof/step3.tikz}
\nonumber \\
&
\equal{\ref{lemma:cs-plus-commutation}}

\input{./ZX_diagrams/new_lemmas/drop_link_proof/step4.tikz}
 
\equal{\ref{lemma:triangle-cycle}}

\input{./ZX_diagrams/new_lemmas/drop_link_proof/step5.tikz}

\equal{\ref{lemma:same-direction-triangles-link-removal} \\ \soo }

\input{./ZX_diagrams/new_lemmas/drop_link_proof/step6.tikz}
\nonumber \\
&
\equal{\ref{lemma:green-dots-with-opposed-triangles} \\\ref{lemma:triangle-cycle} }

\input{./ZX_diagrams/new_lemmas/drop_link_proof/step7.tikz}

\equal{\ref{lemma:cs-plus-commutation}}

\input{./ZX_diagrams/new_lemmas/drop_link_proof/step8.tikz}

\end{align}
\qedhere
\end{proof}

\begin{lem}\label{lemma:some_important_controlled_states}
Diagrams:
\begin{align}
C_{\pm\beta} = 
\input{./ZX_diagrams/automatic_differentiation/controlled_states/rbeta.tikz}
, \quad
C^{(n, m)}_\beta =  \tikzformula{automatic_differentiation/controlled_states/partial_xnm}
\end{align}
are controlled states such that
\begin{align}
&\interp{C_{\pm\beta}}|1\rangle = \interp{\normalizer 
\begin{tikzpicture}
	\begin{pgfonlayer}{nodelayer}
		\node [style={x_dot}] (0) at (0, 0.25) {\tiny$\pm\beta$};
		\node [style=none] (1) at (0, -0.25) {};
	\end{pgfonlayer}
	\begin{pgfonlayer}{edgelayer}
		\draw (0) to (1.center);
	\end{pgfonlayer}
\end{tikzpicture}

}, \quad \interp{C^{(n,m)}_\beta}|1\rangle = \interp{\minusi \normalizer^{\otimes(n+m-1)} \partial_{\ZX}[X(n, m)]}
\end{align}
\end{lem}

\begin{proof}
For the diagram $C_{\pi\beta}$ the claim follows directly from the definition.
For the diagram $C^{(n, m)}_\beta$ we check two cases:

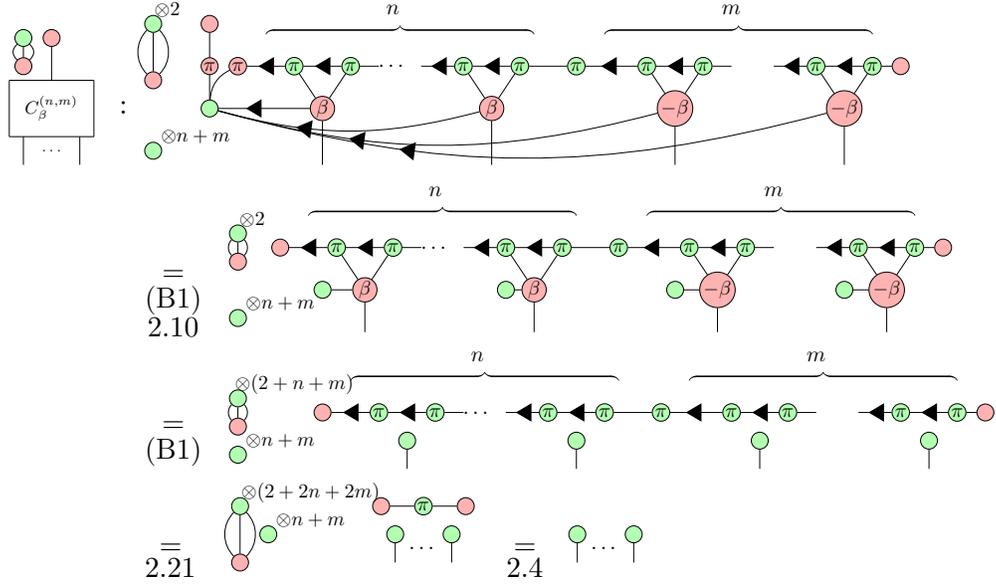
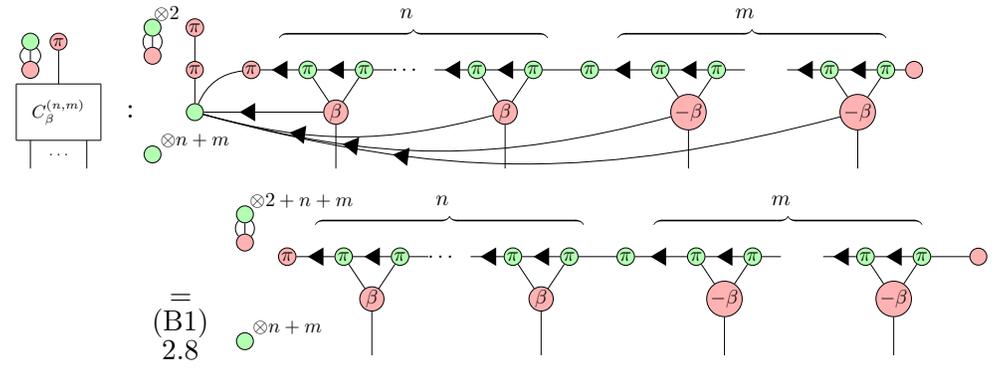
\begin{align}
\tikzformula{new_lemmas/controlled_d_xnm/zero/step0}: 
&\tikzformula{new_lemmas/controlled_d_xnm/zero/step1} \nonumber \\
& \equal{\bo \\ \ref{lemma:pi-triangle-up}}
\tikzformula{new_lemmas/controlled_d_xnm/zero/step2}\nonumber \\
& \equal{\bo}
\tikzformula{new_lemmas/controlled_d_xnm/zero/step3}\nonumber \\
& \equal{\ref{lemma:two-pi-triangle-is-identity}}
\tikzformula{new_lemmas/controlled_d_xnm/zero/step4}\nonumber 
\equal{\ref{lemma:sqrt-2-sqrt-1-over-2}}
\tikzformula{new_lemmas/controlled_d_xnm/zero/step5}\nonumber
\end{align}
\begin{align}
\tikzformula{new_lemmas/controlled_d_xnm/one/step0}:
&\tikzformula{new_lemmas/controlled_d_xnm/one/step1} \nonumber \\
& \equal{\bo \\ \ref{lemma:zero-triangle-up}}
\tikzformula{new_lemmas/controlled_d_xnm/one/step2}\tag*{\qedhere}
\end{align}
\end{proof}

\begin{proof}[claim \ref{claim:rhs_as_controlled_diagram}]

We introduce the diagram:
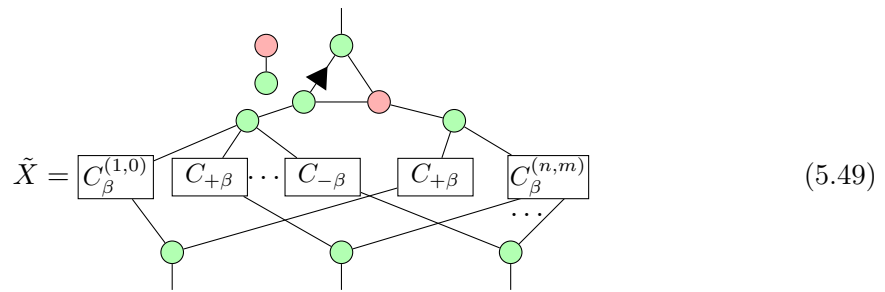
\begin{align}
\tilde{X} = 
\input{./ZX_diagrams/automatic_differentiation/controlled_states/partial_xnm_plus.tikz}
\label{diag:cs_from_claim}
\end{align}
where $C_{\pm \beta}$ and $C_\beta^{(n,m)}$ are defined in the previous lemma (\ref{lemma:some_important_controlled_states}). From these definitions and the lemma (\ref{lemma:sum_and_product_of_cs}) follows that 
\begin{align}
\interp{\tilde{X}}&|1\rangle = \interp{C^{(1,0)}_{\beta}}|1\rangle \otimes \interp{C_{+\beta}}|1\rangle \otimes \dots \otimes \interp{C_{-\beta}}|1\rangle + \interp{C_{+\beta}}|1\rangle \otimes \interp{C^{(n,m)}_{\beta}}|1\rangle \nonumber \\
&=\interp{\minusi \normalizer^{\otimes(n+m)}}\left(\interp{\partial_{\ZX} \left[ X_\beta(1, 0) \right] \otimes X_\beta(n, m)} + \interp{ X_\beta(1, 0) \otimes \partial_{\ZX} \left[ X_\beta(n, m)\right]}\right)
\end{align}
Therefore the claim holds for the controlled state (\ref{diag:cs_from_claim}) and the scalar $c = \tikzformula{automatic_differentiation/constant}$
\qedhere
\end{proof}

\begin{proof}[claim \ref{claim:lhs_and_rhs_diagrams_are_equal}]
Firstly, we simplify the expression $\tilde{X} \circ \ketone$:
\begin{align}
& = \tikzformula{automatic_differentiation/step/step1} \nonumber \\
&\equal{\ref{lemma:pi-push}}
 \tikzformula{automatic_differentiation/step/step2} \nonumber \\
&\equal{}
\tikzformula{automatic_differentiation/step/step2bis} \nonumber \\
&\equal{\ref{lemma:pi-triangle-push} \\ \ref{lemma:pi-push}}
\tikzformula{automatic_differentiation/step/step3} \nonumber \\
&\equal{\ref{lemma:factor-beta}}
\tikzformula{automatic_differentiation/step/step4}  = \circled{1}\nonumber \\
\end{align}

We apply the lemma \ref{lemma:drop_link} to parts in the red frame and obtain:
\begin{align}
\circled{1} & = \tikzformula{automatic_differentiation/step/step5} \nonumber \\
&\equal{\ref{lemma:pi-triangle-push}}
\tikzformula{automatic_differentiation/step/step6}
\end{align}

The claim holds for the scalar $c =
\begin{tikzpicture}
	\begin{pgfonlayer}{nodelayer}
		\node [style={z_dot}] (0) at (-0.25, 0.25) {\small$\pi$};
		\node [style={x_dot}] (1) at (-0.25, -0.25) {\tiny$\frac{\pi}{2}$};
		\node [style={x_dot}] (2) at (0.25, -0.25) {};
		\node [style={z_dot}] (3) at (0.25, 0.25) {};
		\node [style=none] (4) at (1.25, 0.5) {\small$\otimes(n+m-2)$};
	\end{pgfonlayer}
	\begin{pgfonlayer}{edgelayer}
		\draw (0) to (1);
		\draw (3) to (2);
	\end{pgfonlayer}
\end{tikzpicture}

$
\qedhere
\end{proof}

\subsection{Simplified formula for paired spiders}\label{section:diff_with_formulas_2}

Variational quantum algorithms use gradients in the search for optimal parameter values. The objective minimized by these algorithms can be expressed as $\langle \psi(\beta)| H |\psi(\beta)\rangle$ where the diagram for $\langle \psi(\beta)| = \left(|\psi(\beta)\rangle\right)^\dagger$ is obtained out of the diagram for $|\psi(\beta)\rangle$ by flipping upside down followed by the change of signs in spiders. Therefore, parameters in the diagram for $\langle \psi(\beta)| H |\psi(\beta)\rangle$ appear in pairs $ \rminusbeta \quad \rbeta$. 

We suggest a more compact formula for diagrams in what we call \textit{pair-factored form}: $D_2 \circ (D_1 \otimes Y(n))$. In this expression $Y_\beta(n) = \underbrace{\left(\rminusbeta \quad \rbeta \right) \dots \left(\rminusbeta \quad \rbeta \right)}_n$.

\begin{lem}\label{lemma:derivative_for_paired_spiders}
The diagram:
\begin{align}
\partial_P(Y_\beta(n)) = 
\input{./ZX_diagrams/pair_formula/statement.tikz}
\label{eq:derivative_yn} 
\end{align}
satisfies $\interp{\partial_P(Y_\beta(n))} = \partial_M\interp{Y_{\beta}(n)}$.
\end{lem}

The lemma (\ref{lemma:derivative_for_paired_spiders}) can be proven by applying the same approach as in the proof of the formula (\ref{lemma:automatic_derivative}) for individual spiders. We leave the exact proof to the reader.

It is possible to extend \ref{lemma:derivative_for_paired_spiders} to find the derivative for $X_\beta(n, m)$ when $n \neq m$. Indeed, it is always possible to transform a diagram into an equivalent diagram that has the same number of occurrences of $\beta$ and $-\beta$, using the fact that $
\input{./ZX_diagrams/pair_formula/scalar_with_pm_beta.tikz}
$. For instance,  if $n > m$:
$\partial_P\left(X_\beta(n, m)\right) = \sigma \circ \left[
\input{./ZX_diagrams/pair_formula/x_n_m_definition.tikz}
\right]$
where $\sigma$ is some wire permutation and 
\begin{align}
\tikzformula{pair_formula/x_n_m_definition} \equal{\bo \\ \ref{lemma:sqrt-2-sqrt-1-over-2}} \tikzformula{pair_formula/x_n_m_diagram}
\end{align} 

\begin{exa}\label{example:cos_good}
We apply \ref{lemma:derivative_for_paired_spiders} to the same diagram as in   \ref{example:cos_bad}:
\begin{align}
\!\!\!\!\!\!\!\!\partial_P\left(
\begin{tikzpicture}
	\begin{pgfonlayer}{nodelayer}
		\node [style={x_dot}] (17) at (-0.5, 0.25) {\small$-\beta$};
		\node [style={x_dot}] (18) at (0.5, 0.25) {\small$\beta$};
		\node [style={z_dot}] (22) at (0, -0.25) {$\pi$};
	\end{pgfonlayer}
	\begin{pgfonlayer}{edgelayer}
		\draw (17) to (22);
		\draw (22) to (18);
	\end{pgfonlayer}
\end{tikzpicture}

\right) = 
\begin{tikzpicture}
	\begin{pgfonlayer}{nodelayer}
		\node [style={x_dot}] (0) at (-2, 0.5) {};
		\node [style={z_dot}] (1) at (-2, 0) {};
		\node [style={z_dot}] (2) at (-2.25, -0.5) {};
		\node [style=none] (3) at (-1.75, 0.75) {\small$\otimes 3$};
		\node [style={x_dot}] (5) at (-1.5, 0.5) {$\pi$};
		\node [style={z_dot}] (6) at (0, 0.5) {$\pi$};
		\node [style={z_dot}] (7) at (-1, 0.5) {};
		\node [style={z_dot}] (8) at (1, 0.5) {$\pi$};
		\node [style={x_dot}] (12) at (1.5, 0.5) {};
		\node [style={x_dot}] (17) at (-0.5, 0) {\small$-\beta$};
		\node [style={x_dot}] (18) at (0.5, 0) {\small$\beta$};
		\node [style={z_dot}] (27) at (0, -0.5) {$\pi$};
		\node [style={z_dot}] (28) at (-2.5, 0.5) {\small$\pi$};
		\node [style={x_dot}] (29) at (-2.5, 0) {\tiny$\frac{\pi}{2}$};
	\end{pgfonlayer}
	\begin{pgfonlayer}{edgelayer}
		\draw (0) to (1);
		\draw [bend right=45] (0) to (1);
		\draw [bend left=45] (0) to (1);
		\draw [style=plus-edge] (6) to (7);
		\draw [style=plus-edge] (8) to (6);
		\draw (5) to (7);
		\draw (7) to (17);
		\draw (17) to (6);
		\draw (6) to (18);
		\draw (18) to (8);
		\draw (12) to (8);
		\draw (17) to (27);
		\draw (27) to (18);
		\draw (28) to (29);
	\end{pgfonlayer}
\end{tikzpicture}

 = 
\begin{tikzpicture}
	\begin{pgfonlayer}{nodelayer}
		\node [style={x_dot}] (0) at (-1, 0.5) {};
		\node [style={z_dot}] (1) at (-1, 0) {};
		\node [style={z_dot}] (2) at (-1.25, -0.5) {};
		\node [style=none] (3) at (-0.75, 0.75) {\small$\otimes 3$};
		\node [style={z_dot}] (6) at (0, 0.5) {$\pi$};
		\node [style={x_dot}] (17) at (-0.5, 0) {\tiny$\pi{-}\beta$};
		\node [style={x_dot}] (18) at (0.5, 0) {\small$\beta$};
		\node [style={z_dot}] (27) at (0, -0.5) {$\pi$};
		\node [style={z_dot}] (28) at (-1.5, 0.5) {\small$\pi$};
		\node [style={x_dot}] (29) at (-1.5, 0) {\tiny$\frac{\pi}{2}$};
	\end{pgfonlayer}
	\begin{pgfonlayer}{edgelayer}
		\draw (0) to (1);
		\draw [bend right=45] (0) to (1);
		\draw [bend left=45] (0) to (1);
		\draw (17) to (6);
		\draw (6) to (18);
		\draw (17) to (27);
		\draw (27) to (18);
		\draw (28) to (29);
	\end{pgfonlayer}
\end{tikzpicture}

 = 
\begin{tikzpicture}
	\begin{pgfonlayer}{nodelayer}
		\node [style={x_dot}] (0) at (0, 0.25) {};
		\node [style={z_dot}] (1) at (0, -0.25) {};
		\node [style={x_dot}] (17) at (0.75, 0) {\tiny$\pi{-}2\beta$};
		\node [style={z_dot}] (28) at (-0.5, 0.25) {$\pi$};
		\node [style={x_dot}] (29) at (-0.5, -0.25) {\small$\beta$};
		\node [style={z_dot}] (30) at (-1, 0.25) {\small$\pi$};
		\node [style={x_dot}] (31) at (-1, -0.25) {\tiny$\frac{\pi}{2}$};
		\node [style=none] (32) at (0.25, 0.5) {$\otimes 2$};
	\end{pgfonlayer}
	\begin{pgfonlayer}{edgelayer}
		\draw (0) to (1);
		\draw [bend right=45] (0) to (1);
		\draw [bend left=45] (0) to (1);
		\draw (28) to (29);
		\draw (30) to (31);
	\end{pgfonlayer}
\end{tikzpicture}

\end{align}
\end{exa}

We observe that using the formula for the pair-factored form (\ref{eq:derivative_yn}) we obtain a much more compact result than with the inductive procedure (see the example \ref{example:cos_bad}). Even if both approaches lead to diagrams with the same semantics, in practice the diagrams obtained with the formulas (\ref{eq:derivative_yn}) and (\ref{eq:automatic_derivative}) are less verbose. For this reason they are easier to manipulate.
  
\subsection{Discussion}
Contrary to derivatives defined in \cite{zx_differentation_toumi} and \cite{barren_plateau_zx} in our approaches a derivative of a ZX-diagram is another ZX-diagram. 

In the inductive approach we proceed by integration of the product rules using \textit{controlizers}. We also observe that as the derivative of a constant diagram is trivial, it is beneficial to "factor-out" the part that depends on the parameter. We adopted this approach in formulas for diagrams in $\beta$-factored form and pair-factored forms. To derive the formulas, we got the intuition in the desired semantics of the diagrams. Thenceforth, our formulas were rigorously proven by induction. 

A result similar to our simplified formula for $\beta$-factored forms was independently derived in \cite{zx_differention_harny}. The major difference between our formula and the method shown in \cite{zx_differention_harny} is the \textit{considered language}. Indeed, in our work we operate ZX-diagrams while Wang and Yeung consider diagrams from the more expressive \textit{algebraic ZX-calculus}. 

The difficulty to represent derivatives for non-linear diagrams follows from the fact that there is no simple way to represent real numbers in the vanilla ZX-calculus. In the algebraic ZX-calculus this restriction is removed. As a consequence, when an algebraic ZX-diagram is parameterized by an arbitrary derivable function $f(x)$, the differentiated algebraic ZX-diagram is parametrized by $f'(x)$.

\section{Diagrammatic representation of Ising Hamiltonians}\label{section:hamiltonian_diagram}

The addition of diagrams naturally appears in the representation of Hamiltonians. Usually, Hamiltonians are given as a weighted sum of Pauli tensors (see \cite{my_thesis} for an introduction to Hamiltonians):
\begin{align}
H = \sum_{k} \alpha_k \sigma^0_k \otimes \dots \otimes \sigma^n_k, \qquad \alpha_k \in \mathbb{R}, \sigma^q_k =I \text{ or } X \text{ or } Y \text{ or } Z
\end{align}
Pauli tensors have a particularly simple decomposition on elementary matrices. Therefore, the procedure described in \cite{zx_addition_oxford} is well-suited for the construction of diagrammatic representation for such Hamiltonians. 

In addition, the work \cite{zx_addition_oxford} also suggests a way to \textit{exponentiate} diagrams representing Hamiltonians. The exponentiate of a Hamiltonian $H$ is a unitary $e^{i\theta H}$. The exponentiation is done with the aid of the \textit{Cayley-Hamilton theorem}. The Cayley-Hamilton theorem allows to decompose the exponent $e^{i\theta H}$ on a linear combination of $H^k = \underbrace{H \times \dots \times H}_k$ for $k \in [1, \dots, 2^n]$.  

We remark that in a particular case when $H = \sum_i H_i$ is a sum of commuting terms it is usually straightforward to obtain $e^{i\theta H}$ as a product of individual $e^{i \theta H_i}$. However, if some local terms in $H$ don't commute the product decomposition doesn't apply. For instance, this is the case for \textit{modified mixer Hamiltonians} used in the extension of QAOA to constrained optimization problems \cite{QAOA_Anzats}.

In this section, we will also consider an inverse problem. Using \textit{Stone's theorem}, we demonstrate how to obtain a diagram for $H$  given the \textit{derivative} of the diagram for $e^{i \theta H}$. 

The Stone theorem relating the derivative of a unitary group $e^{i\gamma H}$ to its generator $H$ was used before in \cite{zx_differentation_toumi}. However, in \cite{zx_differentation_toumi} the derivatives were defined as formal sums of diagrams. As a consequence, the work \cite{zx_differentation_toumi} didn't suggest a single diagram representation for a Hamiltonian but rather a representation as a sum of diagrams.

We remark that for an Ising Hamiltonian $H$ the diagram $D_{U}(\beta)$ of the linear map $U(\beta) = e^{i \beta H}$ is easy to find. For Hamiltonians with integer coefficients the matrix $U(\beta) = e^{i \beta H}$ belongs to $\Mat(\beta)$. It satisfies the definition of a strongly continuous one-parameter unitary group:

\begin{defi}[Unitary group \cite{zx_differentation_toumi}]\label{def:unitary_group}
A one-parameter unitary group is a unitary matrix $U:n \rightarrow n$ in $\Mat(\beta)$ with $U(0) = id_n$  and $U(\beta)U(\beta') = U(\beta + \beta')$ for all $\beta, \beta' \in \mathbb{R}$. It is strongly continuous when $\lim_{\beta\rightarrow\beta_0}U(\beta) = U(\beta_0)$ for all $\beta_0\in \mathbb{R}$.
\end{defi}

\begin{thm}[Stone (\cite{Stone_theorem})]\label{theorem:stone}
There is a one-to-one correspondence between strongly continuous one-parameter
unitary groups $U: n \rightarrow n$ in $\Mat(\beta)$ and self-adjoint matrices $H:n\rightarrow n$ in $\Mat$. The bijection is given explicitly by $U(\beta) = e^{i\beta H}$ and $H=-i (\partial_M U)(0)$.
\end{thm} 

We use the bijection from Stone's theorem to find the diagram $h \in ZX_{\mathbb{R}}$ such that $\interp{h} = H$. Using the property $U(0) = id_n$ we obtain:
 
\begin{align}
H& = -i [\partial_M U(\beta)](0)= -i \otimes \interp{\left[\partial_{\ZX} D_{U}\right](\beta)}(0) = -i \otimes \interp{\left[\partial_{\ZX} D_{U}\right](0)} \nonumber \\
& = \interp{\minusi \normalizer \left[\partial_{\ZX} D_{U}\right](0)}  = \interp{h}
\end{align}
where the third equality is due to the fact that the evaluation commutes with the standard interpretation.

We give an example of a diagram for an Ising Hamiltonian obtained via our approach.

\begin{exa}
Let $H:2 \rightarrow 2$ be the Hamiltonian given by $H = Z_1 - Z_2 + Z_1 Z_2$. The diagram $D_{U}(\beta)$ for $U(\beta) = e^{i\beta H}$ is:
\begin{align}
D_{U}(\beta) = \tikzformula{h_example/step1} = \tikzformula{h_example/step2}
\end{align}
Using the formula (\ref{eq:automatic_derivative}) we find the derivative of $D_{U(\beta)}$:
\begin{align}
\partial_{\ZX} D_{U(\beta)} &= \tikzformula{h_example/step3} %\nonumber
\end{align}
\vspace{-0.2cm}
\begin{align}
\!\! h &= \minusi \normalizer \left[\partial_{\ZX} D_{U(\beta)}\right]_{\beta \rightarrow 0} = \tikzformula{h_example/result} 
\end{align}
\end{exa}

It is of course possible to obtain a diagram for $H$ using the addition procedure we obtained earlier, rather than going through computing its exponential and then differentiating. However, in the case when $H$ is Ising, the exponential of $H$ is typically easier to obtain than $H$ itself.

\section{Discussions}

In this work,  we have introduced an inductive definition for addition of $\ZX$-diagrams, that we have then used to introduce an inductive definition of the differentiation of $\ZX$-diagrams. Addition and differentiation are essential tools for the development and the study of quantum algorithms, but, as a matter of fact, both of them are leading to large diagrams, even when the initial diagrams are fairly simple. 

In sections \ref{section:diff_with_formulas} and \ref{section:diff_with_formulas_2}, we have shown that instead of simplifying the resulting diagrams \emph{a posteriori}, one can \emph{a priori} put the initial diagrams in an appropriate form. While this approach is not inductive anymore, it seems to ease the differentiation of diagrams in practice. As an application we have shown that our result allows the construction of diagrams for Ising Hamiltonians and for derivatives of parametrized circuits. Therefore, it becomes possible to study variational algorithms entirely within the $\ZX$-calculus. In particular, we can use rewrite rules to simplify such expressions as $\langle\psi(\hat{\beta})|H_f|\psi(\hat{\beta})\rangle$ and $\frac{\partial \langle\psi(\hat{\beta})|H_f|\psi(\hat{\beta})\rangle}{\partial \beta}$. We believe that it will lead to a better understanding of the potential of variational algorithms and of their applications to real-world problems.

\section*{Acknowledgements}

This work was supported in part by the CIFRE EDF/Loria Quantum Computing for
Combinatorial Optimisation, the French National Research Agency (ANR) under the research projects SoftQPro ANR-17-CE25-0009-02 and VanQuTe ANR-17-CE24-0035, 
the PEPR integrated project EPiQ ANR-22-PETQ-0007 part of Plan France 2030, by the STIC-AmSud project Qapla’ 21-STIC-10, and by
the European projects NEASQC (funded from the European Union’s Horizon 2020 research and innovation programme grant agreement No 951821) and HPCQS (European High-Performance Computing Joint Undertaking under grant agreement No 101018180).

\bibliographystyle{alphaurl}
\bibliography{References}

\newcommand{\etalchar}[1]{$^{#1}$}
\begin{thebibliography}{DKPvdW20}

\bibitem[Bac14]{clifford_completeness_Backens}
Miriam Backens.
\newblock The {ZX}-calculus is complete for stabilizer quantum mechanics.
\newblock {\em New Journal of Physics}, 16(9):093021, sep 2014.
\newblock \href {https://doi.org/10.1088/1367-2630/16/9/093021}
  {\path{doi:10.1088/1367-2630/16/9/093021}}.

\bibitem[BK21]{Training_is_hard}
Lennart Bittel and Martin Kliesch.
\newblock Training variational quantum algorithms is {NP}-hard.
\newblock {\em Physical Review Letters}, 127(12), sep 2021.
\newblock \href {https://doi.org/10.1103/physrevlett.127.120502}
  {\path{doi:10.1103/physrevlett.127.120502}}.

\bibitem[CAB{\etalchar{+}}21]{VQA}
M.~Cerezo, Andrew Arrasmith, Ryan Babbush, Simon~C. Benjamin, Suguru Endo,
  Keisuke Fujii, Jarrod~R. McClean, Kosuke Mitarai, Xiao Yuan, Lukasz Cincio,
  and Patrick~J. Coles.
\newblock Variational quantum algorithms.
\newblock {\em Nature Reviews Physics}, 3(9):625--644, Aug 2021.
\newblock \href {https://doi.org/10.1038/s42254-021-00348-9}
  {\path{doi:10.1038/s42254-021-00348-9}}.

\bibitem[CD11]{zx_original}
Bob Coecke and Ross Duncan.
\newblock Interacting quantum observables: categorical algebra and
  diagrammatics.
\newblock {\em New Journal of Physics}, 13(4):043016, apr 2011.
\newblock \href {https://doi.org/10.1088/1367-2630/13/4/043016}
  {\path{doi:10.1088/1367-2630/13/4/043016}}.

\bibitem[CDD{\etalchar{+}}20]{phase_gadget}
Alexander Cowtan, Silas Dilkes, Ross Duncan, Will Simmons, and Seyon Sivarajah.
\newblock Phase gadget synthesis for shallow circuits.
\newblock {\em Electronic Proceedings in Theoretical Computer Science},
  318:213--228, may 2020.
\newblock \href {https://doi.org/10.4204/eptcs.318.13}
  {\path{doi:10.4204/eptcs.318.13}}.

\bibitem[CK17]{dodo_book}
Bob Coecke and Aleks Kissinger.
\newblock {\em Picturing Phases and Complementarity}, pages 510--623.
\newblock Cambridge University Press, 2017.
\newblock \href {https://doi.org/10.1017/9781316219317.010}
  {\path{doi:10.1017/9781316219317.010}}.

\bibitem[CSV{\etalchar{+}}21]{Barren_plateau_nonlocal_cost}
M.~Cerezo, Akira Sone, Tyler Volkoff, Lukasz Cincio, and Patrick~J. Coles.
\newblock Cost function dependent barren plateaus in shallow parametrized
  quantum circuits.
\newblock {\em Nature Communications}, 12(1), mar 2021.
\newblock \href {https://doi.org/10.1038/s41467-021-21728-w}
  {\path{doi:10.1038/s41467-021-21728-w}}.

\bibitem[CW12]{linear_combination}
Andrew~M. {Childs} and Nathan {Wiebe}.
\newblock {Hamiltonian Simulation Using Linear Combinations of Unitary
  Operations}.
\newblock {\em arXiv e-prints}, page arXiv:1202.5822, February 2012.
\newblock \href {http://arxiv.org/abs/1202.5822} {\path{arXiv:1202.5822}}.

\bibitem[dBBW20]{t_count}
Niel de~Beaudrap, Xiaoning Bian, and Quanlong Wang.
\newblock Techniques to reduce $\uppi$/4-parity-phase circuits, motivated by
  the {ZX} calculus.
\newblock {\em Electronic Proceedings in Theoretical Computer Science},
  318:131--149, may 2020.
\newblock \href {https://doi.org/10.4204/eptcs.318.9}
  {\path{doi:10.4204/eptcs.318.9}}.

\bibitem[dBH20]{lattice_surgery_1}
Niel de~Beaudrap and Dominic Horsman.
\newblock The {ZX} calculus is a language for surface code lattice surgery.
\newblock {\em {Quantum}}, 4:218, January 2020.
\newblock \href {https://doi.org/10.22331/q-2020-01-09-218}
  {\path{doi:10.22331/q-2020-01-09-218}}.

\bibitem[DKPvdW20]{pivoting_simplification}
Ross Duncan, Aleks Kissinger, Simon Perdrix, and John van~de Wetering.
\newblock {Graph-theoretic Simplification of Quantum Circuits with the
  {ZX}-calculus}.
\newblock {\em Quantum}, 4:279, jun 2020.
\newblock \href {https://doi.org/10.22331/q-2020-06-04-279}
  {\path{doi:10.22331/q-2020-06-04-279}}.

\bibitem[DP10]{MBQC_circuit}
Ross Duncan and Simon Perdrix.
\newblock Rewriting measurement-based quantum computations with generalised
  flow.
\newblock In Samson Abramsky, Cyril Gavoille, Claude Kirchner, Friedhelm Meyer
  auf~der Heide, and Paul~G. Spirakis, editors, {\em Automata, Languages and
  Programming}, pages 285--296, Berlin, Heidelberg, 2010. Springer Berlin
  Heidelberg.

\bibitem[DP14]{real_clifford_completeness}
Ross Duncan and Simon Perdrix.
\newblock Pivoting makes the {ZX}-calculus complete for real stabilizers.
\newblock {\em Electronic Proceedings in Theoretical Computer Science},
  171:50–62, Dec 2014.
\newblock \href {https://doi.org/10.4204/eptcs.171.5}
  {\path{doi:10.4204/eptcs.171.5}}.

\bibitem[FGG14]{QAOA_original}
Edward Farhi, Jeffrey Goldstone, and Sam Gutmann.
\newblock A quantum approximate optimization algorithm, 2014.
\newblock \href {https://doi.org/10.48550/ARXIV.1411.4028}
  {\path{doi:10.48550/ARXIV.1411.4028}}.

\bibitem[GD18]{color_code}
Liam Garvie and Ross Duncan.
\newblock Verifying the smallest interesting colour code with quantomatic.
\newblock {\em Electronic Proceedings in Theoretical Computer Science},
  266:147--163, feb 2018.
\newblock \href {https://doi.org/10.4204/eptcs.266.10}
  {\path{doi:10.4204/eptcs.266.10}}.

\bibitem[GS17]{parameter_opt_compare}
Gian~Giacomo Guerreschi and Mikhail Smelyanskiy.
\newblock Practical optimization for hybrid quantum-classical algorithms, 2017.
\newblock \href {http://arxiv.org/abs/1701.01450} {\path{arXiv:1701.01450}}.

\bibitem[Had15]{w_spider}
Amar Hadzihasanovic.
\newblock A diagrammatic axiomatisation for qubit entanglement.
\newblock In {\em 2015 30th Annual ACM/IEEE Symposium on Logic in Computer
  Science}, pages 573--584, 2015.
\newblock \href {https://doi.org/10.1109/LICS.2015.59}
  {\path{doi:10.1109/LICS.2015.59}}.

\bibitem[HEMN20]{fault_tolerant_compilation}
Michael Hanks, Marta~P. Estarellas, William~J. Munro, and Kae Nemoto.
\newblock Effective compression of quantum braided circuits aided by
  {ZX}-calculus.
\newblock {\em Physical Review X}, 10(4), nov 2020.
\newblock \href {https://doi.org/10.1103/physrevx.10.041030}
  {\path{doi:10.1103/physrevx.10.041030}}.

\bibitem[HNW18]{completeness_oxford}
Amar Hadzihasanovic, Kang~Feng Ng, and Quanlong Wang.
\newblock Two complete axiomatisations of pure-state qubit quantum computing.
\newblock In {\em Proceedings of the 33rd Annual ACM/IEEE Symposium on Logic in
  Computer Science}, LICS '18, pages 502--511, New York, NY, USA, 2018. ACM.
\newblock \href {https://doi.org/10.1145/3209108.3209128}
  {\path{doi:10.1145/3209108.3209128}}.

\bibitem[Hor11]{surface_code}
Clare Horsman.
\newblock Quantum picturalism for topological cluster-state computing.
\newblock {\em New Journal of Physics}, 13(9):095011, sep 2011.
\newblock \href {https://doi.org/10.1088/1367-2630/13/9/095011}
  {\path{doi:10.1088/1367-2630/13/9/095011}}.

\bibitem[HWO{\etalchar{+}}19]{QAOA_Anzats}
Stuart Hadfield, Zhihui Wang, Bryan O{\textquotesingle}Gorman, Eleanor Rieffel,
  Davide Venturelli, and Rupak Biswas.
\newblock From the quantum approximate optimization algorithm to a quantum
  alternating operator ansatz.
\newblock {\em Algorithms}, 12(2):34, feb 2019.
\newblock \href {https://doi.org/10.3390/a12020034}
  {\path{doi:10.3390/a12020034}}.

\bibitem[ILY21]{generalized_phase_shift_algebraic}
Artur~F. Izmaylov, Robert~A. Lang, and Tzu-Ching Yen.
\newblock Analytic gradients in variational quantum algorithms: Algebraic
  extensions of the parameter-shift rule to general unitary transformations.
\newblock {\em Physical Review A}, 104(6), dec 2021.
\newblock \href {https://doi.org/10.1103/physreva.104.062443}
  {\path{doi:10.1103/physreva.104.062443}}.

\bibitem[JPV18a]{clifford_t_completeness_nancy}
Emmanuel Jeandel, Simon Perdrix, and Renaud Vilmart.
\newblock {A Complete Axiomatisation of the {ZX}-Calculus for Clifford+T
  Quantum Mechanics}.
\newblock In {\em {The 33rd Annual \{ACM/IEEE\} Symposium on Logic in Computer
  Science, \{LICS\} 2018}}, Proceedings of the 33rd Annual ACM/IEEE Symposium
  on Logic in Computer Science, pages 559--568, Oxford, United Kingdom, July
  2018.
\newblock \href {https://doi.org/10.1145/3209108.3209131}
  {\path{doi:10.1145/3209108.3209131}}.

\bibitem[JPV18b]{completeness_nancy}
Emmanuel Jeandel, Simon Perdrix, and Renaud Vilmart.
\newblock {Diagrammatic Reasoning beyond Clifford+T Quantum Mechanics}.
\newblock In {\em {The 33rd Annual Symposium on Logic in Computer Science}},
  Proceedings of the 33rd Annual ACM/IEEE Symposium on Logic in Computer
  Science, pages 569--578, Oxford, United Kingdom, July 2018.
\newblock \href {https://doi.org/10.1145/3209108.3209139}
  {\path{doi:10.1145/3209108.3209139}}.

\bibitem[JPV19]{normal_form_nancy}
Emmanuel Jeandel, Simon Perdrix, and Renaud Vilmart.
\newblock A generic normal form for {ZX}-diagrams and application to the
  rational angle completeness.
\newblock In {\em Proceedings of the 34th Annual ACM/IEEE Symposium on Logic in
  Computer Science}, LICS '19. IEEE Press, 2019.

\bibitem[JPV22]{my_paper_2}
Emmanuel {Jeandel}, Simon {Perdrix}, and Margarita {Veshchezerova}.
\newblock {Addition and Differentiation of {ZX}-diagrams}.
\newblock {\em arXiv e-prints}, page arXiv:2202.11386, February 2022.
\newblock \href {http://arxiv.org/abs/2202.11386} {\path{arXiv:2202.11386}}.

\bibitem[JPVW17]{jeandel2017cyclotomic}
Emmanuel Jeandel, Simon Perdrix, Renaud Vilmart, and Quanlong Wang.
\newblock {ZX-Calculus: Cyclotomic Supplementarity and Incompleteness for
  Clifford+T Quantum Mechanics}.
\newblock In Kim~G. Larsen, Hans~L. Bodlaender, and Jean-Francois Raskin,
  editors, {\em 42nd International Symposium on Mathematical Foundations of
  Computer Science (MFCS 2017)}, volume~83 of {\em Leibniz International
  Proceedings in Informatics (LIPIcs)}, pages 11:1--11:13, Dagstuhl, Germany,
  2017. Schloss Dagstuhl--Leibniz-Zentrum fuer Informatik.
\newblock \href {https://doi.org/10.4230/LIPIcs.MFCS.2017.11}
  {\path{doi:10.4230/LIPIcs.MFCS.2017.11}}.

\bibitem[KB22]{QAD}
B{\'{a}}lint Koczor and Simon~C. Benjamin.
\newblock Quantum analytic descent.
\newblock {\em Physical Review Research}, 4(2), apr 2022.
\newblock \href {https://doi.org/10.1103/physrevresearch.4.023017}
  {\path{doi:10.1103/physrevresearch.4.023017}}.

\bibitem[KE21]{generalized_phase_shift_kyriienko}
Oleksandr Kyriienko and Vincent~E. Elfving.
\newblock Generalized quantum circuit differentiation rules.
\newblock {\em Phys. Rev. A}, 104:052417, Nov 2021.
\newblock \href {https://doi.org/10.1103/PhysRevA.104.052417}
  {\path{doi:10.1103/PhysRevA.104.052417}}.

\bibitem[KvdW19]{MBQC_universality}
Aleks Kissinger and John van~de Wetering.
\newblock Universal {MBQC} with generalised parity-phase interactions and pauli
  measurements.
\newblock {\em Quantum}, 3:134, apr 2019.
\newblock \href {https://doi.org/10.22331/q-2019-04-26-134}
  {\path{doi:10.22331/q-2019-04-26-134}}.

\bibitem[KvdW20]{t_count_2}
Aleks Kissinger and John van~de Wetering.
\newblock {Reducing the number of non-Clifford gates in quantum circuits}.
\newblock {\em Physical Review A}, 102(2), aug 2020.
\newblock \href {https://doi.org/10.1103/physreva.102.022406}
  {\path{doi:10.1103/physreva.102.022406}}.

\bibitem[KvdWK19]{Toffoli_t_count}
Stach Kuijpers, John van~de Wetering, and Aleks Kissinger.
\newblock Graphical fourier theory and the cost of quantum addition, 2019.
\newblock \href {https://doi.org/10.48550/ARXIV.1904.07551}
  {\path{doi:10.48550/ARXIV.1904.07551}}.

\bibitem[MBK21]{finite_difference_vs_phase_shift}
Andrea Mari, Thomas~R. Bromley, and Nathan Killoran.
\newblock Estimating the gradient and higher-order derivatives on quantum
  hardware.
\newblock {\em Physical Review A}, 103(1), jan 2021.
\newblock \href {https://doi.org/10.1103/physreva.103.012405}
  {\path{doi:10.1103/physreva.103.012405}}.

\bibitem[MBS{\etalchar{+}}18]{Barren_plateaus_random}
Jarrod~R. McClean, Sergio Boixo, Vadim~N. Smelyanskiy, Ryan Babbush, and
  Hartmut Neven.
\newblock Barren plateaus in quantum neural network training landscapes.
\newblock {\em Nature Communications}, 9(1), nov 2018.
\newblock \href {https://doi.org/10.1038/s41467-018-07090-4}
  {\path{doi:10.1038/s41467-018-07090-4}}.

\bibitem[MNKF18]{phase_shift_original}
K.~Mitarai, M.~Negoro, M.~Kitagawa, and K.~Fujii.
\newblock Quantum circuit learning.
\newblock {\em Physical Review A}, 98(3), sep 2018.
\newblock \href {https://doi.org/10.1103/physreva.98.032309}
  {\path{doi:10.1103/physreva.98.032309}}.

\bibitem[NW18]{clifford_t_completeness_oxford}
Kang~Feng Ng and Quanlong Wang.
\newblock Completeness of the {ZX}-calculus for pure qubit {Clifford+T} quantum
  mechanics, 2018.
\newblock \href {http://arxiv.org/abs/1801.07993} {\path{arXiv:1801.07993}}.

\bibitem[PMS{\etalchar{+}}14]{VQE_original}
Alberto Peruzzo, Jarrod McClean, Peter Shadbolt, Man-Hong Yung, Xiao-Qi Zhou,
  Peter~J. Love, Al{\'{a}}n Aspuru-Guzik, and Jeremy~L. O'Brien.
\newblock A variational eigenvalue solver on a photonic quantum processor.
\newblock {\em Nature Communications}, 5(1), jul 2014.
\newblock \href {https://doi.org/10.1038/ncomms5213}
  {\path{doi:10.1038/ncomms5213}}.

\bibitem[Pre18]{Preskill_NISQ}
John Preskill.
\newblock Quantum computing in the {NISQ} era and beyond.
\newblock {\em Quantum}, 2:79, aug 2018.
\newblock \href {https://doi.org/10.22331/q-2018-08-06-79}
  {\path{doi:10.22331/q-2018-08-06-79}}.

\bibitem[SBG{\etalchar{+}}19]{phase_shift_basic}
Maria Schuld, Ville Bergholm, Christian Gogolin, Josh Izaac, and Nathan
  Killoran.
\newblock Evaluating analytic gradients on quantum hardware.
\newblock {\em Physical Review A}, 99(3), mar 2019.
\newblock \href {https://doi.org/10.1103/physreva.99.032331}
  {\path{doi:10.1103/physreva.99.032331}}.

\bibitem[SH22]{tobias}
Tobias {Stollenwerk} and Stuart {Hadfield}.
\newblock {Diagrammatic Analysis for Parameterized Quantum Circuits}.
\newblock {\em arXiv e-prints}, page arXiv:2204.01307, April 2022.
\newblock \href {http://arxiv.org/abs/2204.01307} {\path{arXiv:2204.01307}}.

\bibitem[Sto32]{Stone_theorem}
M.~H. Stone.
\newblock On one-parameter unitary groups in hilbert space.
\newblock {\em Annals of Mathematics}, 33(3):643--648, 1932.
\newblock URL: \url{http://www.jstor.org/stable/1968538}.

\bibitem[TYd21]{zx_differentation_toumi}
Alexis {Toumi}, Richie {Yeung}, and Giovanni {de Felice}.
\newblock Diagrammatic differentiation for quantum machine learning.
\newblock {\em arXiv e-prints}, page arXiv:2103.07960, March 2021.
\newblock \href {http://arxiv.org/abs/2103.07960} {\path{arXiv:2103.07960}}.

\bibitem[vdW20]{zx_introduction}
John van~de Wetering.
\newblock {ZX}-calculus for the working quantum computer scientist, 2020.
\newblock \href {http://arxiv.org/abs/2012.13966} {\path{arXiv:2012.13966}}.

\bibitem[Ves22]{my_thesis}
Margarita Veshchezerova.
\newblock {\em {Quantum algorithms for energy management optimization
  problems}}.
\newblock Theses, {Universit{\'e} de Lorraine}, December 2022.
\newblock URL: \url{https://hal.univ-lorraine.fr/tel-04105922}.

\bibitem[Vil19]{vilmart_optimal_axioms}
Renaud Vilmart.
\newblock A near-optimal axiomatisation of {ZX}-calculus for pure qubit quantum
  mechanics.
\newblock In {\em Proceedings of the 34th Annual ACM/IEEE Symposium on Logic in
  Computer Science (LICS)}, 2019.
\newblock URL: \url{https://arxiv.org/abs/1812.09114}, \href
  {http://arxiv.org/abs/arXiv:1812.09114} {\path{arXiv:arXiv:1812.09114}}.

\bibitem[Wan20]{normal_form_algebraic}
Quanlong Wang.
\newblock Algebraic complete axiomatisation of {ZX}-calculus with a normal form
  via elementary matrix operations, 2020.
\newblock \href {https://doi.org/10.48550/ARXIV.2007.13739}
  {\path{doi:10.48550/ARXIV.2007.13739}}.

\bibitem[WIWL22]{generalized_phase_shift_pennylane}
David Wierichs, Josh Izaac, Cody Wang, and Cedric Yen-Yu Lin.
\newblock General parameter-shift rules for quantum gradients.
\newblock {\em {Quantum}}, 6:677, March 2022.
\newblock \href {https://doi.org/10.22331/q-2022-03-30-677}
  {\path{doi:10.22331/q-2022-03-30-677}}.

\bibitem[WSW{\etalchar{+}}21]{Meta-learning-for-parameters}
Max Wilson, Rachel Stromswold, Filip Wudarski, Stuart Hadfield, Norm Tubman,
  and Eleanor Rieffel.
\newblock Optimizing quantum heuristics with meta-learning.
\newblock {\em Quantum Machine Intelligence}, 3, 06 2021.
\newblock \href {https://doi.org/10.1007/s42484-020-00022-w}
  {\path{doi:10.1007/s42484-020-00022-w}}.

\bibitem[WY21]{elementary_matrix}
Quanlong Wang and Richie Yeung.
\newblock Representing and implementing matrices using algebraic {ZX}-calculus,
  2021.
\newblock \href {https://doi.org/10.48550/ARXIV.2110.06898}
  {\path{doi:10.48550/ARXIV.2110.06898}}.

\bibitem[WY22]{zx_differention_harny}
Quanlong Wang and Richie Yeung.
\newblock Differentiating and integrating {ZX} diagrams, 2022.
\newblock \href {http://arxiv.org/abs/2201.13250} {\path{arXiv:2201.13250}}.

\bibitem[YWS]{zx_addition_oxford}
Richie Yeung, Quanlong Wang, and Razin~A. Shaikh.
\newblock {\em How to sum and exponentiate Hamiltonians in {ZXW} calculus}.
\newblock URL:
  \url{https://www.qplconference.org/proceedings2022/QPL_2022_paper_85.pdf}.

\bibitem[ZG21]{barren_plateau_zx}
Chen Zhao and Xiao-Shan Gao.
\newblock Analyzing the barren plateau phenomenon in training quantum neural
  networks with the {ZX}-calculus.
\newblock {\em {Quantum}}, 5:466, June 2021.
\newblock \href {https://doi.org/10.22331/q-2021-06-04-466}
  {\path{doi:10.22331/q-2021-06-04-466}}.

\end{thebibliography}

\end{document}